\documentclass[galley,usenatbib]{mn2e}
\usepackage{myaasmacros}
\usepackage{mathrsfs}
\usepackage{amsmath}
\usepackage{graphicx}

\renewcommand{\vec}[1]{\ensuremath{\boldsymbol{#1}}}

\newcommand{\Glass}{{\sc Glass}}
\newcommand{\PixeLens}{{\sc PixeLens}}
\newcommand{\Rmap}{\ensuremath{R_\mathrm{map}}}
\newcommand{\Rpix}{\ensuremath{R_\mathrm{pix}}}
\newcommand{\M}{\ensuremath{\mathscr{M}}}
\newcommand{\E}{\ensuremath{\mathscr{E}}}
\newcommand{\eps}{\ensuremath{\varepsilon}}

\newcommand{\Mddd}{\ensuremath{M}}
\newcommand{\pixrad}{\ensuremath{\mathrm{\tt pixrad}}}
\newcommand{\url}[1]{\tt #1}

\newcommand{\Msun}{\ensuremath{\mathrm{M}_\odot}}
\newcommand{\tabref}[1] {Table~\ref{#1}}
\newcommand{\figref}[1] {Figure~\ref{#1}}
\newcommand{\eqnref}[1] {Eq.~(\ref{#1})}
\newcommand{\eqnrefp}[1] {(Eq.~\ref{#1})}
\newcommand{\secref}[1] {\S\ref{#1}}
\newcommand{\appref}[1] {Appendix~\ref{#1}}
\newcommand{\e}[1]{\ensuremath{\times 10^{#1}}}

\newcommand{\mockAA}{{\sc star1.0-dmCore}}
\newcommand{\mockAC}{{\sc star1.0-dmCusp}}
\newcommand{\mockBB}{{\sc star1.5-dmCore}}
\newcommand{\mockBC}{{\sc star1.5-dmCusp}}

\newcommand{\cth}[1][]{\cos^{#1}\theta}
\newcommand{\sth}[1][]{\sin^{#1}\theta}
\def\dth{d\theta}

\newcommand\plotthree[3]{{%
 \centering
 \leavevmode
 \columnwidth=.30\textwidth
 \includegraphics[width={\columnwidth}]{#1}%
 \hfil
 \includegraphics[width={\columnwidth}]{#2}%
 \hfil
 \includegraphics[width={\columnwidth}]{#3}%
}}%

\title[Lens Recovery with \Glass]{Gravitational Lens Recovery with \Glass: Measuring the mass profile and shape of a lens}
\author[J.~P.~Coles~et~al.]{%
Jonathan P.~Coles,$^{1,3}$\thanks{jonathan@exascale-computing.eu}%
\newauthor%
Justin~I.~Read$^2$,%
\newauthor%
and Prasenjit~Saha$^3$%
\\
$^1$Exascale Research Computing Lab, Campus Teratec, 2 Rue de la Piquetterie, 91680 Bruyeres-le-Chatel, France \\
$^2$Department of Physics, University of Surrey, Guildford, Surrey, GU2 7XH, United Kingdom \\
$^3$Physik-Institut, Universit\"at Z\"urich, 190 Winterthurerstrasse, 8057, Z\"urich, Switzerland
}

\begin{document}
\maketitle

\begin{abstract}
We use a new non-parametric gravitational modelling tool -- \Glass{} -- to
determine what quality of data (strong lensing, stellar kinematics, and/or
stellar masses) are required to measure the circularly averaged mass profile of
a lens and its shape. \Glass{} uses an under-constrained adaptive grid of mass
pixels to model the lens, searching through thousands of models to marginalise
over model uncertainties. Our key findings are as follows: (i) for pure lens
data, multiple sources with wide redshift separation give the strongest
constraints as this breaks the well-known mass-sheet or steepness degeneracy;
(ii) a single quad with time delays also performs well, giving a good recovery
of both the mass profile and its shape; (iii) stellar masses -- for lenses
where the stars dominate the central potential -- can also break the steepness
degeneracy, giving a recovery for doubles almost as good as having a quad with time delay data, or
multiple source redshifts; (iv) stellar kinematics provide a robust measure of
the mass at the half light radius of the stars $r_{1/2}$ that can also break
the steepness degeneracy if the Einstein radius $r_E \neq r_{1/2}$; and (v)
if $r_E \sim r_{1/2}$, then stellar kinematic data can be used to probe the
stellar velocity anisotropy $\beta$ -- an interesting quantity in its own
right. Where information on the mass distribution from lensing and/or other
probes becomes redundant, this opens up the possibility of using strong lensing
to constrain cosmological models. 
\end{abstract}


\section{Introduction}\label{sec:intro}

Strong gravitational lenses are rare. Since the discovery of the first lens
Q0957+561~\citep{1979Natur.279..381W}, just $\sim400$ have been discovered to
date\footnote{See, e.g., {\url{http://masterlens.astro.utah.edu}} for a
catalogue.}.  However, this number is expected to increase to several thousand
over the next ten years as new surveys, both
ground-based\footnote{\url{http://pan-starrs.ifa.hawaii.edu}}$^{\rm,}$\footnote{\url{http://www.darkenergysurvey.org}}
and space-based\footnote{\url{http://www.euclid-ec.org}} -- together with a
community of citizen-science volunteers examining the image data for
candidates\footnote{\url{http://spacewarps.org}} -- come online. 

Since lensing depends only on gravity, strong lenses offer a unique window onto
dark matter and cosmology \citep{2010CQGra..27w3001B,2013LRR....16....6A}.
However, extracting dark matter properties or cosmological constraints from
these lensing data will require sophisticated modelling. In particular, with an
unprecedented data set imminent, it is prudent to look again at systematic
errors in the lens models to determine what quality of data (in particular
complementary data from stellar/gas kinematics, lens time delays and/or stellar
mass constraints) are required to address problems of interest. It is towards
that goal that this present work is directed.

To see why lens modelling details are of crucial importance, let us recall the
essential quantities that appear in lensing (see also \S\ref{sec:theory} for a
more detailed exposition). First we have the distances. Let $D_L$, $D_S$,
$D_{LS}$ be the angular-diameter distances to the lens, source, and from lens
to source; these are all proportional to $c/H_0$ but have factors that depend
on the particular choice of cosmology\footnote{Here, $c$ is the speed of light
in vacuo and $H_0$ is the Hubble parameter.}.  Typically:
\begin{equation}
D_L \approx z_L \frac c{H_0}\ \hbox{and}\ \frac{D_S}{D_{LS}} \sim 1.
\end{equation}
where $z_L$ is the redshift of the lens. 
For multiple images, the sky-projected density must exceed the critical lensing
density in some region:
\begin{equation}
\Sigma_\mathrm{crit} = \frac{c^2}{4\pi G D_L} \sim \frac{1\rm\,kg\,m^{-2}}{z_L}
\end{equation}
where $G$ is Newton's gravitational constant. 
The angular separation between the lensed images is of order
the Einstein radius $\theta_E$, which is related to
the mass by:
\begin{equation}
\theta_E \sim \frac{R_G}{D_L} \frac{D_{LS}}{D_S}
\end{equation}
where $R_G = GM/c^2$ (with $M$ the projected mass enclosed within $\theta_E$)
is the gravitational radius. If the source is a quasar or otherwise rapidly
variable, a time delay $\Delta t$ in the variability will be present where:
\begin{equation}
\Delta t \sim R_G/c
\end{equation}
So in principle, one can not only measure the mass of the lens, one can use the
dependence on the cosmology-dependent $D$ factors to extract the cosmological
model and all its parameters.  \cite{1937ApJ....86..217Z} drew attention to the
former, and \cite{1964MNRAS.128..307R,1966MNRAS.132..101R} pointed out the
latter, all long before lenses were discovered.  The difficulty with actually
doing this, however, became apparent soon after the discovery of the first lens
by \cite{1979Natur.279..381W}.  In the first ever paper on lens modelling,
\cite{1981ApJ...244..736Y} found that many plausible mass distributions could
reproduce the data.

\cite{1981ApJ...244..736Y} were remarkably prescient about the subsequent
development of lens modelling.  First, they introduced the technique of
choosing a parametric form for the lensing mass and then fitting for the
parameters, which is still the most common strategy \citep[see for
example][]{2010GReGr..42.2151K,2011A&ARv..19...47K}.  Second, they pointed out
the non-uniqueness of lens models -- lensing degeneracies.  Third, they
suggested combining lensing data with stellar kinematics and X-rays, to reduce
the effect of the degeneracies.  Later work, as well as following up these
suggestions, has introduced some further new ideas.  Five of these are
important for the present work:
\begin{enumerate}
\item {\bf Free-form modelling:} In `free-form' or non-parametric
  modelling, there is no specified parametric form for the mass
  distribution.  There are still assumptions (or priors) on the mass
  distribution, such as smoothness or being centrally concentrated
  \citep{1997MNRAS.292..148S,2005MNRAS.360..477D,2009A&A...500..681M,2010ApJ...723.1678C}
  but these are much less restrictive than parametric forms.  A
  particularly elegant prior is implemented by
  \cite{2006MNRAS.367.1209L}, requiring that the mass distribution to
  be non-negative and no extra images allowed. To be concrete, we
  define from here on:
  
  \begin{quote}
  {\it Non-parametric, or `free-form' $\equiv$ more parameters than data constraints (i.e. deliberately under-constrained)}
  \end{quote}
  
  Being under-constrained, it is then {\it necessary} to explore model
  degeneracies rather than finding a single `best-fit'
  solution. Free-form models are more commonly used with cluster
  lenses
  \citep{2006ApJ...652L...5S,2009ApJ...690..154S,2009A&A...500..681M,2014MNRAS.437.2642S},
  but can be used with galaxy lenses as well, where their less
  restrictive assumptions can be important.  For example, in
  time-delay galaxy lenses, parametric model measures of the Hubble parameter $H_0$ have historically been at 
  tension with independent measures 
  \citep[e.g.,][]{2002astro.ph..4043K,2002ApJ...578...25K}; these are
  resolved once the less restrictive assumptions of free-form models
  are permitted \citep{2007ApJ...667..645R}.  Hybrid methods, using a
  mass grid on top a parametric model, have also been explored
  \citep[e.g.,][]{2010MNRAS.408.1969V}.

\item {\bf Model ensembles:} Model ensembles, exploring a diverse range of
    possible mass distributions that nonetheless all fit the data, are a way of
    combating the non-uniqueness of models.  Such ensembles are possible in
    parametric models \citep[e.g.,][]{1999AJ....118...14B,2010Sci...329..924J,
    2014MNRAS.444..268R,2014arXiv1405.0222J,2014arXiv1405.0011C}, but are more
    common in free-form models, where -- since such models are deliberately
    under-constrained -- they become vital
    \citep{2000AJ....119..439W,2009ApJ...690..154S,2012MNRAS.425.3077L}.

\item {\bf Stellar kinematic constraints:} This was first suggested by
  \citet{2002MNRAS.337L...6T} as a means to break lensing
  degeneracies. The idea is that stellar kinematics can provide an
  independent estimate of the Einstein radius, via the virial theorem:
\begin{equation}
  \frac{\langle v^2_\mathrm{los}\rangle}{c^2} \approx
  \frac{\theta_E}{6\pi} \frac{D_S}{D_{LS}}
\label{eqn:virial}
\end{equation}
where $\langle v^2_\mathrm{los}\rangle$ is the line of sight stellar velocity
dispersion, and the above relation becomes exact for isothermal lenses. This
can then be used to probe cosmological parameters if lenses are known to be
isothermal \citep[e.g.,][]{2012MNRAS.424.2864C}; or to break the steepness
degeneracy in the more general situation (see \S\ref{sec:kinematics}). The
technique has since been applied to many lenses
\citep[e.g.,][]{2006ApJ...649..599K,2008ApJ...682..964B}.  Going further, the
use of two-dimensional kinematics \citep{2011MNRAS.415.2215B} is especially
interesting.

\item {\bf Stellar mass constraints:} The stellar mass in a lens can be
    inferred from photometry and compared with the total mass
    \citep[e.g.,][]{1998ApJ...509..561K,2000ApJ...543..131K,2003ApJ...587..143R,2005ApJ...623L...5F,2008MNRAS.383..857F,2011ApJ...740...97L}.
    Since the inferred stellar mass depends on the assumed IMF, lenses in which
    stellar mass dominates can be used to derive upper bounds on the stellar
    $M/L$ \citep{2010MNRAS.409L..30F}. Lower bounds on stellar $M/L$ have also
    recently been claimed by fitting $\Lambda$CDM semi-analytic models to the
    tilt of the fundamental plane \citep{2013MNRAS.428.3183D}. 
    
\item {\bf Testing modelling strategies:} Using mock data to see how well a
    given model can recover simulated lenses is increasingly being recognised
    as essential.  Simple blind tests have appeared in earlier work \citep[for
    example, Figure 2 in][]{2000AJ....119..439W}, but more recently, tests
    against dynamically simulated galaxies or clusters are favoured
    \citep{2007ApJ...667..645R,2007MNRAS.380.1729L,2009A&A...500..681M,2009MNRAS.393.1114B,2010ApJ...723.1678C}.
\end{enumerate}

There are three further key modelling ideas in the literature that we will not
touch upon in this present work: to use X-ray intensity and temperature
profiles as a mass constraint \citep[e.g.,][]{2013ApJ...765...25N}; and to model
multiple lenses simultaneously, with one or more cosmological parameters
variable but shared between the lenses. This latter strategy has been used to
constrain $H_0$ from time delay lenses
\citep{2006ApJ...652L...5S,2008ApJ...679...17C,2010ApJ...712.1378P} and
recently the cosmological parameters $\Omega$ as well
\citep{2014MNRAS.437..600S}.  Third, it is in principle possible to estimate
the $\Omega$ parameters even from a single lens, if there are lensed sources at
multiple redshifts \citep{2014MNRAS.437.2461L} or by using additional priors 
\citep{2010Sci...329..924J,2014ApJ...788L..35S}.

In this paper, we introduce a new non-parametric lens modelling framework  --
\Glass{} (Gravitational Lensing AnalysiS Software). This shares some aspects
with an earlier code \PixeLens{}~\citep{Saha2004,2008ApJ...679...17C}. However,
\Glass{} -- which contains all new code written from the ground up --
significantly improves upon \PixeLens{} in several key ways: 

\begin{enumerate}
\item At the heart of \Glass{} is a new uniform sampling algorithm for high
    dimensional spaces \citep{2012MNRAS.425.3077L}. This allows for large
    ensembles of $>10,000$ models to be efficiently generated. 
\item \Glass{} provides a modular framework that allows new priors to be added
    and modified easily.
\item The basis functions approximating a model can be easily changed (in this
    paper, we assume pixels as in \PixeLens). 
\item With so many models in the final ensemble, we can afford to apply
    non-linear constraints (for example stellar kinematic data; or the removal of
    models with spurious extra images) to accept/reject models in a post-processing
    step.
\item The central region of the mass map can have a higher resolution to more
    efficiently capture steep models.
\item Stellar density can be used as an additional constraint on the models. 
\item Point or extended mass objects can be placed in the field.
\end{enumerate}
As a first application, we use \Glass{} on mock data to determine which
combination of lensing, stellar mass and/or stellar kinematic constraints best
constrain the projected mass profile and shape of a gravitational lens. We will
apply \Glass{} to real lens data in a series of forthcoming papers. 

This paper is organised as follows. In \secref{sec:glass}, we describe the
\Glass{} code. In \secref{sec:theory}, we review the key elements of lensing
theory, stellar population synthesis, and stellar dynamics we will need. In
\secref{sec:mockdata}, we describe our mock data. In \secref{sec:results}, we
present our results from applying \Glass{} to these mock data. Finally, in
\secref{sec:conclusions} we present our conclusions. 

\section{Theory}\label{sec:theory}

\subsection{Lensing essentials}\label{sec:lensing_basic}

In the following summary, we follow \cite{1986ApJ...310..568B} with
some differences in notation, in particular putting back the
speed of light $c$ and the gravitational constant $G$.

The lens equation:
\begin{equation}
\begin{aligned}
    \vec\beta &= \vec\theta - \frac{D_{LS}}{D_S}\vec\alpha(\vec\theta) \\
\vec\alpha(\vec\theta) &= \frac{4G}{c^2D_L} \int \Sigma(\vec\theta')
                          \frac{(\vec\theta - \vec\theta')}
                          {\ |\vec\theta - \vec\theta'|^2} \, d^2\vec\theta'
\end{aligned}
\label{eqn:lens_equation}
\end{equation}
maps an observed image position $\vec\theta$ to a source position
$\vec\beta$.  Using the thin lens approximation, the lens can be thought of as
a projected surface density $\Sigma$ which diverts the path of a photon
instantaneously through the bending angle $\vec\alpha$.  The $D$ factors, as in
the previous section, are angular diameter distances, which depend on the
cosmological density-parameters $\Omega$, the redshifts $z_L,z_S$ of the lens
and the source, and the Hubble parameter $H_0$, thus
\begin{equation}
D_{LS} = \frac c{H_0} \frac{1+z_S}{1+z_L} \int_{z_L}^{z_S}
                      \frac{dz}{\sqrt{\Omega_m(1+z)^3 + \Omega_\Lambda}}
\end{equation}
and $D_L \equiv D_{0,L}$, $D_S \equiv D_{0,S}$.  One way to understand the lens
equation is via Fermat's principle. We can think of light as travelling only
along extremum paths where lensed images occur.  Such paths occur at the
extrema of the photon {\it arrival time} $t(\vec\theta)$ that depends on the
geometric path the photon takes and the general relativistic gravitational time
dilation due to a thin lens at redshift $z_L$:
\begin{equation}
\begin{aligned}
\frac{ct(\vec\theta)}{(1+z_L)D_{L}}
&= {\textstyle\frac12} |\vec\theta - \vec\beta|^2
   \cdot \frac{D_{S}}{D_{LS}} \\
&- \frac{4GD_L}{c^2}
   \int \Sigma(\vec\theta') \ln |\vec\theta-\vec\theta'| \, d^2\vec\theta'
\label{full arrival time}
\end{aligned}
\end{equation}
We can simplify the above equation by introducing a dimensionless time $\tau$
and density $\kappa$: 
\begin{equation}
\tau(\vec\theta) = \frac{ct(\vec\theta)}{(1+z_L)D_{L}} \quad ; \quad 
\kappa(\vec\theta) \equiv \frac{\Sigma(\vec\theta)}{\Sigma_\mathrm{crit}}
\end{equation}
and hence rewrite \eqnref{full arrival time} as:
\begin{equation}
\tau(\vec\theta) = {\textstyle\frac12} |\vec\theta - \vec\beta|^2
                   \cdot \frac{D_{S}}{D_{LS}}
                 - \frac1\pi \int \kappa (\vec\theta')
                   \ln|\vec\theta - \vec\theta'| d^2\vec\theta'
\label{arrival time2}
\end{equation}
The scaled arrival time $\tau$ is like a solid angle. It is of order the area
(in steradians) of the full lensing system. The expression $|\vec\theta -
\vec\beta|^2$ is of order the image-separation squared, and the other terms are
of similar size.  For this reason, is convenient to measure $\tau$ in
arcsec$^{2}$.

Lensing observations provide information only at $\vec\theta$ where there are
images.  Hence, the arrival-time surface $\tau(\vec\theta)$ is not itself
observable.  Its usefulness lies in that observables can be derived from it.
An image observed at $\vec\theta_1$ implies that $\nabla\tau(\vec\theta_1)=0$.
A measurement of time delays between images at $\theta_1$ and $\theta_2$
implies that $t(\vec\theta_1)-t(\vec\theta_2)$ is known.  Interestingly, both
these types of observations give constraints that are linear in $\kappa$ and
$\vec\beta$.

The rather complicated dependence of lensing observables on the mass
distribution $\kappa(\vec\theta)$ has an important consequence: very different
mass distributions can result in similar observables.  This is the phenomenon
of lensing degeneracies.  While the non-uniqueness of lens models noted by
\cite{1981ApJ...244..736Y} already hinted at degeneracies, their existence was
first derived by \cite{1985ApJ...289L...1F}.  The most important is the
so-called mass-sheet degeneracy, which is that image positions remain invariant
if $\tau(\vec\theta)$ is multiplied by an arbitrary constant.  
This corresponds to rescaling the surface density at the images
$\kappa(\vec\theta)$. In fact there are infinitely many degeneracies
\citep{2000AJ....120.1654S} because any transformation of the arrival-time
surface away from the images has no effect on the lensing observables.  In
particular, there are degeneracies that involve the shape of the mass
distribution \citep{2006ApJ...653..936S,2014A&A...564A.103S}.  Degeneracies
tend to be suppressed if there are sources at very different redshifts or
`redshift contrast' \citep{1998AJ....116.1541A,2009ApJ...690..154S}, because
the presence of different factors of $D_S/D_{LS}$ in the image plane makes it
more difficult to change the mass distribution and the arrival-time surface
without affecting the lensing observables. But degeneracies are still present
with multiple source redshifts \citep{2008MNRAS.386..307L,2014A&A...568L...2S}.

\subsection{Stellar populations} 

For many galaxy lenses, the gravitational potential in the inner region is
dominated by the stellar mass.  Stellar mass can be estimated by combining
photometry and colours with models of the stellar populations.  Such estimates
are reasonably robust, even if the star-formation history is very uncertain:
given a stellar-population model \citep[such as][]{2003MNRAS.344.1000B} and an
initial mass function (IMF), the stellar mass can be inferred to 0.1 to 0.2 dex
using just two photometric bands \citep[see, e.g., Figure~1 in][]{2008MNRAS.383..857F}.
By comparing the lensing-mass and stellar-mass profiles in elliptical galaxies,
it is possible to extract the radial dependence of the baryonic vs dark-matter
fraction \citep{2005ApJ...623L...5F,2008MNRAS.383..857F,2011ApJ...740...97L}.

The major uncertainty at present in the stellar mass is probably the IMF.  In
the lensing galaxy of the Einstein Cross, the IMF cannot be much more
bottom-heavy than \cite{2003PASP..115..763C}, because otherwise the stellar
mass would exceed the lensing mass \cite{2010MNRAS.409L..30F}.  More massive
galaxies, however, do appear to have more of their stellar mass in low-mass
stars.  This is indicated by molecular spectral features characteristic of low
mass stars
\citep{2004ApJ...614L.101C,2012ApJ...747...69C,2013MNRAS.429L..15F}.
The \cite{2003PASP..115..763C} IMF would, however, still provide a robust
lower limit on the stellar mass and hence, also a limit on the total
mass.  Accordingly, \Glass{} allows a constraint of the form
\begin{equation} 
M(\vec\theta) \geq M_\mathrm{stel}(\vec\theta)
\end{equation} 
on the total mass.

\subsection{Stellar kinematics}\label{sec:kinematics} 

Another useful constraint follows from the velocity of stars within the lensing
galaxy. Assuming spherical symmetry, stars obey the projected Jeans equations
\citep[e.g.,][]{2008gady.book.....B}: 

\begin{equation}
\sigma_p^2(R) = \frac{2}{I(R)}\int_R^\infty dr \left(1-\beta \frac{R^2}{r^2}\right) \frac{\nu \sigma_r^2 r}{\sqrt{r^2 - R^2}};
\label{eqn:sphericaljeans}
\end{equation}
\begin{equation} 
\sigma_r^2(r) = \frac{r^{-2\beta}}{\nu}\int_r^\infty r'^{2\beta} \nu \frac{G\Mddd(r')}{r'^2}dr'
\end{equation} 
where $\sigma_p$ is the projected velocity dispersion of the stars as a
function of projected radius $R$; $I(R)$ is the surface density of the stars;
$\nu(r)$ is the three dimensional stellar density; $\sigma_{r,t}(r)$ are the
radial and tangential velocity dispersions, respectively; $\beta(r) = 1 -
\sigma_t^2/2\sigma_r^2 = \mathrm{const.}$ is the velocity anisotropy (here
assumed to be constant, and not to be confused with $\vec\beta(\vec\theta)$
from lensing); $G$ is Newton's gravitational constant; and $\Mddd(r)$ is the
mass profile that we would like to measure. By convention, we always write $R$
for a projected radius, and $r$ for a 3D radius.

It is immediately clear from \eqnref{eqn:sphericaljeans} that, even assuming
spherical symmetry, we have a degeneracy between the enclosed mass profile
$\Mddd(r)$ and the velocity anisotropy $\beta(r)$. This can be understood
intuitively since $\beta(r)$ measures the relative importance of radial versus
circular orbits and is intrinsically difficult to constrain given only one
component of the velocity vector for each star. Nonetheless, $\beta(r)$ can be
constrained given sufficiently many stars, since radial Doppler velocities
sample eccentric orbits as $r\rightarrow 0$ and tangential orbits as
$r\rightarrow \infty$ \citep[e.g.,][]{2002MNRAS.330..778W}. It can also be
estimated if an independent measure of $\Mddd(r)$ is available -- for example
coming from strong lensing. 

While $\Mddd(r)$ is difficult to measure from stellar kinematics alone, the
mass within the half light radius is robustly recovered
\citep[e.g.,][]{2009ApJ...704.1274W,2010MNRAS.406.1220W,2012ApJ...754L..39A}
since stellar systems in dynamic quasi-equilibrium obey the virial theorem
(equation \ref{eqn:virial}). This means that stellar kinematics can break the
steepness degeneracy if $r_{1/2} \neq r_E$, where $r_E = D_L \theta_E$ is the
physical Einstein radius. We test this expectation in \secref{sec:results}.

We describe our numerical solution of \eqnref{eqn:sphericaljeans} in
\secref{sec:glasskinematics} and present tests applied to mock data in
\secref{sec:results}. 

\section{Numerical Methods}\label{sec:glass}

\subsection{A new lens modelling framework: \Glass}

\Glass{} is the Gravitational Lensing AnalysiS Software. It extends and
develops some of the concepts from the free form modelling tool
\PixeLens{}~\citep{Saha2004,2008ApJ...679...17C}, but with all new code.  The
most compute intensive portion was written in C but Python was chosen because
of its flexibility as a language and for its large scientific library support.
The flexibility allows \Glass{} to have quite sophisticated behavior while at
the same time simplifying the user experience and reducing the overall
development time. One of the striking features is that the input file to
\Glass{} is itself a Python program.  Understanding Python is not necessary for
the most basic use, but this allows a user to build complex analysis of a model
directly into the input file. \Glass{} may furthermore be used as an external
library to other Python programs.  {\it The software is freely available for
download or from the first author.}%
\footnote{\url{http://www.jpcoles.com}}

The key scientific and technical improvements are:
\begin{enumerate}
  \setcounter{enumi}{0}
  \item A new uniform sampling algorithm for high dimensional spaces.
\end{enumerate}
At the heart of \Glass{} lies a new algorithm for sampling the high dimensional
linear space that represents the modelling solution space. This algorithm was
described and tested in \cite{2012MNRAS.425.3077L}; it is multi-threaded
allowing it to run efficiently on many-cored machines.
\begin{enumerate}
  \setcounter{enumi}{1}
  \item A modular framework that allows new priors to be added and modified easily.
\end{enumerate}
Each prior is a simple function that adds linear constraints that operate on
either a single lens object or the entire ensemble of objects. \Glass{} comes
with a number of useful priors (the default ones will be described in \secref{sec:discrete}), but a
user can write their own directly in the input file, or by modifying the source
code.
\begin{enumerate}
  \setcounter{enumi}{2}
  \item The basis functions approximating a model can be changed. 
\end{enumerate}
\Glass{} currently describes the lens mass as a collection of pixels, but the
code has been designed to support alternative methods. In particular, there are
future plans to develop a module using Bessel functions. This will require a
new set of priors that operate on these functions.
\begin{enumerate}
  \setcounter{enumi}{3}
  \item Non-linear constraints can be imposed in an automated post-processing step. 
\end{enumerate}
Once \Glass{} has generated an ensemble of models given the linear constraints,
any number of post processing functions can be applied. Not only can these
functions be used to derive new quantities from the mass models, they can also
be used as a filter to accept or reject a model based on some non-linear
constraint. For example, we can reject models that have spurious extra images
(\secref{sec:glassextraimages}), or models that do not match stellar kinematic
constraints (\secref{sec:glasskinematics}). The plotting functions within
\Glass{} will correctly display models that have been accepted or rejected.
\begin{enumerate}
  \setcounter{enumi}{4}
  \item The central region can have a higher resolution to capture steep models. 
\end{enumerate}
With the default basis set of pixels, the mass distribution of the lens is
described by a uniform grid. However, in the central region of a lensing galaxy
where the mass profile may rise steeply, the center pixel uses a higher
resolution. This allows the density to increase smoothly but still allow for a
large degree of freedom within the inner region without allowing the density to
be arbitrarily high. 
\begin{enumerate}
  \setcounter{enumi}{5}
  \item Stellar density can be used as an additional constraint.  
\end{enumerate}
The mass in inner regions of galaxies is often dominated by the stellar component
which one can estimate using standard mass-to-light models. This data can be added
to the potential as described later in \secref{stellar mass}. By using the stellar
mass one can place a lower bound on the mass and help constrain the inner most
mass profile.
\begin{enumerate}
  \setcounter{enumi}{6}
  \item Point or extended mass objects can be placed in the field.
\end{enumerate}
A shear term can be added to the potential, as shown later in \eqnref{shear},
to account for mass external to the modelled region. This is useful to capture
the gross effects of a distant neighbour, since there is a degeneracy between
the ellipticity of a lens and its shear field (the greater the allowed shear, the
more circular the lens may be). \Glass{} also allows further analytic potential
components to be included. These can be used to model substructure or multiple
neighbours close to the main lens. The substructure may have only a small
effect if the lens is a single galaxy, but if the lens is a group or cluster then a
potential can be added for each of the known member galaxies. A few standard functions are
already included in \Glass{} including those for a point mass, a power law
distribution, or an isothermal (a particular case of the power law).

\subsection{Analysis Tools}\label{sec:tools}
\Glass{} is not only a modeling tool but also an analysis engine. \Glass{}
provides many functions for viewing and manipulating the computed models.
These functions can either be called from a program written by the user or by
using the program \textsc{viewstate.py} included with \Glass. There is also a
tool, \textsc{lenspick.py} for creating a lens, either analytically or from an
$N$-body simulation file. To load the simulation data, \Glass{} relies on the
\textsc{Pynbody} library \citep{pynbody} and can thus load any file supported
by that package.

\subsection{Pixelated models}\label{sec:discrete}
For this paper, we will restrict ourselves to using a pixelated basis set as
used by \PixeLens{} \citep{Saha2004,2008ApJ...679...17C}, but note that it is
straightforward to add other basis function expansions to \Glass. The algorithm
for generating models in \Glass{} samples a convex polytope in a high
dimensional space whose interior points satisfy both the lens equation and
other physically motivated {\it linear} priors \citep{2012MNRAS.425.3077L}.
A limitation of our sampling strategy is that only linear constraints may
be applied when building the model ensemble; however, non-linear
constraints can be applied in post-processing (see
\secref{sec:glassextraimages} and \secref{sec:glasskinematics}). We
therefore formulate all of our equations as equations linear in the
unknowns. We describe the density distribution $\kappa$ as a set of
discrete grid cells or pixels $\kappa_i$ and rewrite the potential
\eqnrefp{lensing potential} as:
\begin{equation}
  \psi(\vec\theta) = \sum_n \kappa_n Q_n(\vec\theta)
  \label{discrete potential}
\end{equation}
where the sum runs over all the pixels and $Q_n$ is the integral of the
logarithm over pixel $n$. The exact form for $Q$ is described in \appref{Q
derivation}.  We can find the discretized lens equation by simply taking the
gradient of the above equations. 

The pixels only cover a finite circular area with physical radius $\Rmap$ and
pixel radius $\Rpix$ with the central cell centered on the lensing galaxy. To
account for any global shearing outside this region from, e.g., a neighboring
galaxy, we also add to \eqnref{discrete potential} two shearing terms:
\begin{equation}
\label{shear}
\gamma_1(\theta_x^2 - \theta_y^2) + 2\gamma_2\theta_x\theta_y\quad.
\end{equation}
We can continue adding terms to account for other potentials. For instance, we
may want to impose a base potential over the field, or add potentials from the
presence of other galaxies in the field. \Glass{} already includes potentials
for a point mass or an exponential form, but custom potentials are
straightforward to add and can be included directly in the input file.  If the
stellar density $\kappa_s$ has been estimated we can use this as a lower bound
where the stellar potential is a known constant of the form \eqnref{discrete
potential}, e.g., $\kappa_n = \kappa_{\mathrm{dm},n} + \kappa_{s,n}$ for a
two-component model.

\subsubsection{Priors}

The lens equation and the arrival times alone are typically not enough to form
a closed volume in the solution space. We therefore require additional linear
constraints -- {\it priors}. Some of these are `physical' in the sense that
they are unarguable -- for example demanding that the mass density is
everywhere positive; others are more subjective, for example demanding that the
mass map is smooth over some region. Such `regularisation' priors may be
switched off for all or some of the mass map if the data are sufficiently
constraining. 

The priors built in to \Glass{} are similar to those used in
\PixeLens{}~\citep{2008ApJ...679...17C}. The physical priors are always used by
default; the regularisation priors are used sparingly -- i.e. only if the data
are not sufficiently constraining to obtain sensible solutions without them:  

\vspace{2mm}
\noindent
{\bf Physical priors}
\begin{enumerate}
\item The density must be non-negative everywhere.
\item Image parity is enforced.
\end{enumerate}

\vspace{1mm}
\noindent
{\bf Regularisation priors}
\begin{enumerate}
\item The local gradient everywhere must point within $45^{\circ}$ of the center.
\item The azimuthally averaged density profile must have a slope everywhere $\le 0$.
\item The density is inversion symmetric.
\end{enumerate}
For typical lens data, the regularisation priors are very important for
creating physically sensible solutions. Prior (i) demands that the peak in the
mass density is at the centre of the mass map. Secondary `plateaus' in the mass
map are possible, but not secondary peaks. Note that this prior still
successfully allows merging galaxy systems to be correctly captured, provided
that the two galaxies are not equally dense in projection (see, for example the
\PixeLens{} model of the merger system B1608 in \citealt{2007ApJ...667..645R});
and for the successful detection of `meso-structure' in strong lensing galaxy
clusters \citep{2007ApJ...663...29S}. Prior (ii) is arguably a physical prior
since a positive slope in the azimuthally averaged density profile would be unstable
\citep[e.g.,][]{2008gady.book.....B}. Note that this prior does not preclude
successful modelling of mergers or substructure unless the total projected mass
in substructure is comparable to the projected mass of the host in an azimuthal
annulus \citep{2007ApJ...667..645R,2007ApJ...663...29S}. Prior (iii) is only
used for doubles that ought to be inversion symmetric and quads where inversion
symmetry is clear from the image configuration.

Finally, we remind the reader that all of the regularisation priors can be
switched off or changed/improved depending on the data quality available. For
clusters, substructure can be explicitly modelled by adding analytic potentials
at the known locations of galaxies; furthermore the above priors can be relaxed
in regions of the mass map where the data are particularly constraining (for
example near the images). We will apply \Glass{} to a host of strong lensing
clusters in forthcoming work, where we will explicitly test the prior on mock
data that has significant substructure.

\subsection{Building the model ensemble} 

In the simplest form, a single model for a lens is a tuple $\M = (\vec\kappa,
\vec\beta, \gamma_1, \gamma_2)$. A single model represents a single point in
the solution space polytope. Using the MCMC sampling strategy described in
\cite{2012MNRAS.425.3077L} we uniformly sample this space. Collectively, the
sampled models are referred to as an ensemble $\E = \{\M_i\}$, where we usually
generate $|\E| \sim 1000$ models. One can choose to further process these
models to impose priors that may be difficult to enforce during the modeling
process. For instance, non-linear constraints, or simply filtering of models
that do not meet some criteria can be excluded, or weighted against as
discussed previously.  In this paper, we do not exclude any models and treat
all models as equally likely.

The time to generate the model ensemble is mostly a function of the size of the
parameter space. The MCMC algorithm has a ``warm-up'' phase where it estimates
the size and shape of each dimension in the solution space.  Once this has been
completed, the models are sampled very quickly. In fact, there is little
difference between generating 1,000 or 10,000 models, although we find little
statistical difference after 1,000 models. For the mock lenses, the typical
``warm-up'' time was about 4s, and the modelling time was 20s using a parallel
shared-memory machine with 40 cores. The ability to rapidly generate so many
models is what allows us to then accept/reject models to apply non-linear
constraints (see \secref{sec:glassextraimages} and
\secref{sec:glasskinematics}). This is a key advantage over our earlier
pixelated strong lens tool \PixeLens. 

\subsection{Raytracing}\label{Raytracing}
\Glass{} can also determine the position of images and time delays from
particle-based simulation output given a source position $\vec\beta$. This is
used to generate the lens configurations used in the parameter study.  The
particles are first projected onto a very high resolution grid representing the
lens plane. The centers $\vec\theta_i$ of each of the grid cells are mapped
back onto the source plane using \eqnref{eqn:lens_equation}. If the location on
the source plane $\vec\beta_i$ is within a user specified
$\eps_\mathrm{accept}$ of $\vec\beta$ then $\vec\theta_i$ is accepted and
further refined using a root finding algorithm until the distance to
$\vec\beta$ is nearly zero. If multiple points converge to an
$\eps_\mathrm{root}$ of each other then only one point is taken.  Care must be
taken that the grid resolution is high enough that the resulting image position
error is below the equivalent observational error. Time delays are then
calculated in order of the arrival time at each image \eqnrefp{tau}.

\subsection{Removing models with extra images}\label{sec:glassextraimages} 
While linear constraints are applied in \Glass{} by the nature of the sampling
algorithm, non-linear constraints must be applied in post-processing. Models
that are inconsistent with such constraints must then be statistically
discarded via a likelihood analysis. An example of such a non-linear constraint
is the spurious presence of unobserved images. This `null-space' prior was
first proposed and explored by \citet{2006MNRAS.367.1209L} and found to be
extremely powerful. We find that our gradient prior in \Glass{} (see
\secref{sec:discrete}), performs much of the same function as
\citeauthor{2006MNRAS.367.1209L}'s null-space prior, but some models can still
rarely turn up spurious images. We reject these in a post-processing step,
where we sweep through the model ensemble applying the ray tracing algorithm
described in \ref{Raytracing}.

\subsection{A post-processing module for stellar kinematics}\label{sec:glasskinematics} 
Similarly to the null-space constraint (\secref{sec:glassextraimages}), stellar
kinematic constraints constitute a non-linear prior on the mass map and must be
applied in post-processing. We sweep through the model ensemble performing an
Abel deprojection to determine $\Mddd(r)$ from the projected surface density
$\Sigma(R)$ assuming spherical symmetry
\citep[e.g.,][]{2008gady.book.....B,2008MNRAS.390.1647B}: 
\begin{eqnarray} 
    \Mddd(r) & = & M_\mathrm{p}(<r) - 4r^2 \int_0^{\pi/2} \Sigma\left(x\right) \left[\frac{1}{\cth[2]} \right. \nonumber \\ 
    & & \left. - \frac{\sth}{\cth[3]} \arctan\left(\frac{\cth}{\sth}\right) \right] \dth
\end{eqnarray}
where 
\begin{equation}
    M_\mathrm{p}(<r) = 2\pi \int_0^r R \Sigma(R) dR
\end{equation}
is the projected enclosed mass evaluated at 3D radius $r$; and $x = r/\cth$. 

This de-projection algorithm was tested on triaxial figures in
\citet{2006ApJ...652L...5S}. They found that for triaxialities typical of our
current cosmology, the method works extremely well unless the triaxial figure
is projected directly along the line of sight such that we see the galaxy or
galaxy cluster `down the barrel'. Such a situation is unlikely, but in any case
avoidable since the resultant figure appears spherical in projection. This
leads to the seemingly counter-intuitive result that the kinematic constraints
-- that rely on the above de-projection -- are most secure for systems that do
not appear spherical in projection (unless independent data can confirm the
three dimensional shape is indeed very round). 

We use the deprojected mass to numerically solve \eqnref{eqn:sphericaljeans}
for constant $\beta(r)$, assuming either $\beta(r) = 1$ or $\beta(r) = 0$ at
all radii to bracket the two extremum situations. Where the data are good
enough, these two may be distinguished giving dynamical information about
$\beta(r)$. In more typical situations, however, we seek to simply marginalise
over the effect of $\beta(r)$, using the stellar kinematics as a robust measure
of $\Mddd(r_{1/2})$ (see \secref{sec:kinematics}). 

\section{The mock data}\label{sec:mockdata}

We now present a study of four mock galaxies with known analytic forms. These
are used to verify that \Glass{} is able to correctly recover the mass profile,
and -- more importantly -- to determine what type and quality of data best
constrain the mass profile and shape of a lens.

\subsection{The triaxial N-body mock galaxies}

We generate four two-component mock galaxies, where the dark matter and stellar
profiles are allowed to be both steep and shallow.  The enclosed mass of the
stars and dark matter are both fixed to be $M_{*,\mathrm{DM}} =
1.8\e{10}$\,M$_\odot$ at the stellar scale radius $a_* = 2$\,kpc, such that the
stars and dark matter contribute equally to the total mass at $a_*$. The dark
matter scale length is fixed for all models at $a_\mathrm{DM} = 20$\,kpc.
These values were chosen to closely resemble the lensing galaxy PG1115+080
\citep{1980Natur.285..641W}. We place the galaxy at a redshift of $z_L = 0.31$
for lensing.  Throughout, we assume a cosmology where $H_0^{-1}=13.7$ Gyr,
$\Omega_M=0.28$, and $\Omega_\Lambda=0.72$. The critical lensing density is
$\kappa_\mathrm{crit}\sim 1.8\e{9}$\Msun/kpc$^2$.

The galaxies were generated as three dimensional particle distributions as in
\citet{2009MNRAS.395.1079D}. Each component follows the profile:
\begin{equation} 
  \rho(\tilde r) = \frac{M}{4\pi a^3}(3-\gamma){(\tilde r/a)^{-\gamma}(1 + \tilde r/a)^{\gamma-4}} 
  \label{Dehnen profile} 
\end{equation} 
where $a$ is the component scale radius mentioned in \tabref{mock galaxy
params}; $\tilde r^2 = (x/\lambda_1)^2 + (y/\lambda_2)^2 + (z/\lambda_3)^2$ is the
ellipsoidal radius; and the axis ratios are $\lambda_1:\lambda_2:\lambda_3 =
6:4:3$.  In the case where the central density profile index $\gamma$ is unity
(and in the limit of spherical symmetry), this is the Hernquist profile
\citep{1990ApJ...356..359H}.  The four combinations of profile indices are
shown in \tabref{mock galaxy params}.

\begin{table}
\begin{tabular}{llllllll}
Galaxy & $\gamma_\star$ & $M_\star$ & $\gamma_\mathrm{DM}$ & $M_\mathrm{DM}$ & $\Rmap$ \\
\hline
\mockAA & 1 & 4 & 0.05 & $11^{2.95}$ & 50 kpc\\ 
\mockAC & 1 & 4 & 1 & $11^2$ & 50 kpc \\ 
\mockBB & 1.5 & $2^{1.5}$ & 0.16 & $11^{2.84}$ & 50 kpc \\ 
\mockBC & 1.5 & $2^{1.5}$ & 1 & $11^2$ & 10 kpc 
\end{tabular}
\caption{Profile parameters for the four mock galaxies. The name indicates whether 
  the galaxy is centrally dark matter or stellar dominated with a shallow or cuspy 
  dark matter density profile.  Masses are in units of
  $1.8\e{10}\Msun$. The scale lengths for all lenses are
  $(a_\star,a_\mathrm{DM})=(2,20)$\,kpc. $\Rmap$ is the 2D projected radius
used to generate the lens configurations. In the case of \mockBC, the profile is 
sufficiently steep that the profile could be truncated at $\Rmap = 10$ kpc.}
\label{mock galaxy params}
\end{table}

\begin{figure*}
\includegraphics[width=0.33\textwidth]{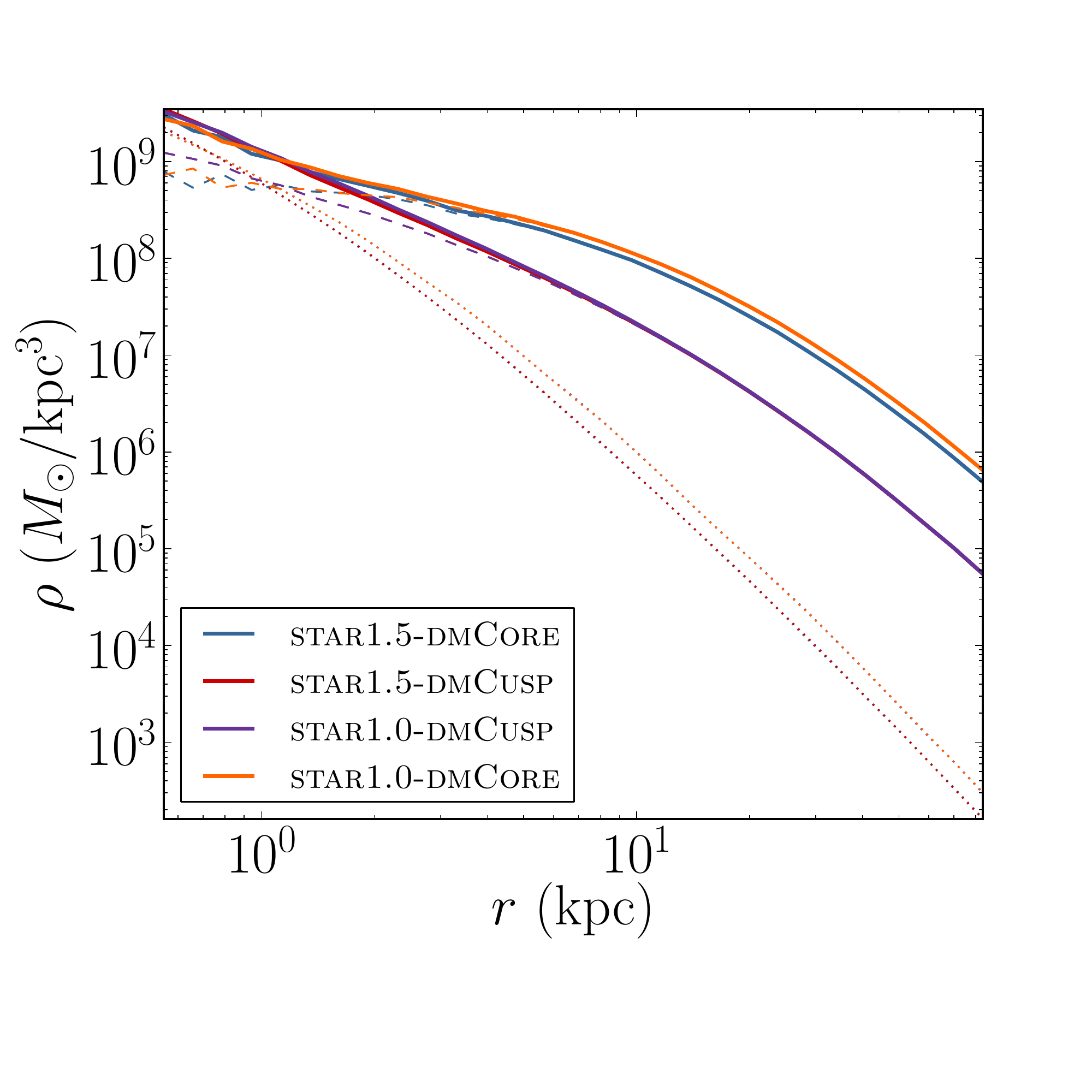} 
\includegraphics[width=0.33\textwidth]{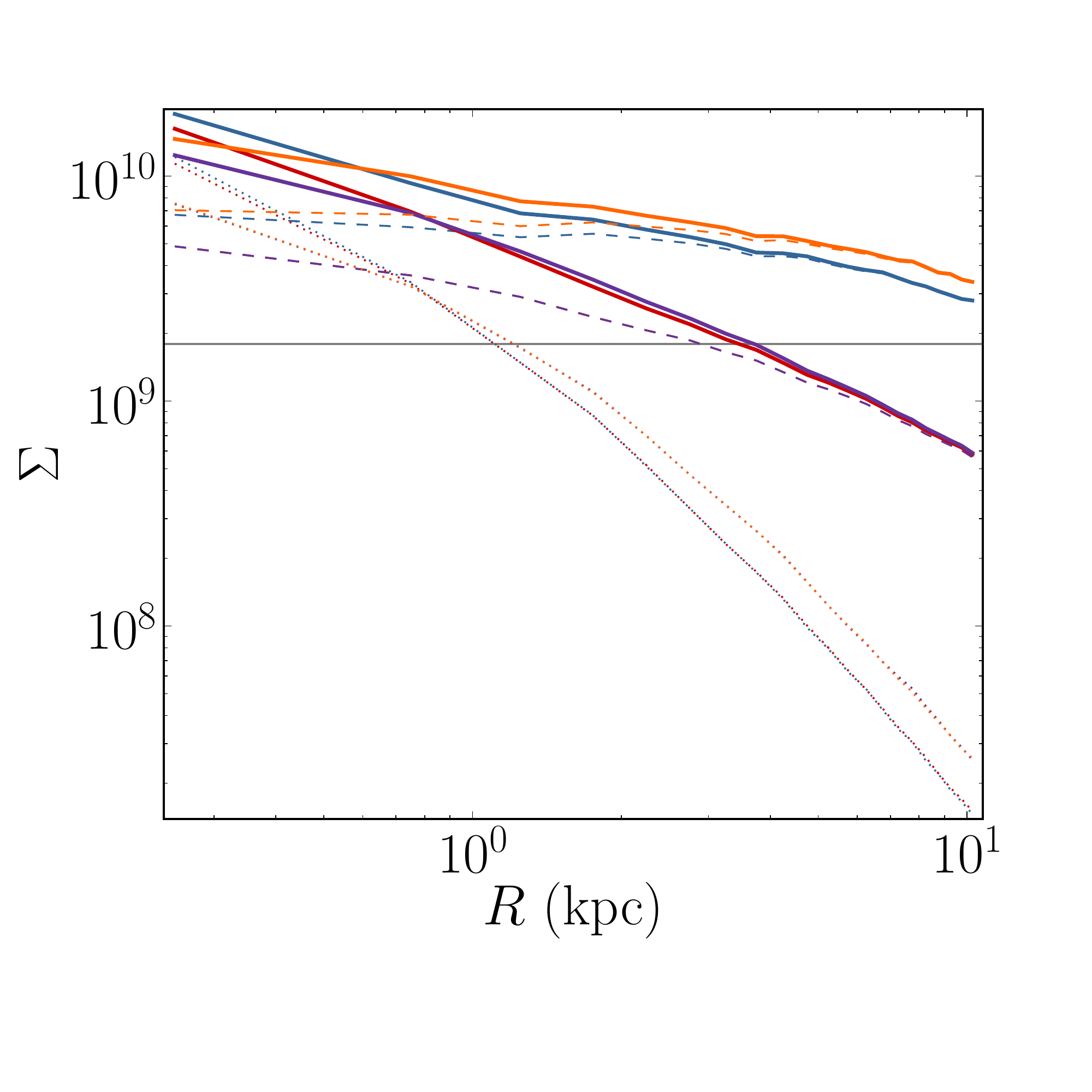} 
\includegraphics[width=0.33\textwidth]{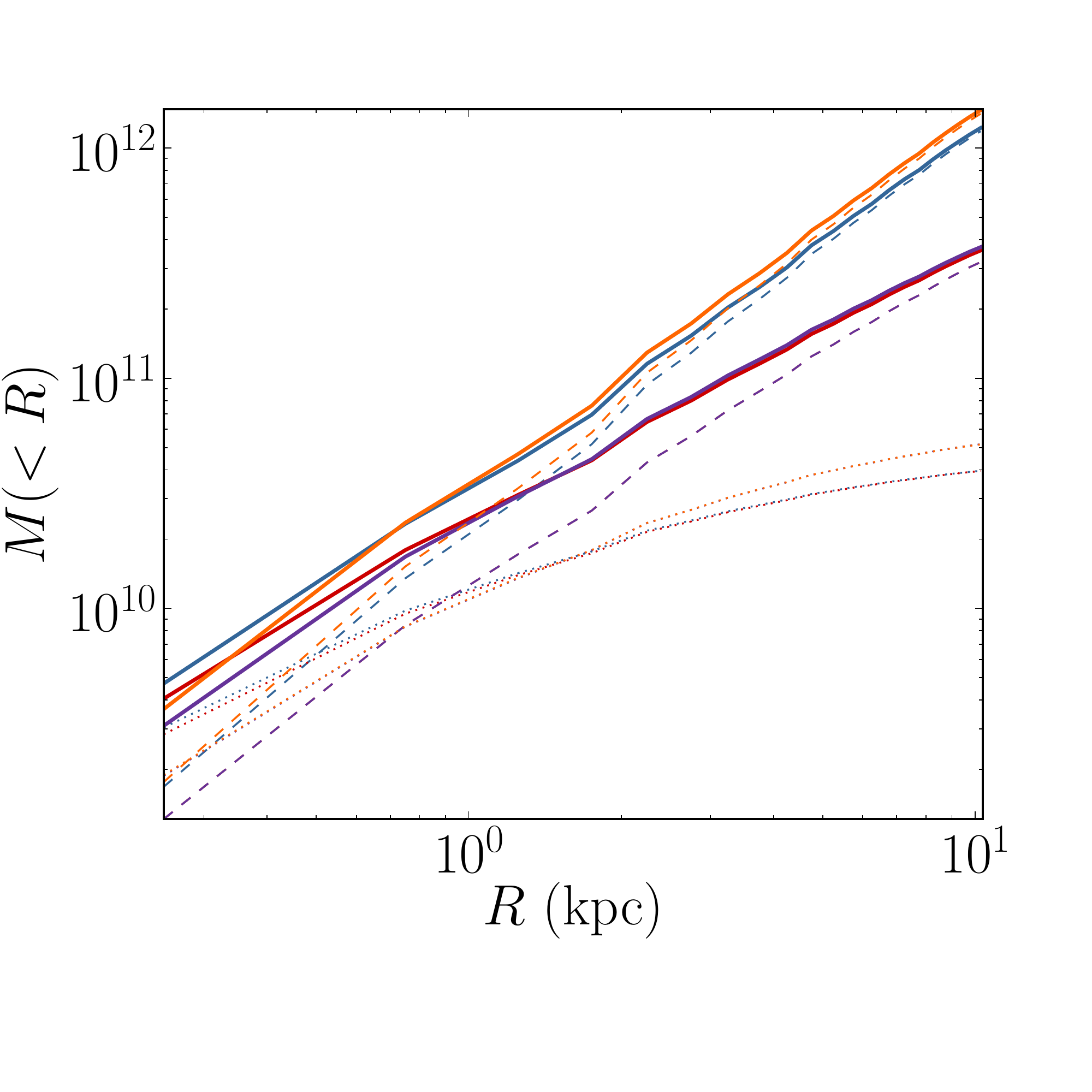}
\caption{
Profiles of the four mock galaxies showing the stellar (dotted) and dark matter (dashed) components and the total (solid).
\textbf{Left:} 
The spherically averaged density. The stars in models \mockBB{} and \mockBC{} contribute significantly to the central potential. 
\textbf{Middle:} 
The radially averaged two-dimensional projected density.
The critical lensing density at $z_L=0.31$, $\kappa_\mathrm{crit}\sim 1.8\e{9}$\Msun/kpc$^2$, is marked by the horizontal line. 
\textbf{Right:}
The enclosed projected mass.
}
\label{mock galaxies}
\end{figure*}

In \figref{mock galaxies}, we show the 3D radial density, the 2D projected
density, and the 2D enclosed mass for each galaxy.

\subsection{Lens configurations}\label{sec:lensconfig}

\begin{figure*}
\includegraphics[width=0.85\textwidth]{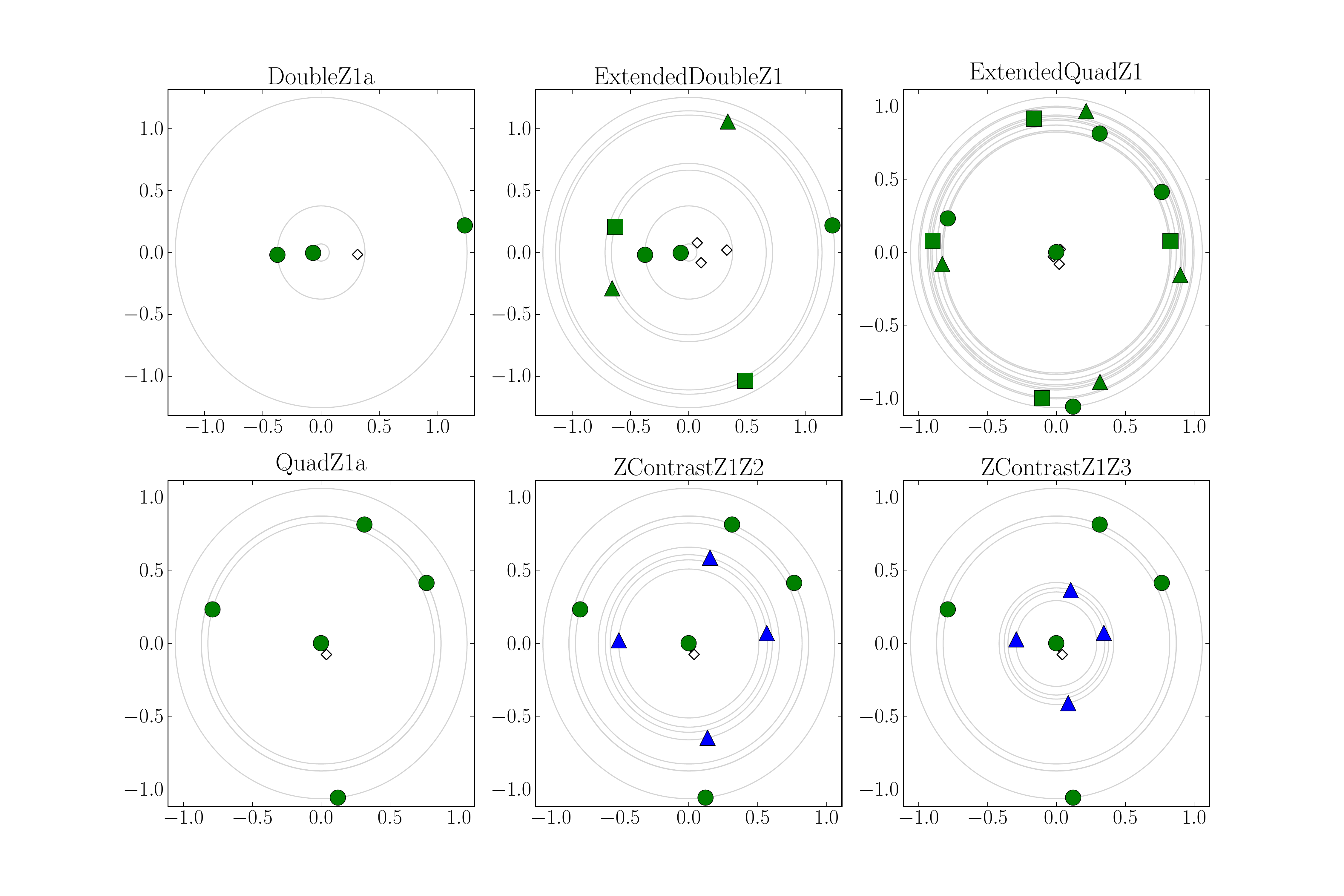}
\caption{The lens configurations for the six test cases using the \mockBC{}
mock galaxy. The other mock galaxies produce similar results. Here, the central
image is shown, although not all tests include it. 
The naming convention indicates the redshift of the
sources with Z1=1.72, Z2=0.72, and Z3=0.51.  The central image only belongs to
the Z1 source to avoid over-constraining the models (see
\secref{sec:lensconfig} for further details). 
Small diamonds identify the location of the source(s) and images of the same shape share a
common source. The extended source examples have been constructed so that the
images will form arclets.  The maximum separation of the sources in the source plane
is 2.23 kpc in the extended double and 0.92 kpc in the extended quad. Grey
circles are a visual aid to help determine radial separation between images.
The axes are in arcseconds.}
\label{arrival surfaces}
\end{figure*}

For each of the four galaxies, we used the raytracing feature of \Glass{}
described in \secref{Raytracing} to construct 6 basic lensing morphologies:

\begin{enumerate}
\item one double and one extended double;
\item one quad and one extended quad;
\item two 2-source quads with varying redshift contrast.
\end{enumerate}
The `extended' configurations use multiple point sources at the same redshift
to simulate an extended source that will produces an arc-like image.
\figref{arrival surfaces} shows the lens configurations for the \mockBC\
galaxy. 
The configurations for the other galaxies are similar.  The labels Z1, Z2, Z3
within the names refer to the redshift of the sources. We have chosen Z1=1.72,
Z2=0.72, and Z3=0.51 so that the radial distribution of the images is roughly
equally spaced. For all mocks, we do not apply any external shear field.  Only
the central image of the Z1 source is used to avoid over-constraining the
models, otherwise all the central images would fall within the central pixel
and no solution exists that satisfies all locations simultaneously for one
pixel value.

Each of these configurations were modelled with and without time delays; with
and without a central image; and with and without the stellar mass as a lower
bound, for a total of 48 test cases. (The central image is typically highly
demagnified. For galaxy lenses it is very difficult to find since it lies along
the sight line to the bright lensing galaxy; in clusters, however, such images
have been seen -- e.g., \citealt{2005PASJ...57L...7I}). We assumed for all our
tests that the lensing mass was radially symmetric (Prior vi). For our mock
data, this is known to be true; it is most often the case with real galaxies,
unless there is an obvious observed asymmetry. (We explore the effect of
switching off the symmetry prior in \appref{no_symm_prior}. For the
quads, the difference is small; for the doubles -- as expected -- the results
are significantly degraded without this prior.) We use, by default, 8 pixels
from the centre to the edge of the mass map; the central pixel was further
refined into $5\times5$ pixels to capture any steep rise in the profile (two of
the four mock galaxies have a steeply rising inner profile). We demonstrate
that our results are robust to changing the grid resolution in 
\appref{pix_convergence_test}.

In all cases -- despite applying no external shear to the mock lenses -- we
allow a broad range of external shear in our lens model reconstructions.
\Glass{} correctly returns a small or zero shear in all cases. It is possible
that more complex shear fields present in real lensing galaxies could introduce
further degeneracies beyond those discussed here. However, any such shear field
can, at least in principle, be constrained by data (e.g., combining weak lensing
constraints, or assuming that the shear field correlates with visible galaxies
-- e.g., \citealt{2009A&A...500..681M,2011ApJ...726...84W}).

\section{Results}\label{sec:results}

\subsection{Radial profile recovery}

\begin{figure*}
  \plotthree{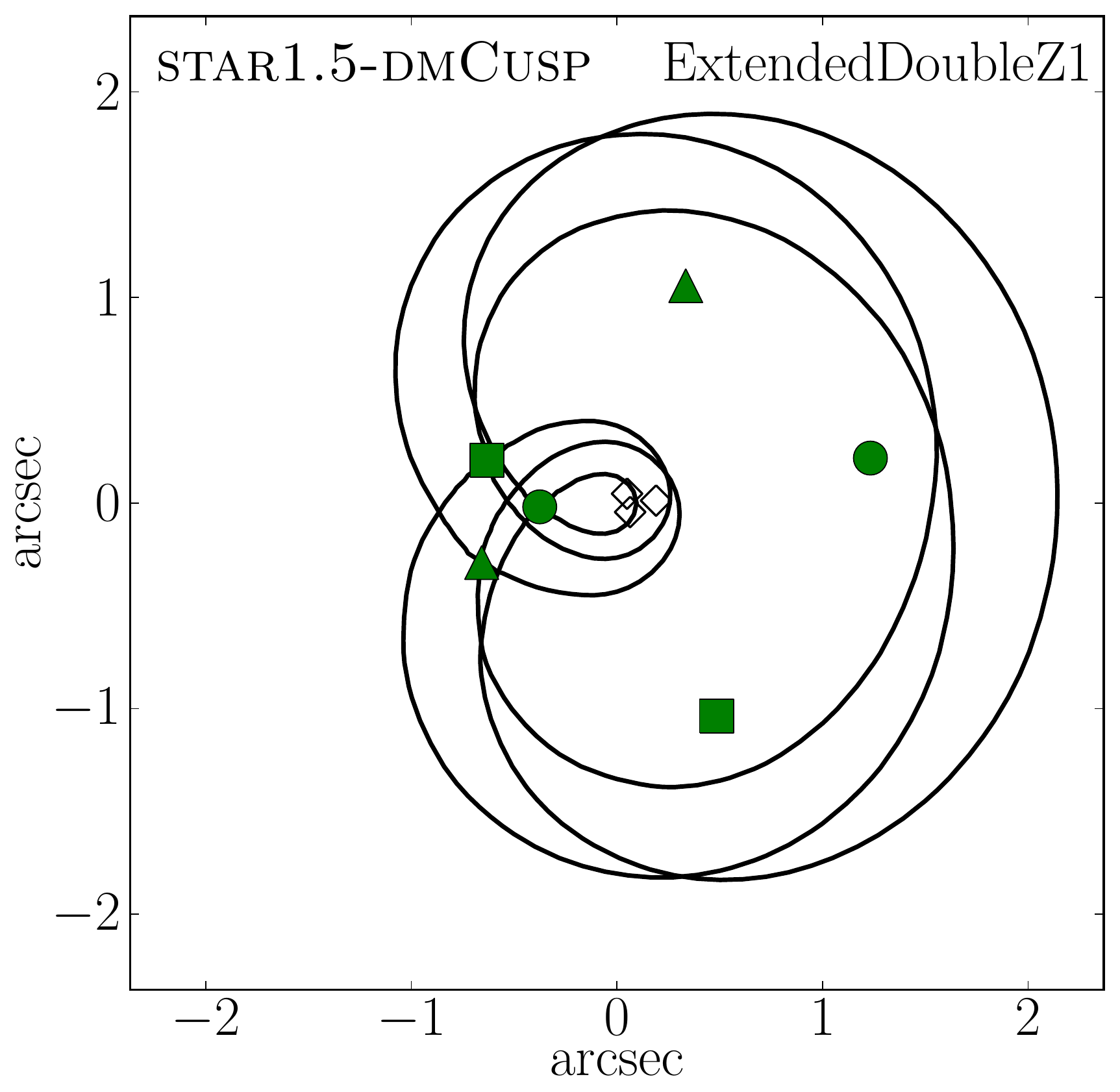} {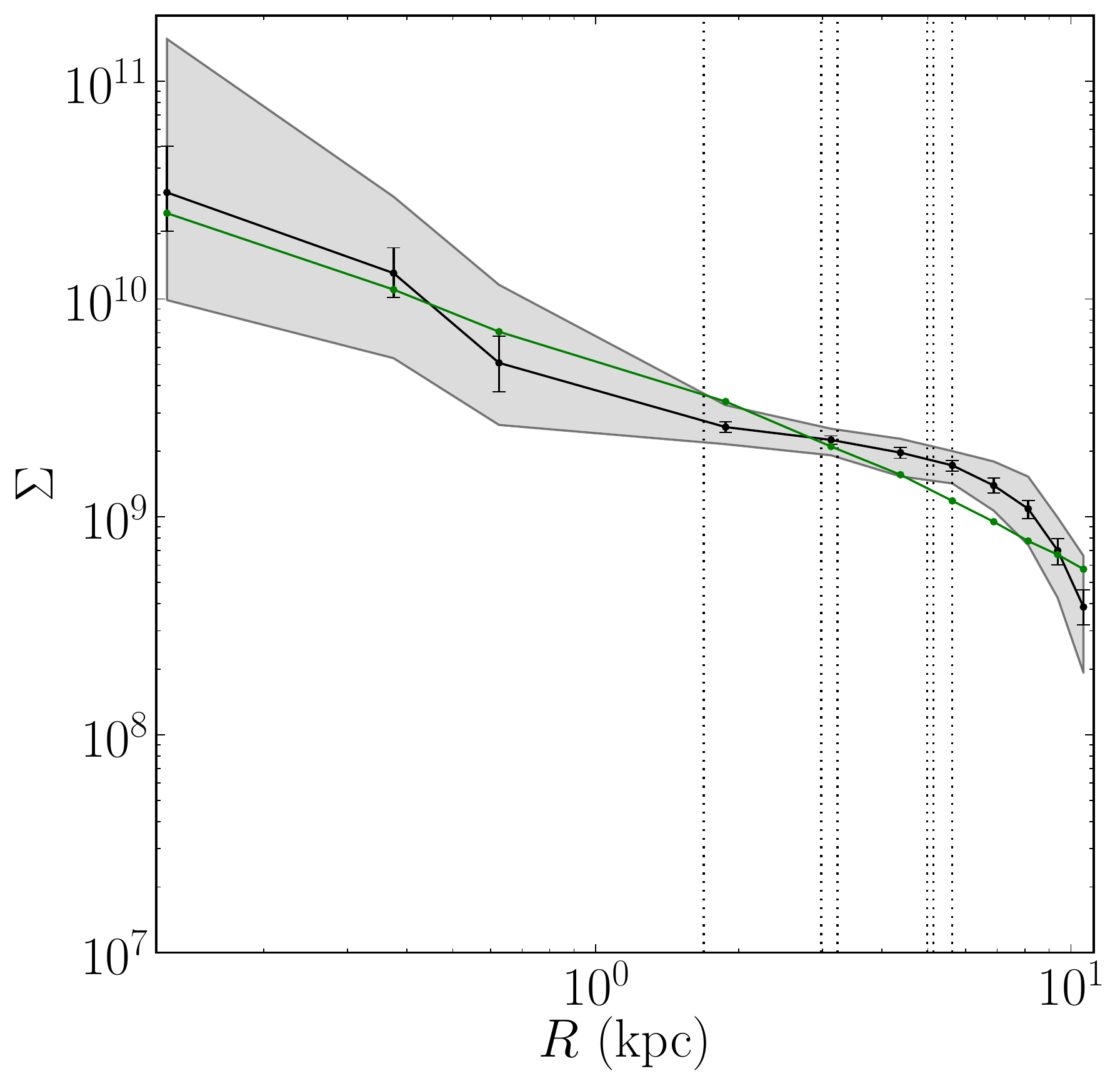} {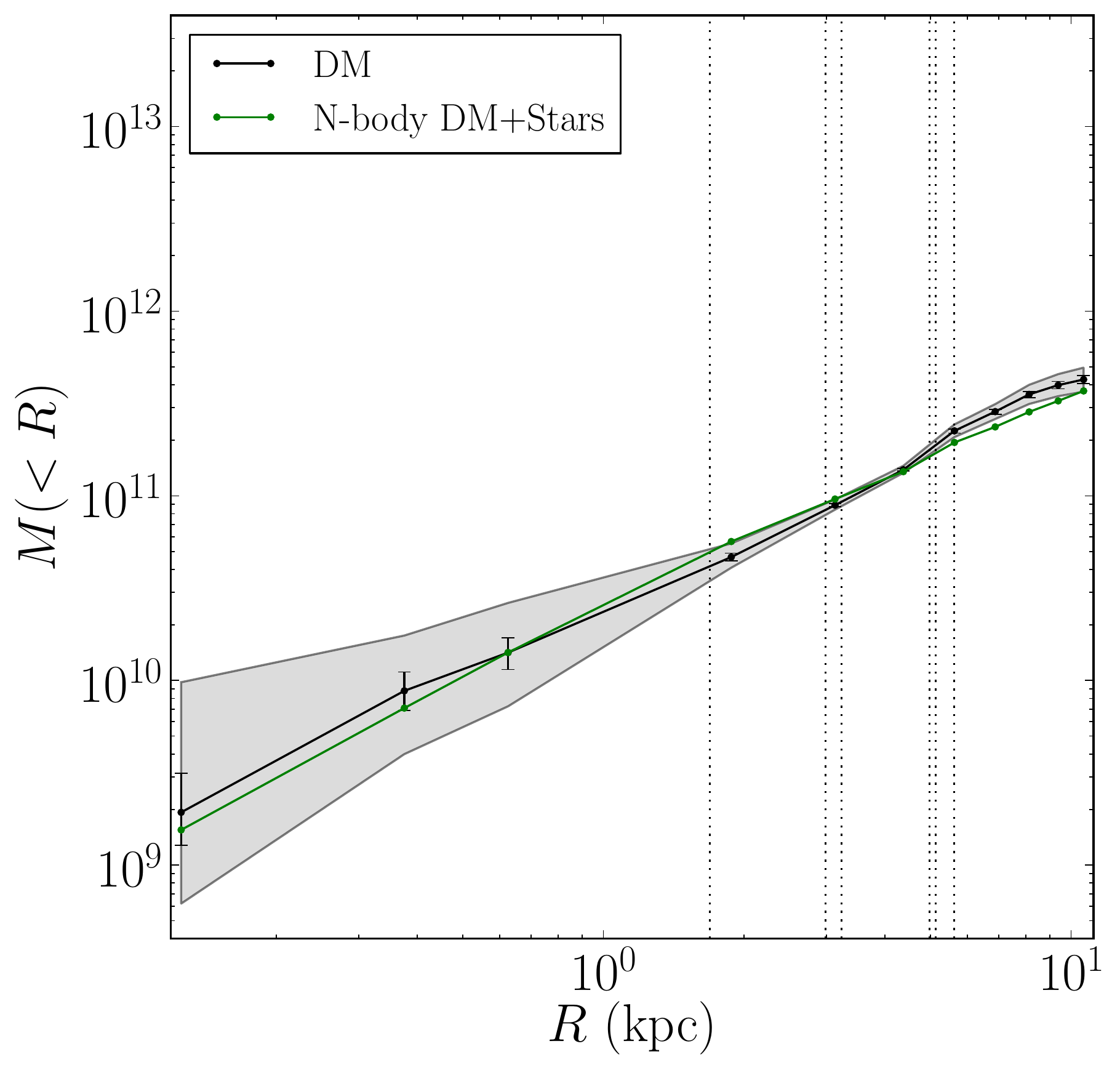}
  \plotthree{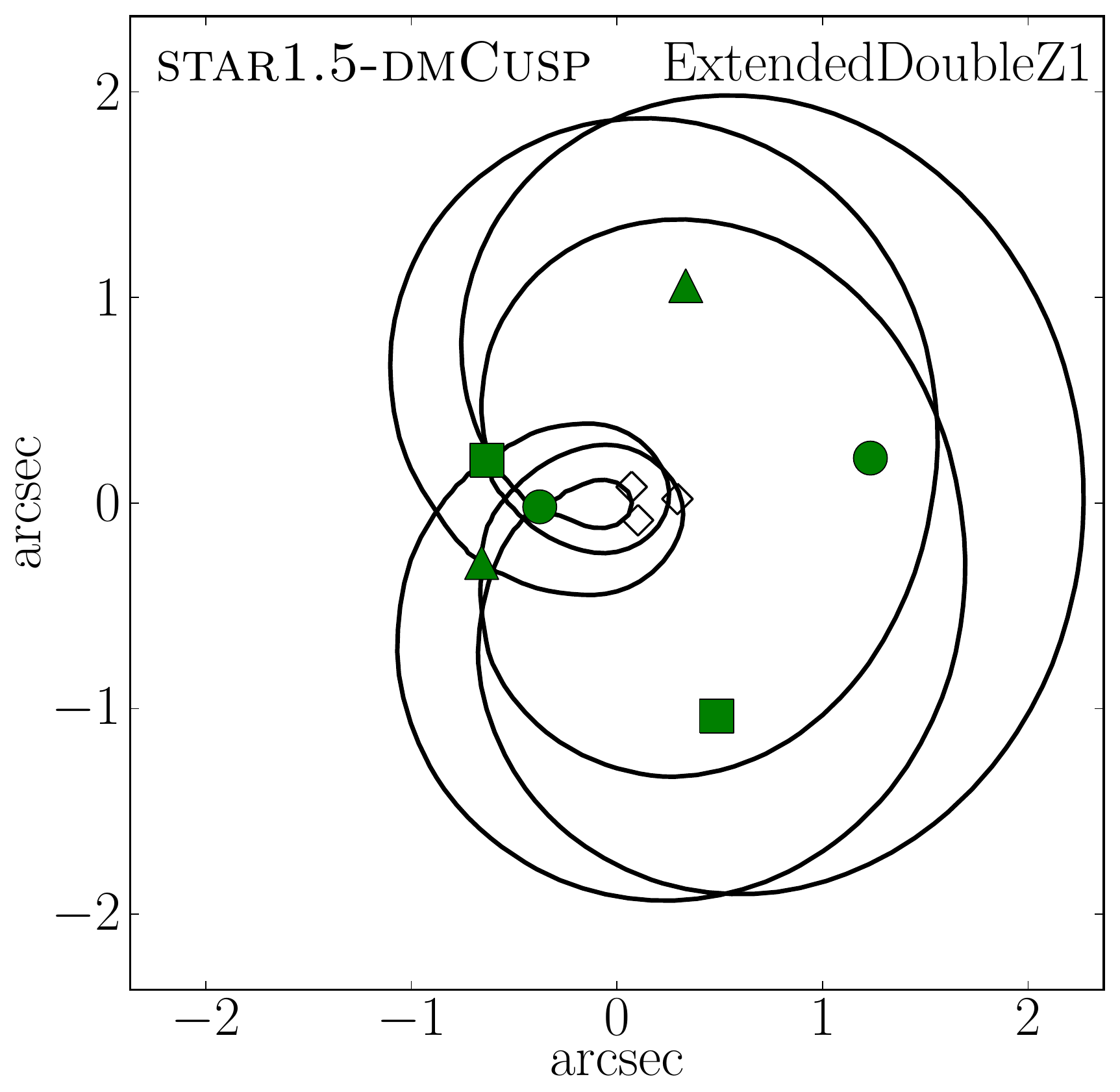} {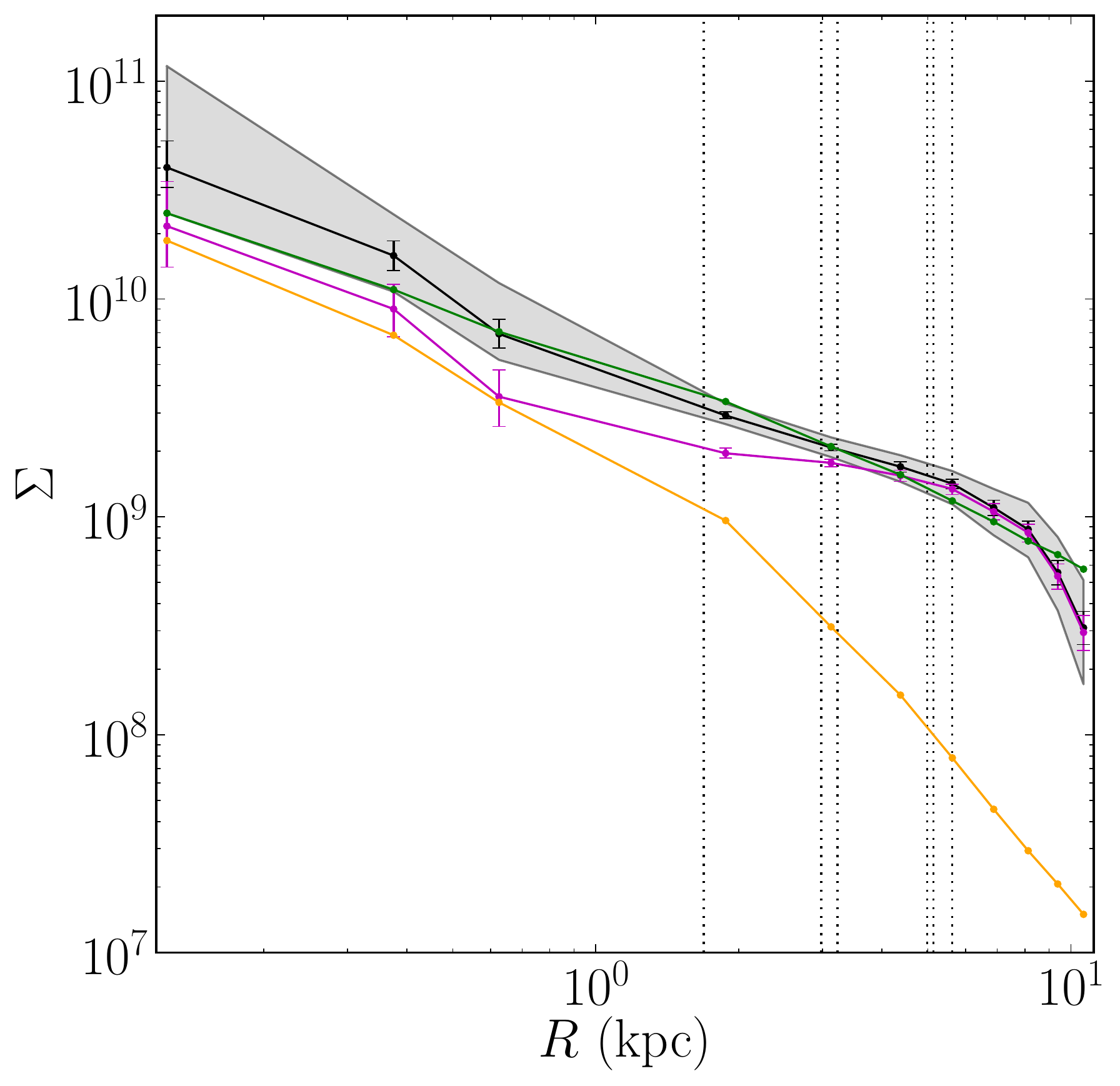} {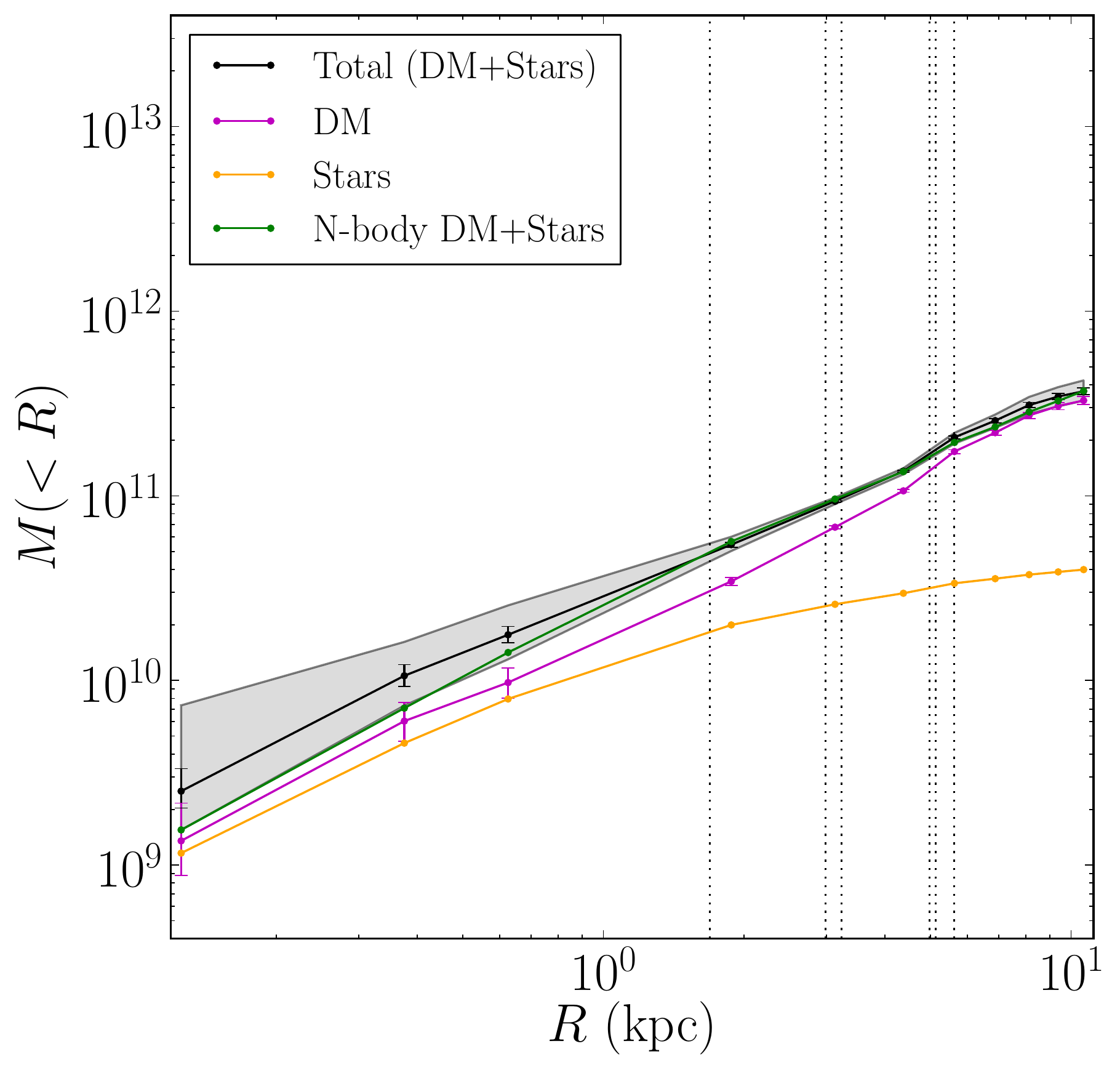}
\caption{
    Two reconstructions of the mock galaxy \mockBC\ for an extended
    double without stellar mass (\textbf{Top}) and with stellar mass
    (\textbf{Bottom}).  No time delays were assumed.  The improved constraints
    on the mass distribution when a lower bound is given by the stellar mass is
    evident in the reduced range of allowable models.  
\textbf{Left:}
The ensemble average arrival time surface with just the iso-contours for the
saddle points drawn. The central diamonds show the reconstructed source
positions.
\textbf{Middle:}
The surface density of the dark matter (DM; magenta); the stars (yellow); and the total (black).
The original $N$-body mass model (with stars) used to create the lens is shown in green.
The vertical lines mark the radial positions of the images. The higher
resolution feature of \Glass{} has been used on the central pixel allowing the
steep profile to be captured.
\textbf{Right:}
The cumulative mass. The error bars on all plots are $1\sigma$; the grey bands show the full range of models.}
\label{reconstruction}
\end{figure*}

\figref{reconstruction} shows some example reconstructions of the radial
profile of our mock lenses. The left column shows the ensemble average arrival
time surface with images marked as circles and the inferred source positions
as diamonds. The centre column shows the radial density profile. The error
bars cover a $1\sigma$ range around the median; the grey bands show the full
ensemble range. The true density profile from the mock data is also plotted for
comparison. The vertical lines mark the radial position of the images. The
right column shows the enclosed mass. From top to bottom, the rows correspond
to an extended double for \mockBC; an extended double with stellar mass
constraints for \mockBC; a quad with time delay data for \mockBC; and a quad
with time delays for \mockAA. \figref{2d mass reconstruction} shows an example
2D reconstruction for \mockBC\ for a quad; we discuss shape recovery further in
\secref{sec:shape}.

As expected, the accuracies and precisions are best in the range of radii with
lensed images where the most information about the lens is present. Even in the
weakly constrained case of the extended double where the radial profile is
poor, the true enclosed mass $M(<R)$ is well recovered at the image radii and
our ensemble always encompasses it. We have verified this is the case in all of
our tests, although for brevity we have not included the plots here. In all
cases, there is a dip in the profile at large $R$ due to the cut off in mass in
the lensing map. This is of little importance, though, as there is no lensing
information there. 

Notice that the extended double (top row) gives the poorest constraints, as
expected. Adding stellar mass (second row) significantly improves the
constraints, for this example where the stars contribute significantly to the
potential. Moving to a quad with time delays gives constraints almost as strong
as the double with stellar mass, but note that {\it focussing only on the
goodness of the fit can be misleading.} In the third row of
\figref{reconstruction}, we obtain a better recovery than in the bottom row for
{\it precisely the same data quality}. This occurs because the \Glass{} prior
favours steeper models consistent with \mockBC, but not \mockAA. It is the
\Glass{} prior, rather than the data that is driving the good recovery for
\mockBC\ in this example. This emphasises the importance of using a wide range
of mock data tests to determine the role of data versus prior in strong
lensing.

\begin{figure*}
  \plotthree{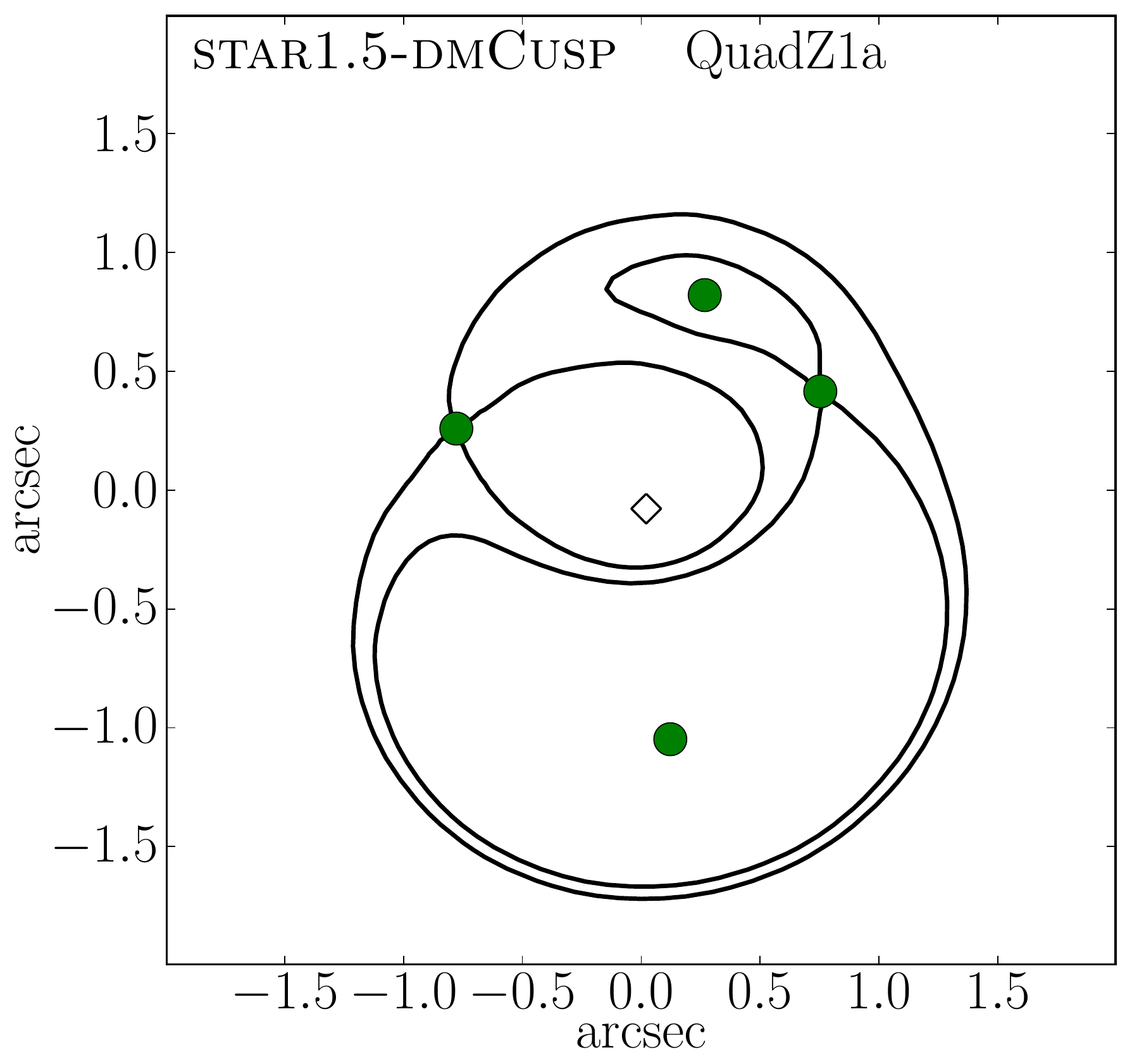} {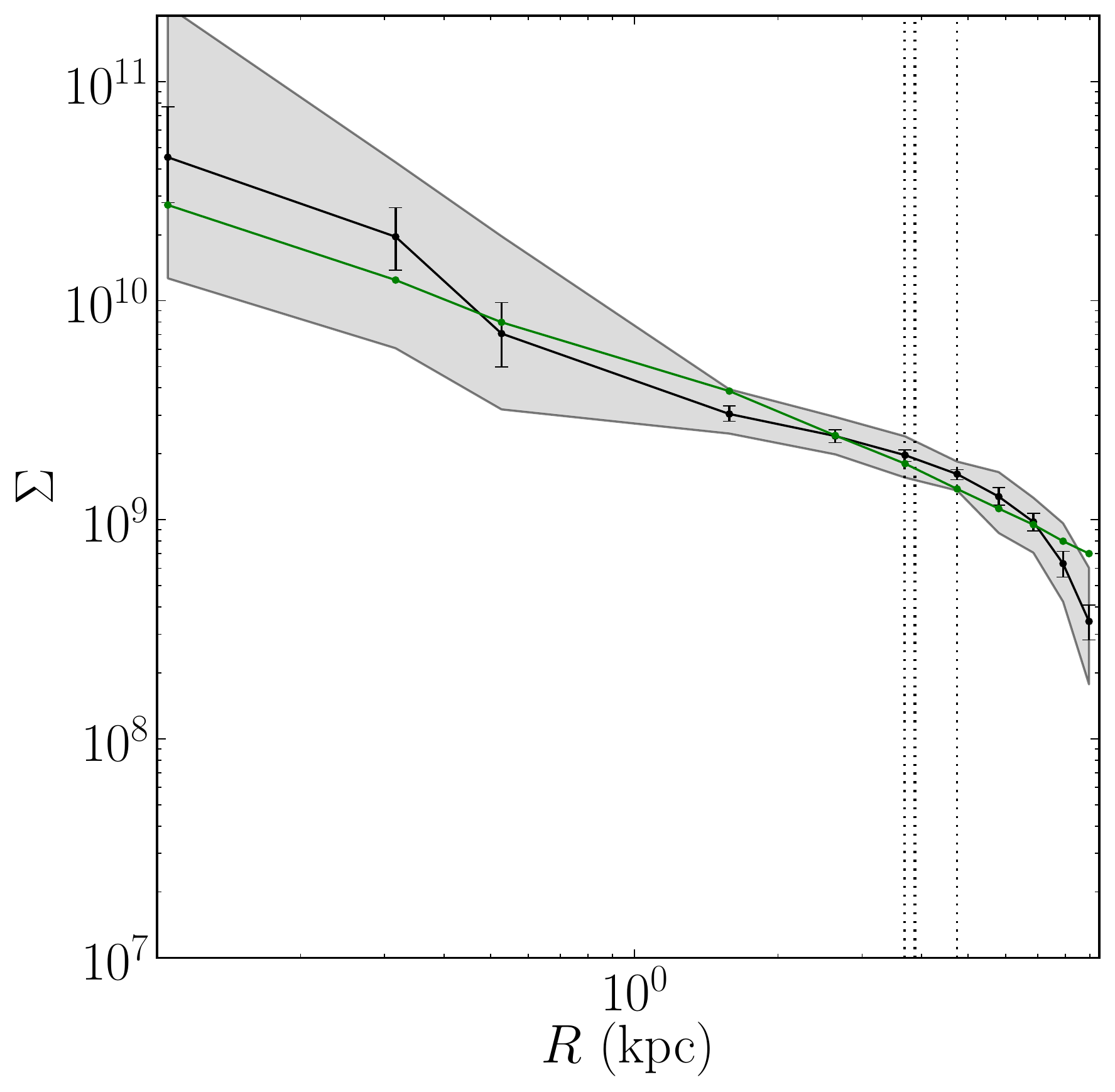} {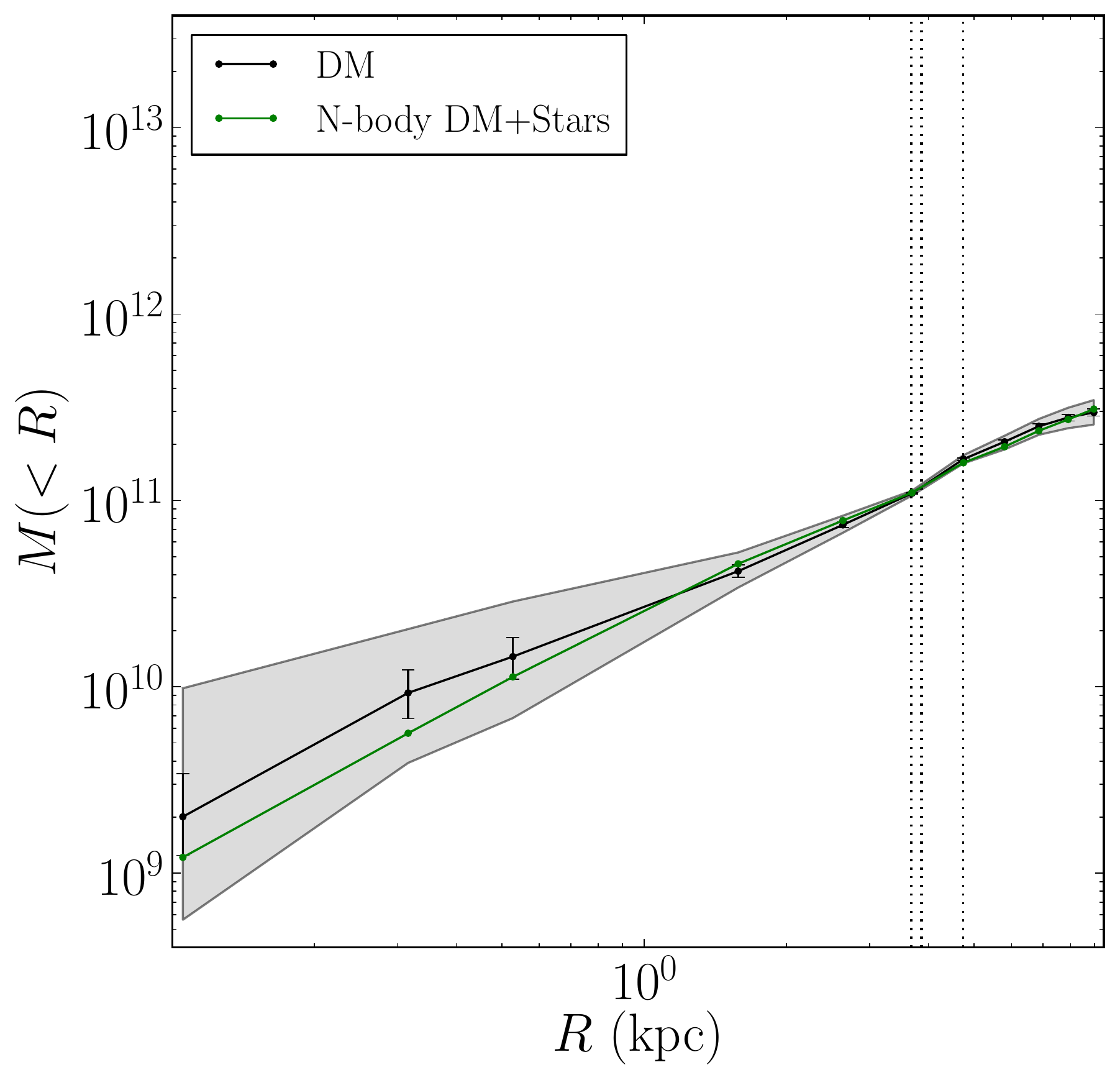}
  \plotthree{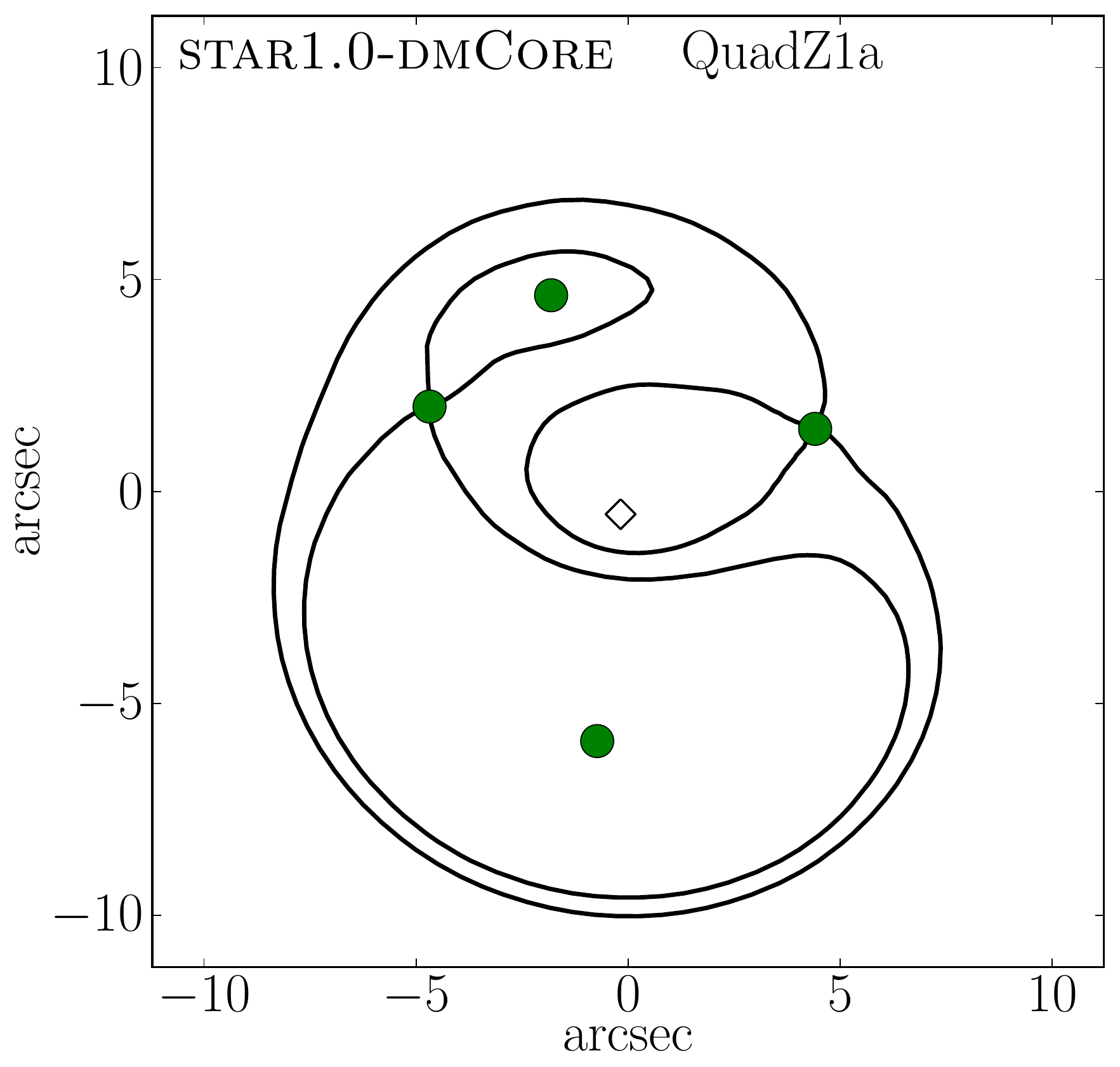} {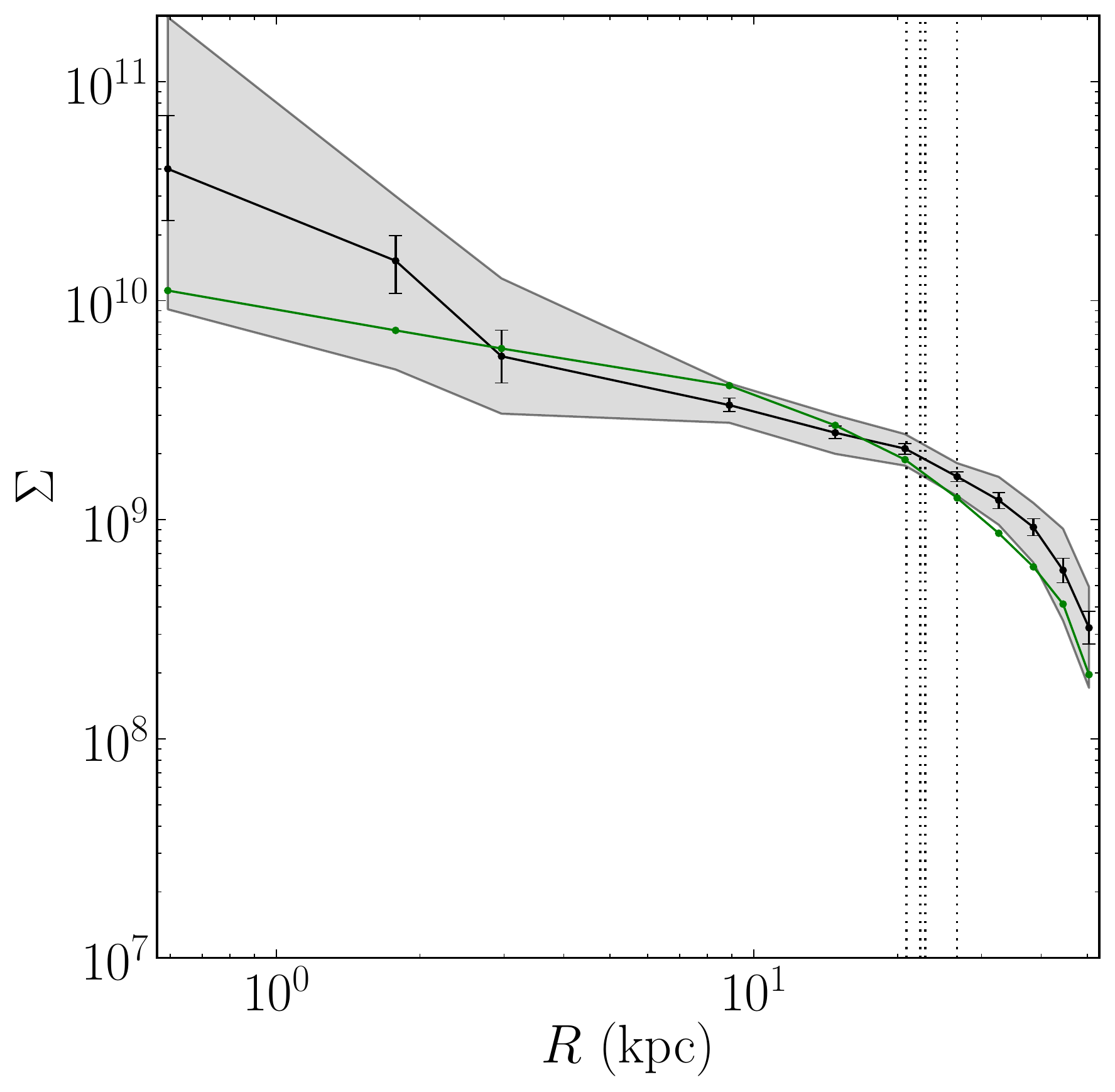} {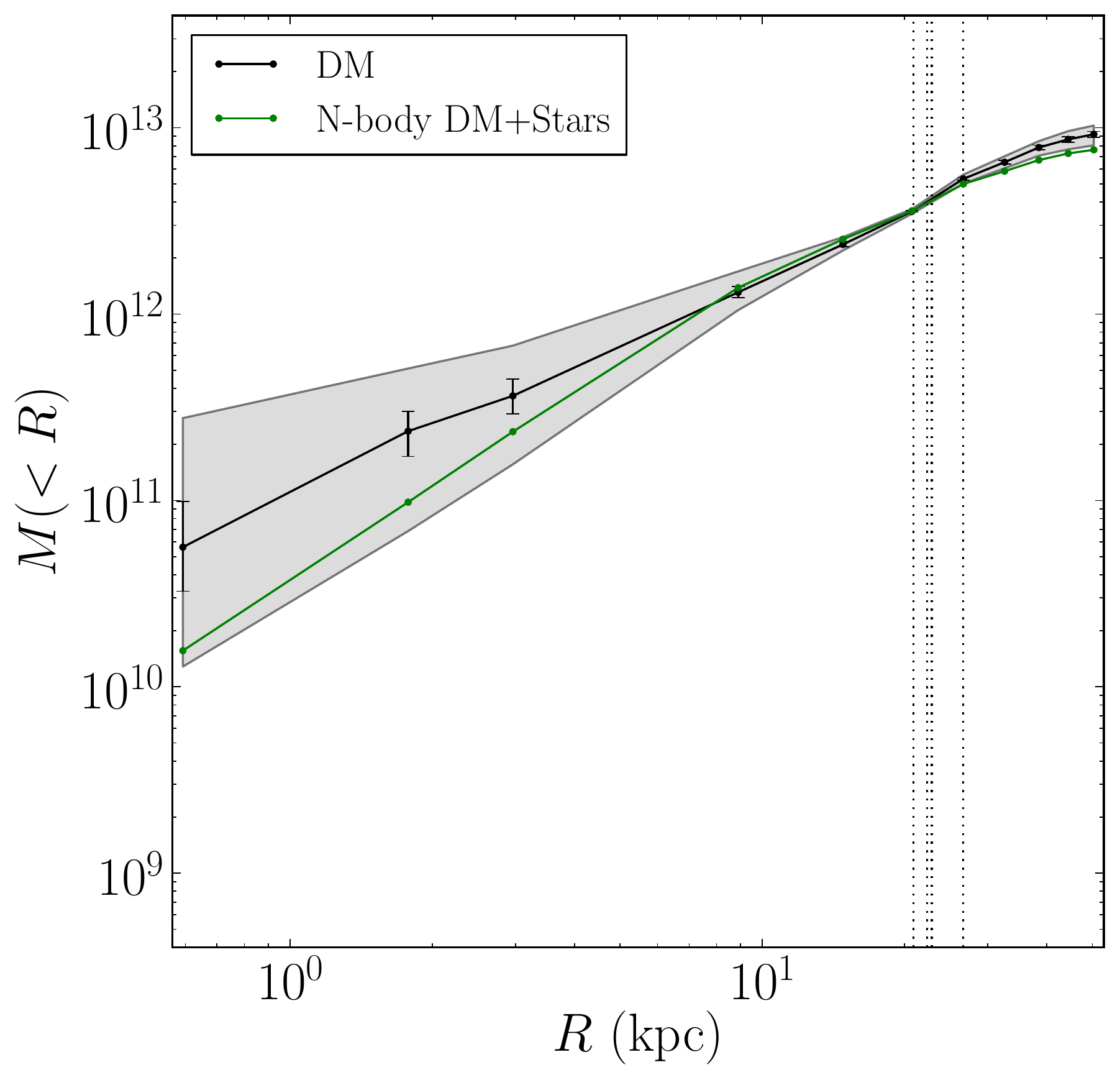}
\caption{
    Two further reconstructions similar to \figref{reconstruction}.
    \textbf{Top:} The mock galaxy \mockBC but including time delays for a
    single quad and no stellar mass. With the added information from the quad,
    the outer regions of the lens are the image radii are better constrained.
    \textbf{Bottom:} A quad with time delays, but using the \mockAA{} mock
    galaxy. This galaxy has a shallower stellar density index, and a core in
    the dark matter. Due to the priors used in \Glass, the modelling favours
    steep solutions without additional information.
}
\label{reconstruction 2}
\end{figure*}

\begin{figure*}
\includegraphics[width=0.33\textwidth]{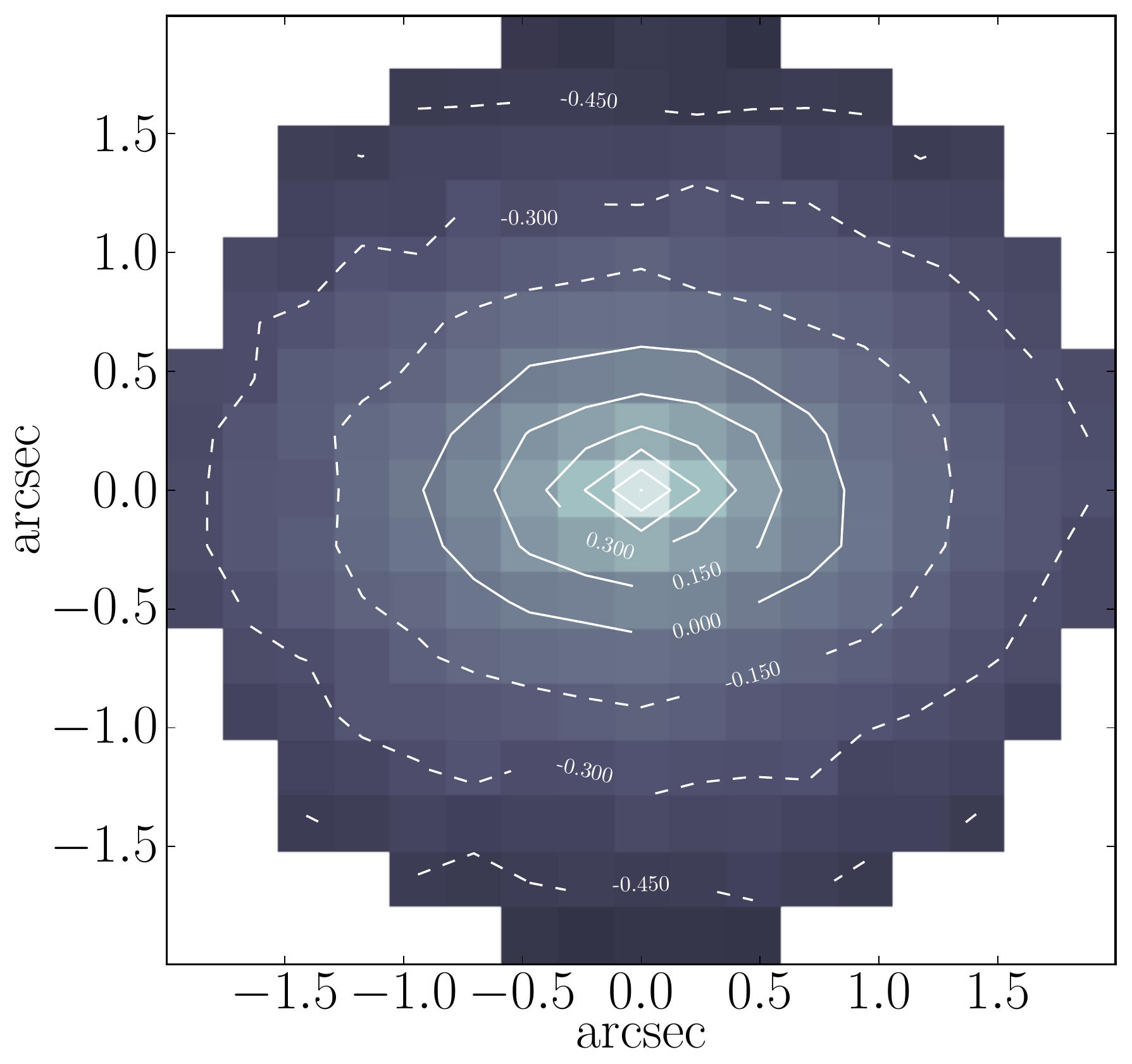}
\includegraphics[width=0.33\textwidth]{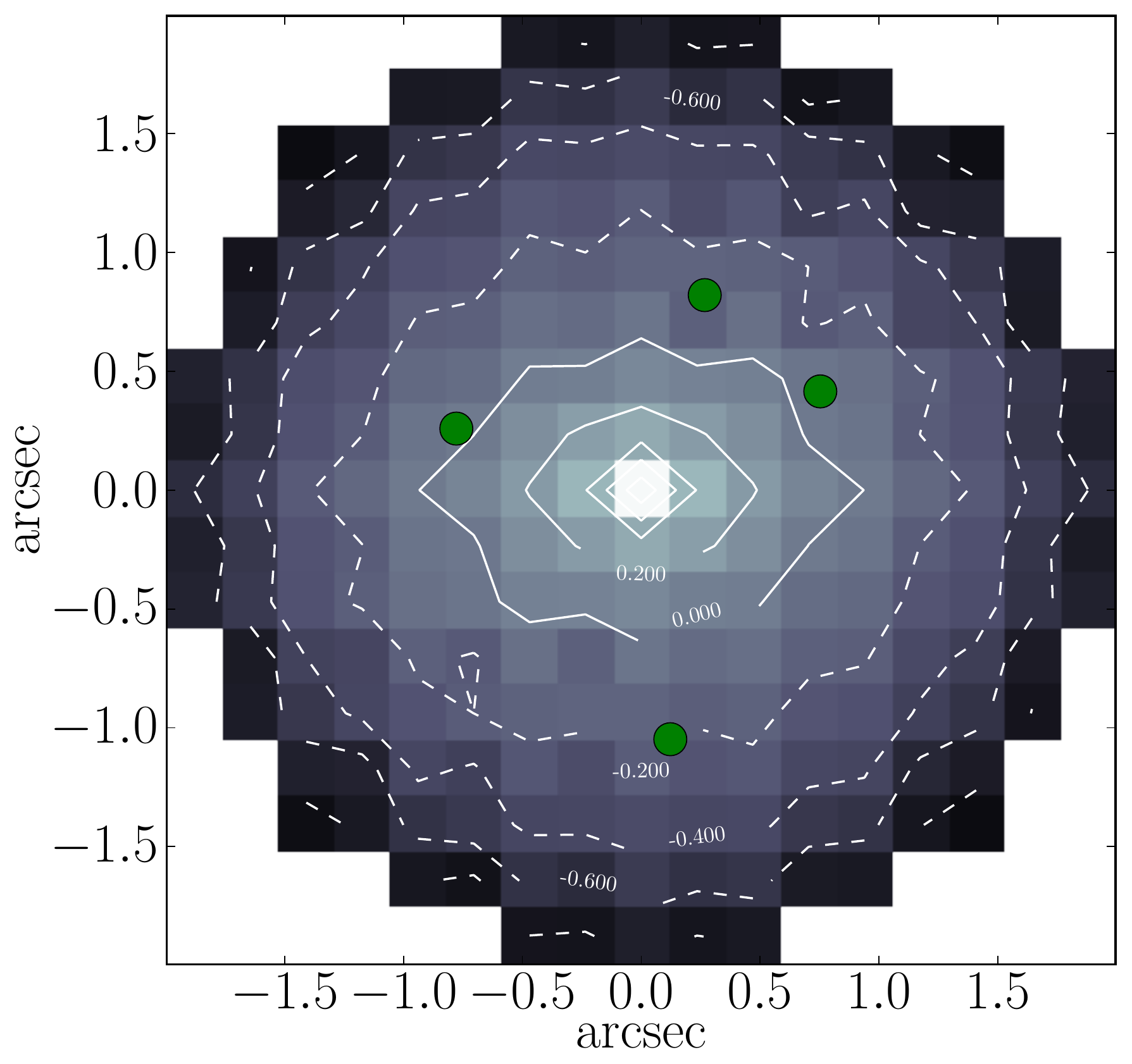}
\caption{ \textbf{Left:} The mock data distribution for \mockBC{} projected
onto a coarse grid.  \textbf{Right:} The recovered ensemble average $\kappa$
distribution for the single quad with time delays. The contours are logarithmic
base 10 values, where level 0 corresponds to the critical lensing density. Contours
below the critical lensing density are drawn with dashed lines.}
\label{2d mass reconstruction}
\end{figure*}

\begin{figure*}
\includegraphics[width=0.49\textwidth]{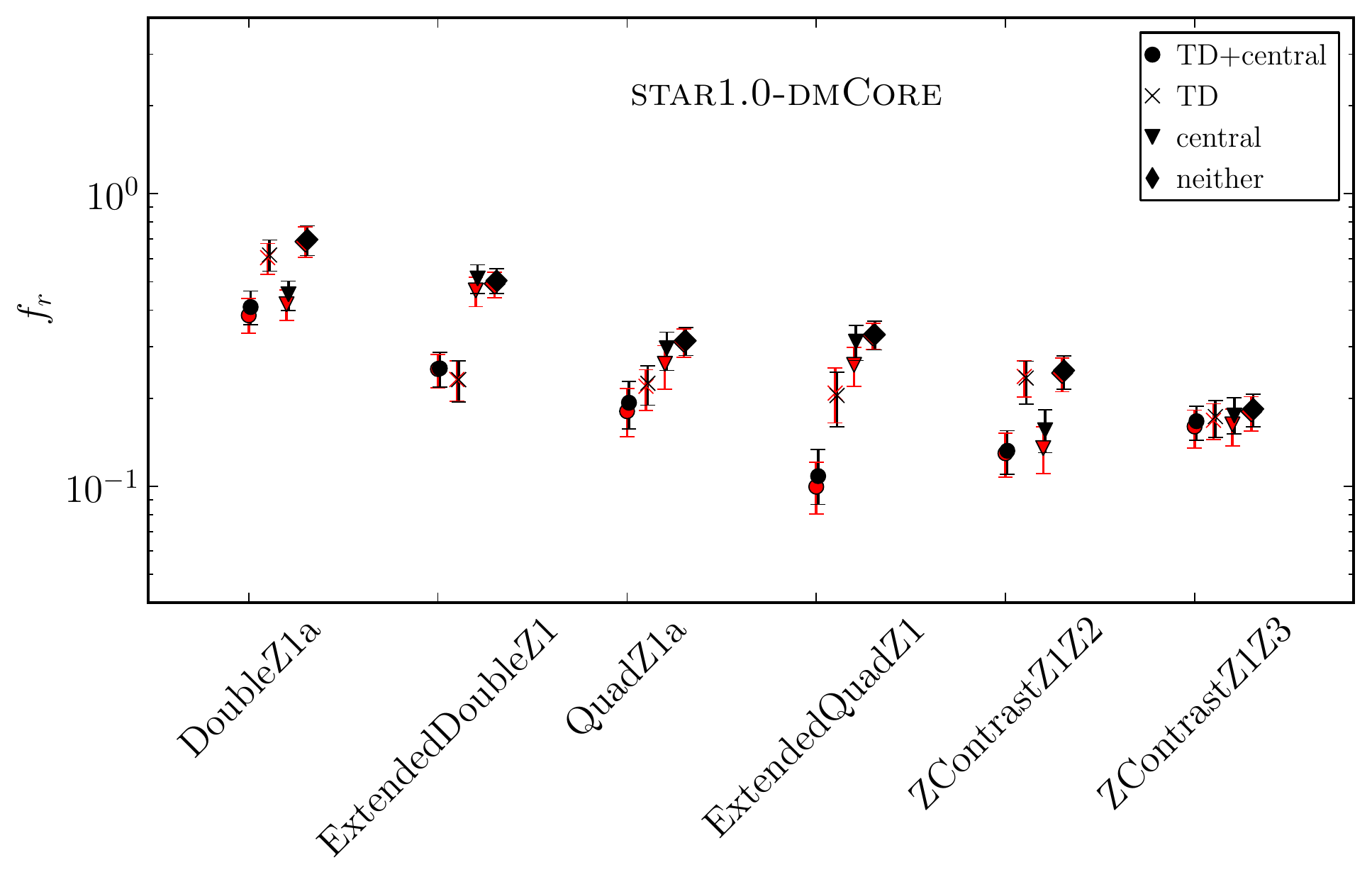}
\includegraphics[width=0.49\textwidth]{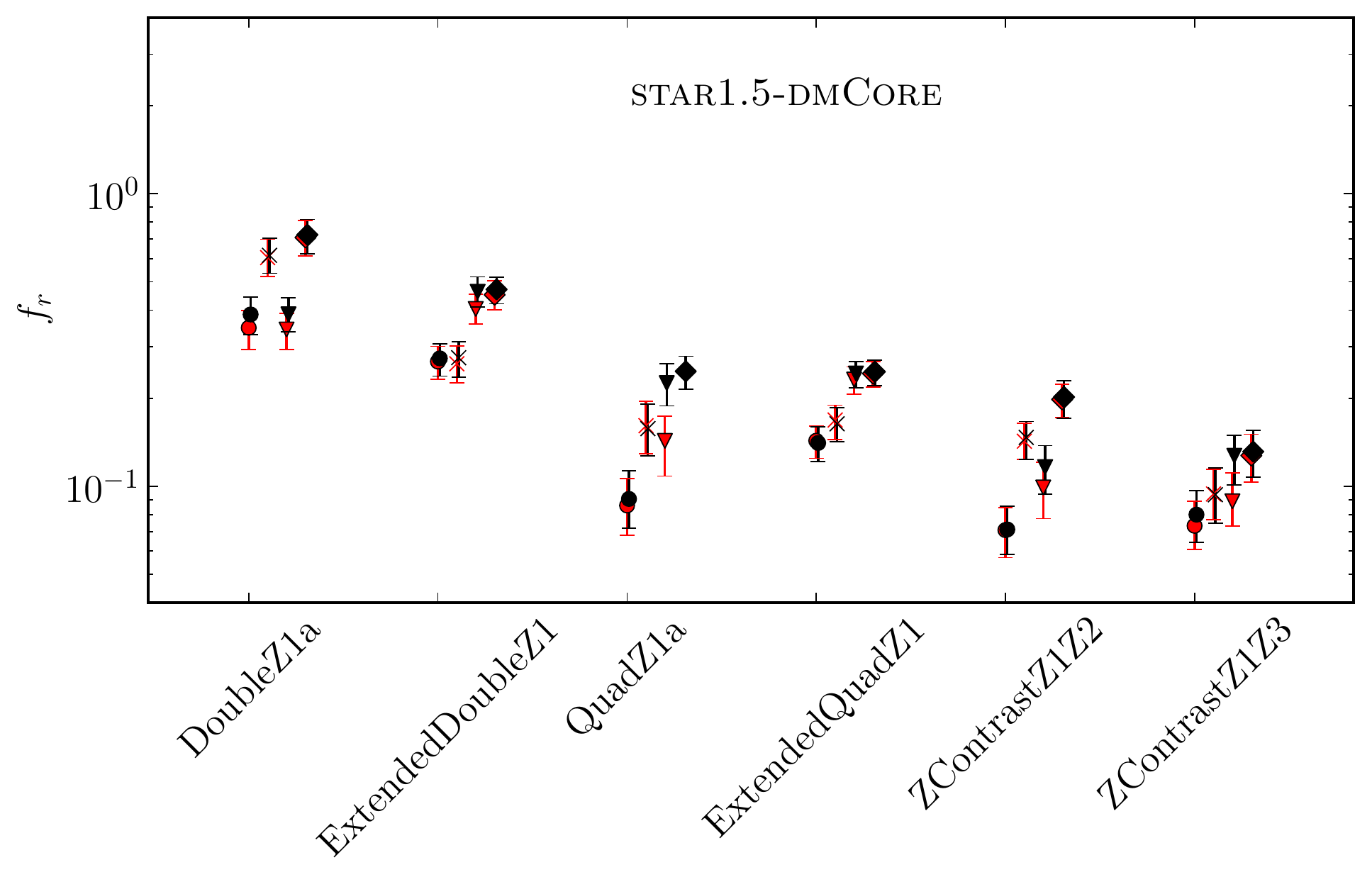}\\
\includegraphics[width=0.49\textwidth]{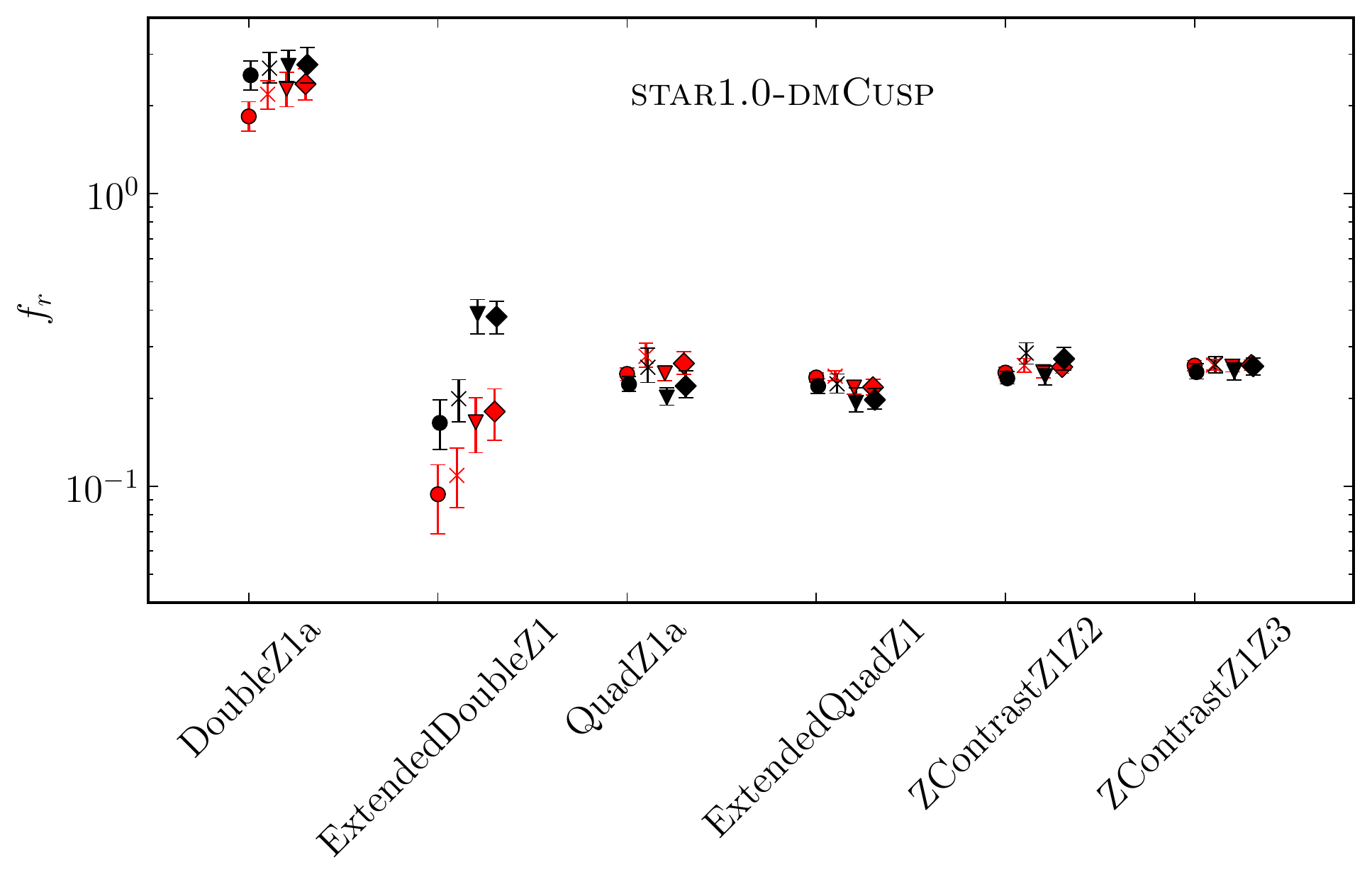}
\includegraphics[width=0.49\textwidth]{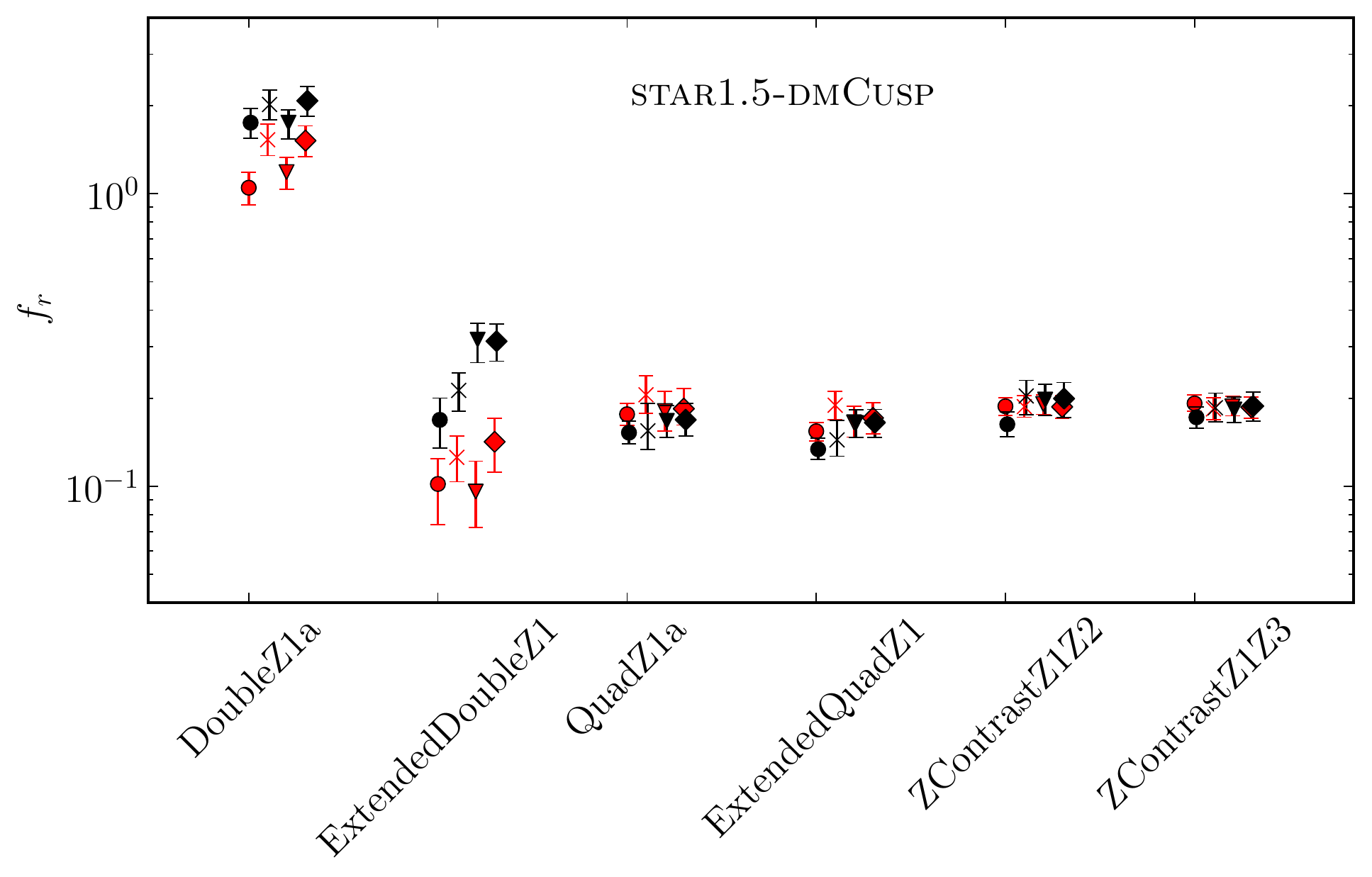}
\caption{Our main results showing the quality of the radially averaged model recovery \eqnrefp{ferror R} for all our test cases. 
Within each panel are six groups of results for each of six lens morphologies. Each morphology
considered the presence of time delays (TD) and a central image (central). The black markers are for tests
that did not include the stellar mass as a lower bound constraint, while the red markers
indicate where the stellar mass has been included. Error bars show the $1\sigma$ range of the model ensemble.}
\label{main results}
\end{figure*}

\begin{figure*}
\includegraphics[width=0.49\textwidth]{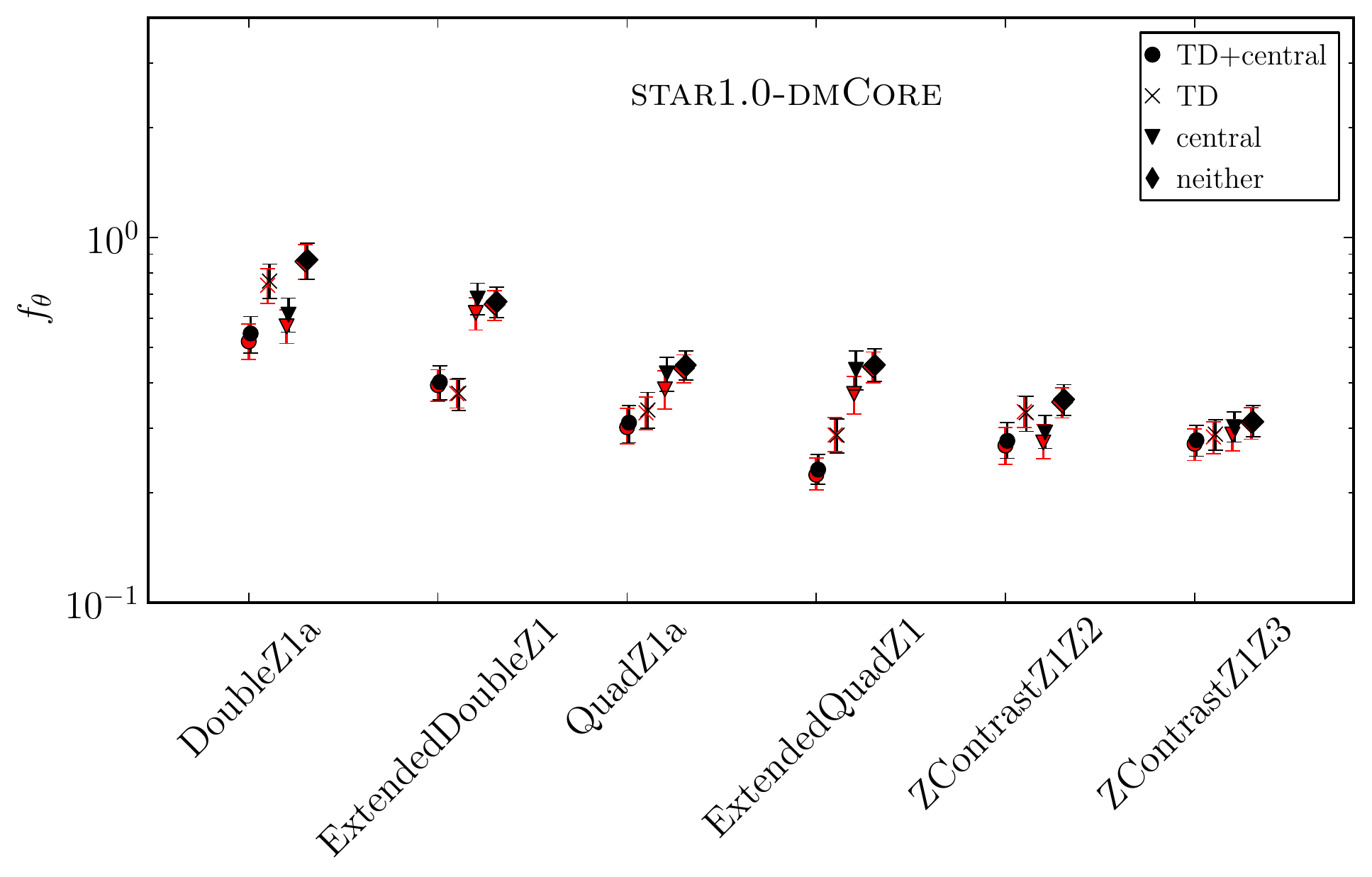}
\includegraphics[width=0.49\textwidth]{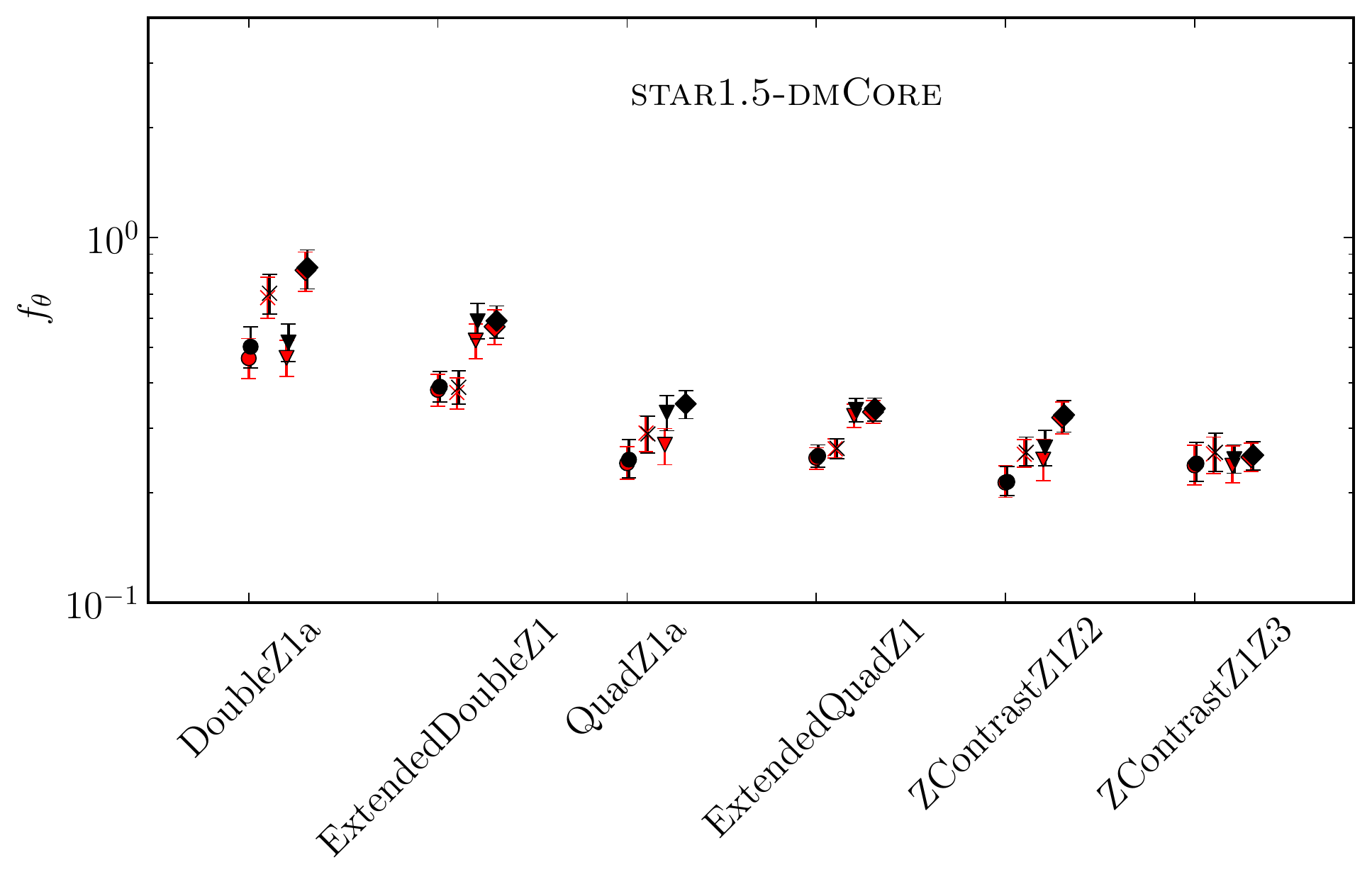}\\
\includegraphics[width=0.49\textwidth]{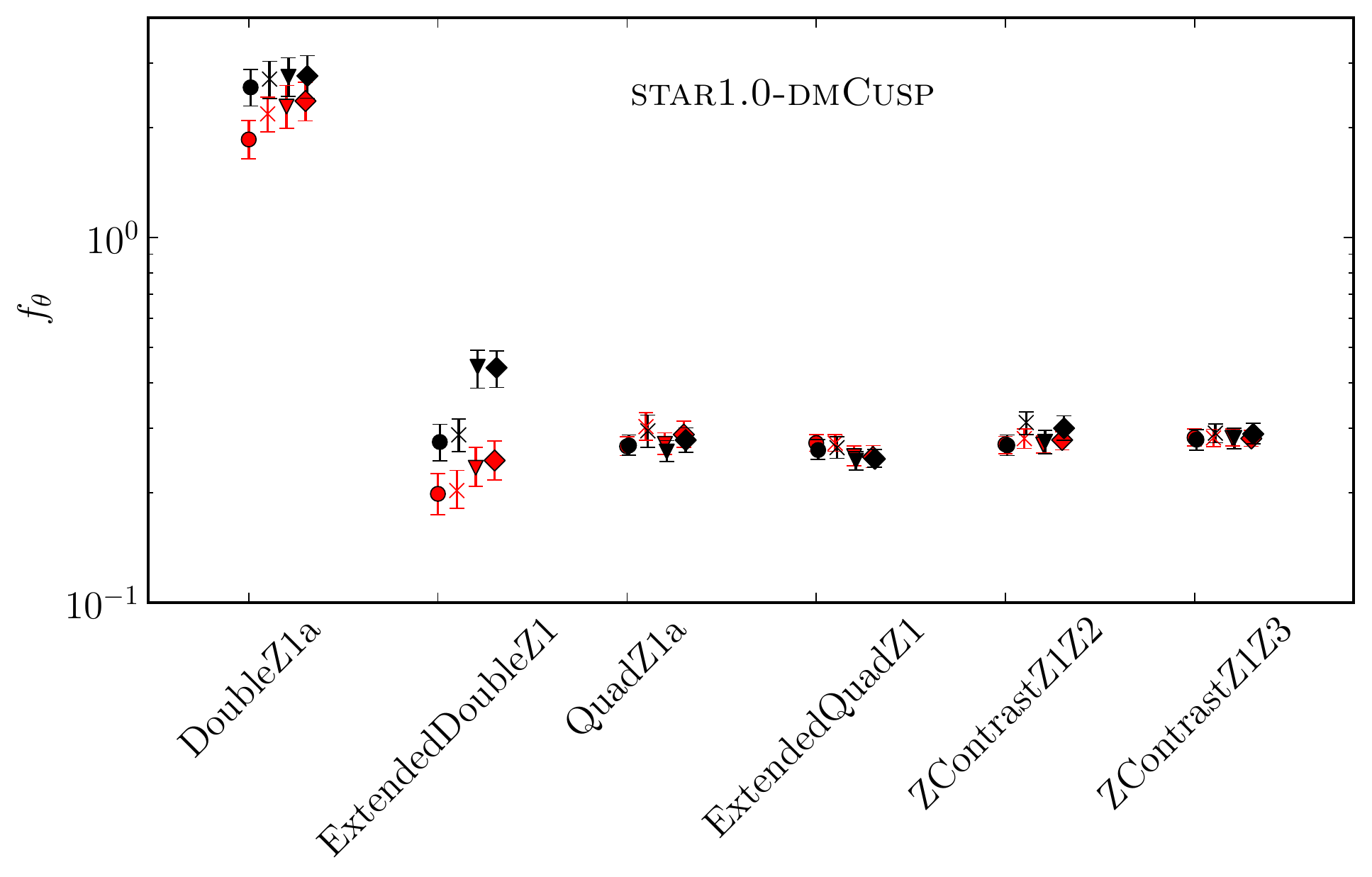}
\includegraphics[width=0.49\textwidth]{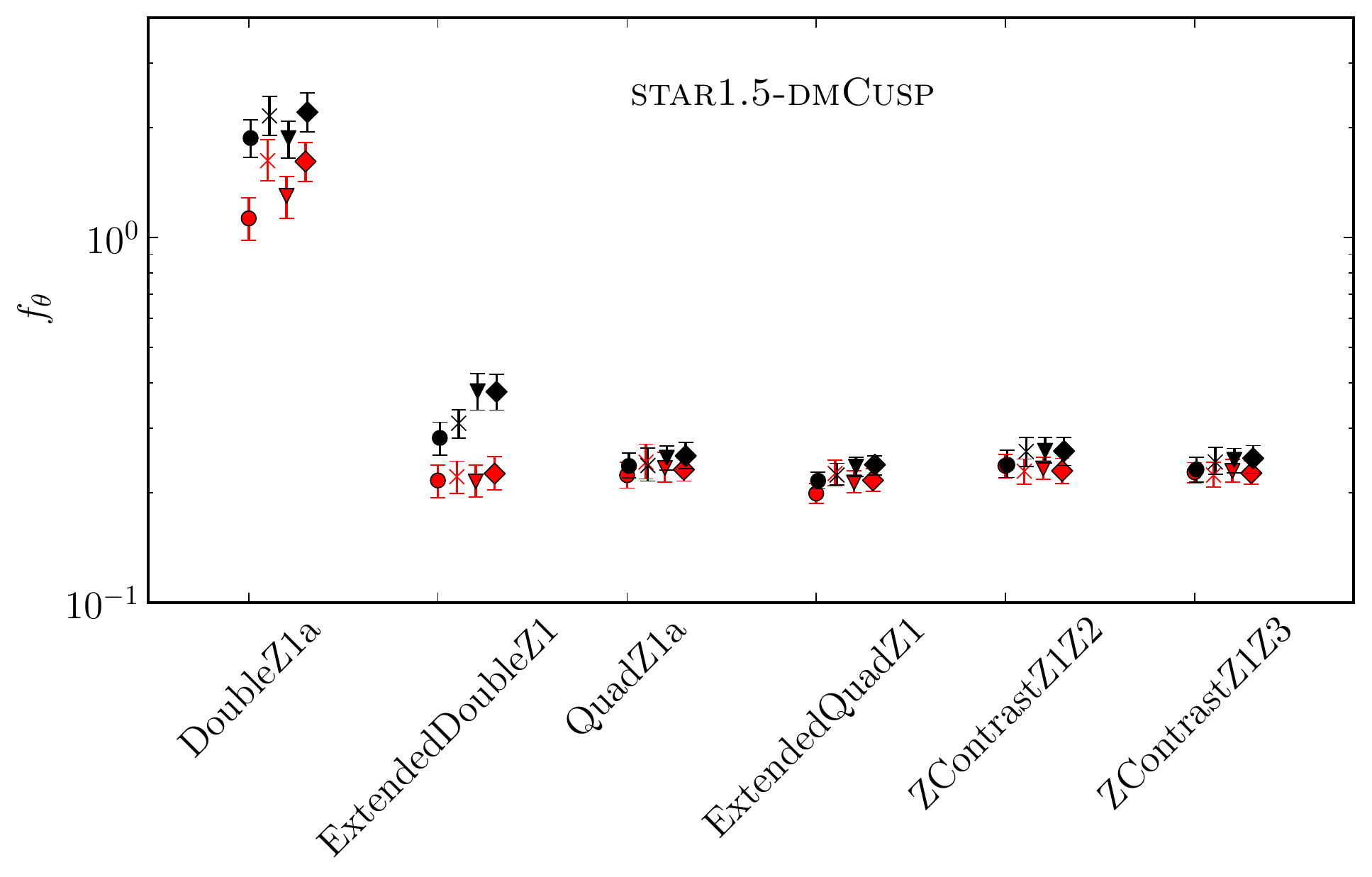}
\caption{Similar to \figref{main results} but for the fractional error
in the pixel-wise recovery \eqnrefp{ferror theta} of all models. The colours and labels are the same
as previously. Error bars show the $1\sigma$ range of the model ensemble.}
\label{main results pixel-wise}
\end{figure*}

\figref{main results} and \figref{main results pixel-wise} show the results for
our full mock data ensemble. Each subplot corresponds to a different mock
galaxy, as marked. We show the fractional error of the mass distribution for
each of the test configurations with (red) and without (black) stellar mass. In
\figref{main results} we define the error:
\begin{equation} \label{ferror R}
  f_R = \frac {\sum_i \left|M(i) - \widehat M(i)\right| } {\sum \widehat M(i)}
\end{equation}
based on the mass $M(i)$ of each pixel ring $i$ and the mass $\widehat M$ from the mock galaxy. 
In \figref{main results pixel-wise} the error is defined over all the pixels $\vec\theta$:
\begin{equation} \label{ferror theta}
f_\theta = \frac {\sum_{\vec\theta} \left|M(\vec\theta) - \widehat M(\vec\theta)\right| } {\sum_{\vec\theta} \widehat M(\vec\theta)}
\end{equation}
Since both error measurements consider the mass of each pixel, we are
implicitly weighting the recovered density by the varying size of the pixels.
The value $f_R$ emphasises the error one would see from radial profiles, while
$f_\theta$ is useful as a measure of how well each individual pixel is recovered.
For both $f_R$ and $f_\theta$, we only consider mass up to one pixel length passed the
outermost image, since there is no longer any lensing information beyond that
point. This means we typically use 8 bins, linearly spaced, ignoring the
outermost 3 bins.  The spacing changes, however, at the border between the high
resolution region in the middle.

The abundance of strong lensing data increases from left to right within each
plot. As a result, there is a general trend for the reconstruction quality to
increase (and therefore for $f$ to decrease). When both time delays and a
central image are present (TD+central), the quality is highest. A double is known to
provide very little constraint on the mass distribution. This is particularly
evident in galaxies \mockAC{} and \mockBC{} where the mass profile is steepest
and the reconstruction of the double is poorest. However, the addition of an
arc from the extended source is sufficient to correct this. Notice that, as in
\figref{reconstruction}, the recovery for \mockBC\ quickly saturates; there is
little improvement as the data improves beyond a single quad. This occurs
because the \Glass{} sample prior in the absence of data favours steep models
like \mockBC\ over shallower models like \mockAA\ (see also
\figref{reconstruction}). 

\subsection{Shape recovery}\label{sec:shape}

\figref{main results pixel-wise} already gives us important information about
how well we can recover the {\it shape} of a lens. The trends are very similar
to the radial profile recovery in \figref{main results}, suggesting that if the
radial profile is well-recovered then, typically, the shape is too. A notable
exception is for the \mockBC\ models where adding stellar mass constraints aids
the shape recovery, but little-improves the radial mass profile. A visual
example of the shape recovery is given in \figref{2d mass reconstruction}.

\begin{figure*}
\includegraphics[width=0.49\textwidth]{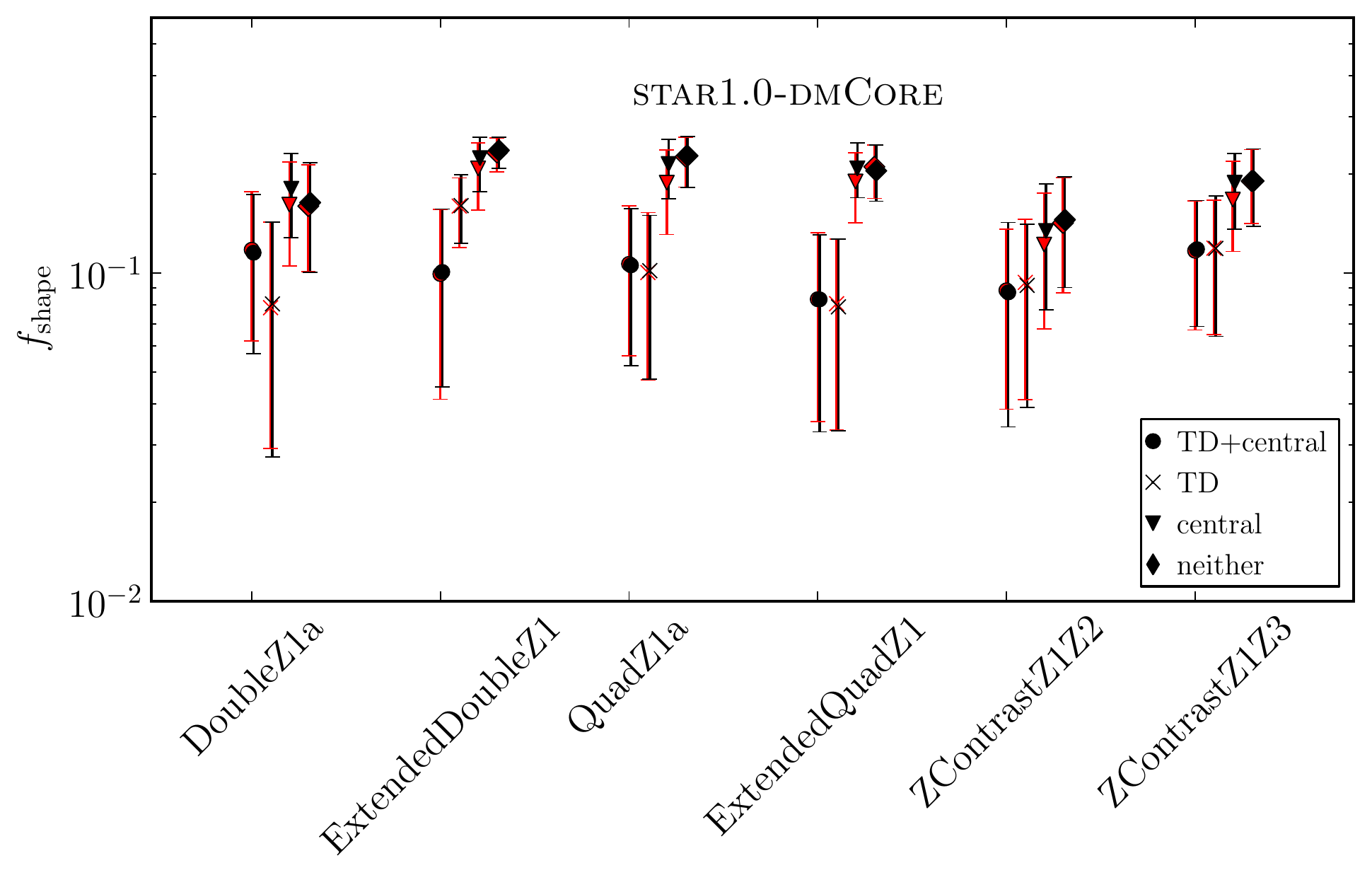}
\includegraphics[width=0.49\textwidth]{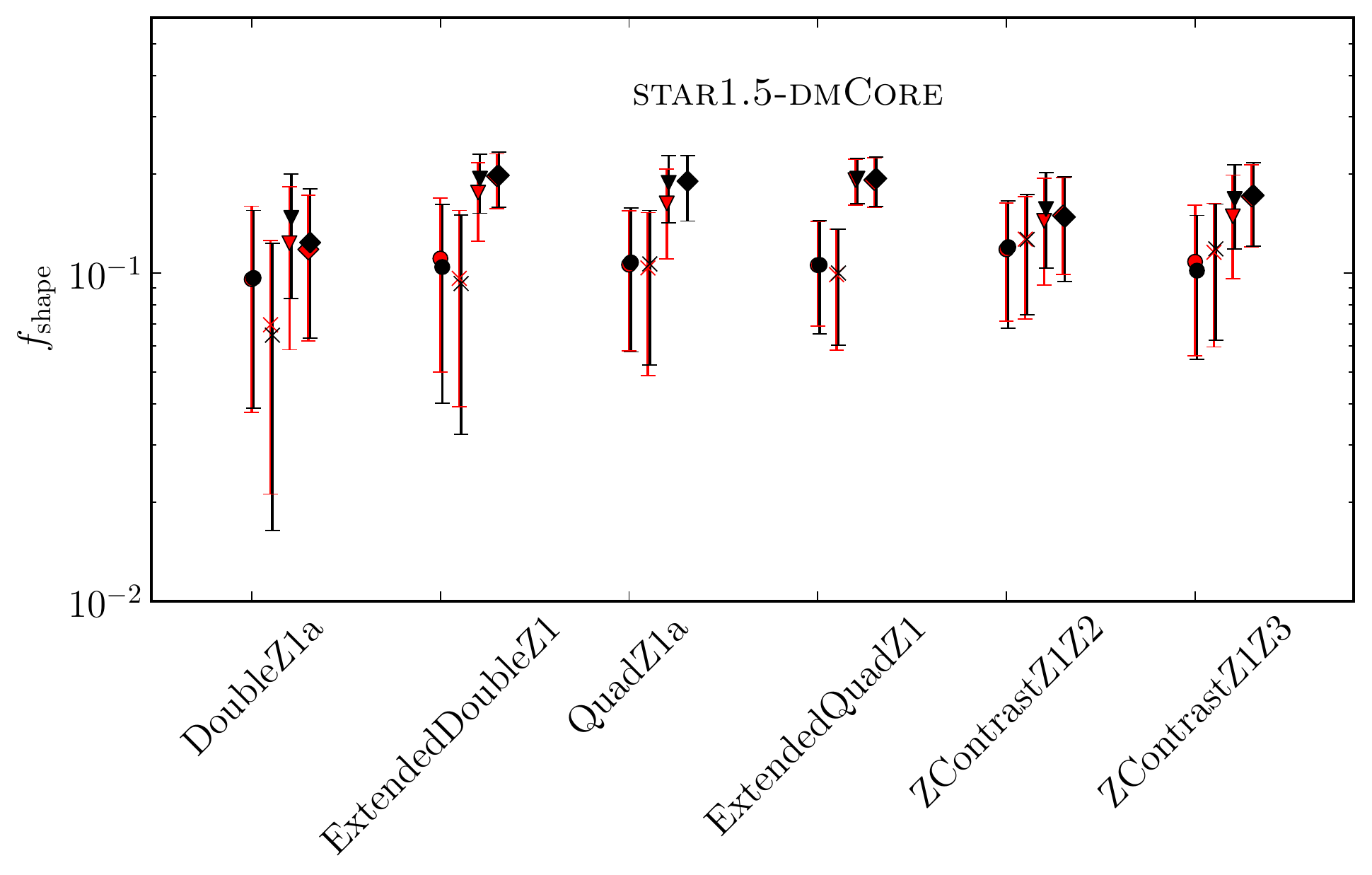}\\
\includegraphics[width=0.49\textwidth]{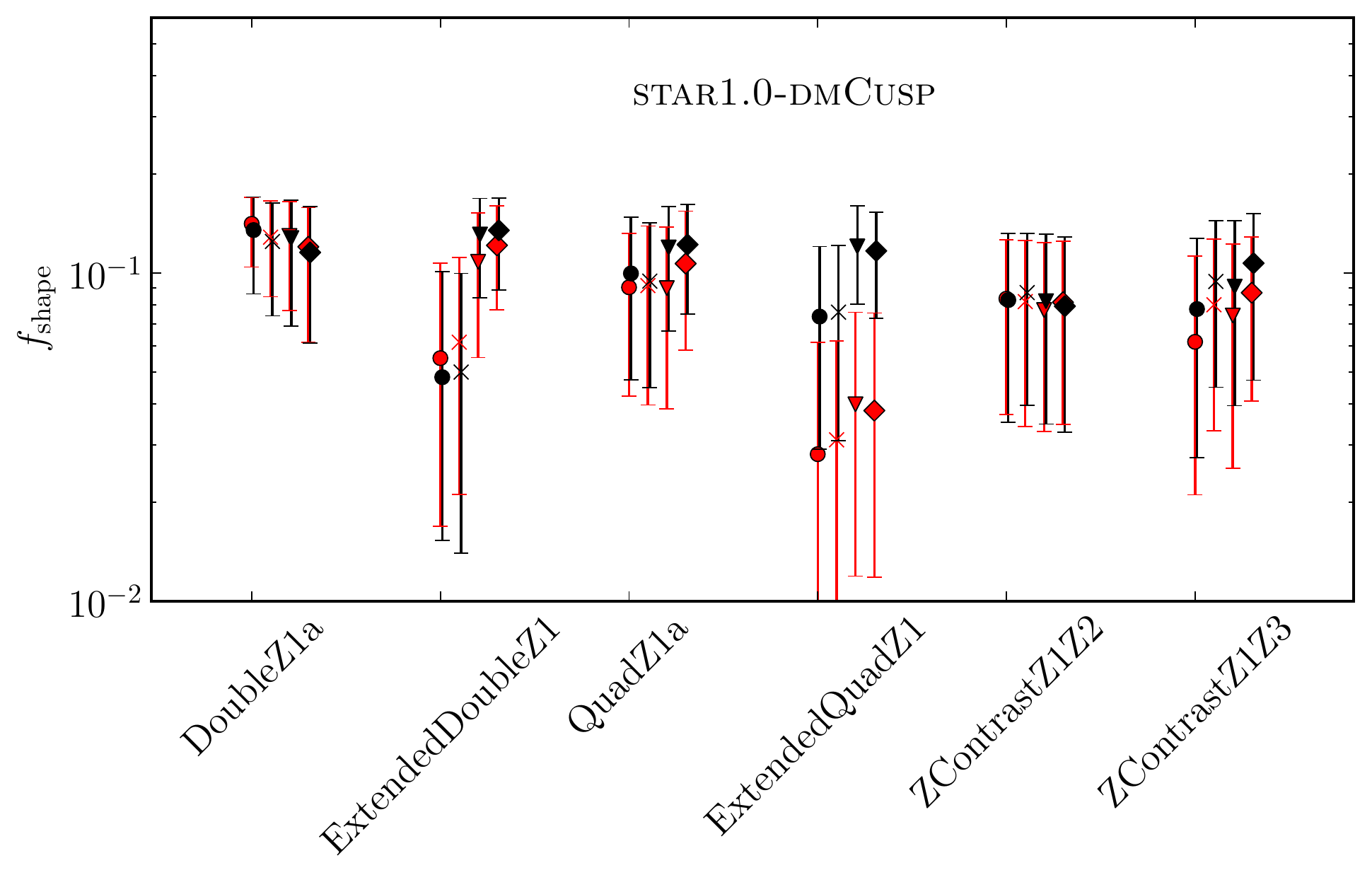}
\includegraphics[width=0.49\textwidth]{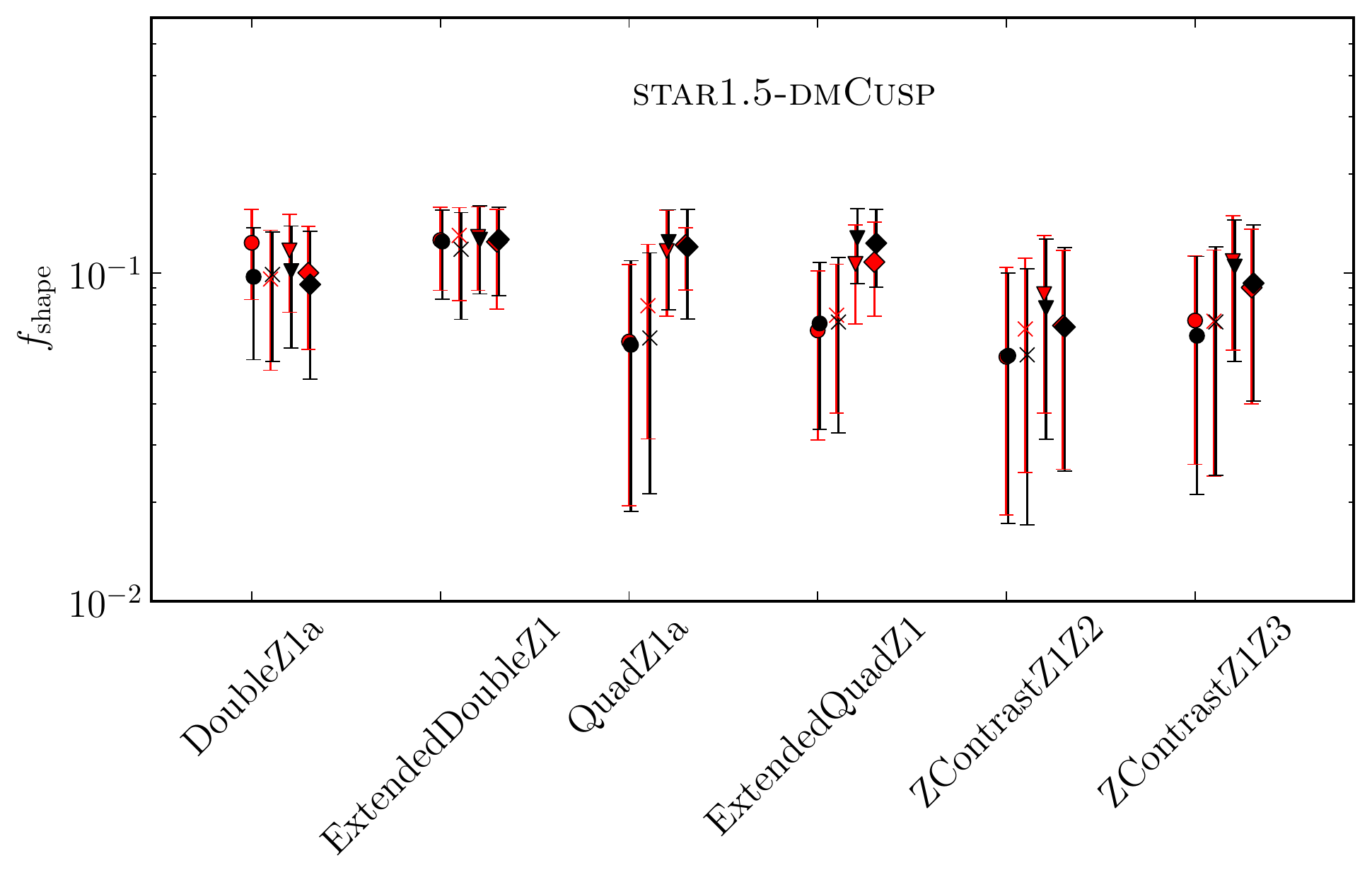}
\caption{Here we demonstrate our ability to recover the shape of the lensing mass. The shape ratio
$\lambda_1/\lambda_2$ is measured from the principal components $\lambda_1, \lambda_2$ of the mass up to the outermost image.
We plot the distribution of fractional error compared with the shape of the mock galaxies \eqnref{ferror shape}.}
\label{shape results}
\end{figure*}

We can also more directly probe the recovery of the shape of the mass
distribution by considering the ratio of the major and minor axes $\lambda_1,
\lambda_2$ of the inertia ellipse. If they are equal, the mass is distributed
uniformly on the projected disc. The more dissimilar they are, the more
elliptical the mass distribution. We define the global measure of lens shape as:
\begin{equation} 
    s \equiv \lambda_1/\lambda_2
\end{equation} 
where $\lambda_1$ and $\lambda_2$ are the eigenvalues of the 2D inertia tensor:
\begin{equation}
\left(
\begin{matrix}
   \sum_{\vec\theta} M(\vec\theta) \theta^2_y 
    & - \sum_{\vec\theta} M(\vec\theta) \theta_x \theta_y \\
    - \sum_{\vec\theta} M(\vec\theta) \theta_x \theta_y
    & \sum_{\vec\theta} M(\vec\theta) \theta^2_x
\end{matrix}
\right)
    \label{eqn:inertia}
\end{equation}
We always take $\lambda_1$ to be the largest value. As with $f_R$ and
$f_\theta$, we only consider mass up to one radial position passed the
outermost image and compute the fractional error as:
\begin{equation} \label{ferror shape}
  f_\mathrm{shape} = \left|s - \widehat s\right|  / \widehat s
\end{equation}
where $\widehat s$ is the shape of the mock galaxy. The distribution of
$f_\mathrm{shape}$ for each mock galaxy and each test case is shown in
\figref{shape results}.  Interestingly, for this global shape parameter
recovery it appears more important to have time delay data and/or a central
image (TD,TD+central) than to have a quad or multiple sources with wide redshift
separation. In all cases, the stellar mass little-aids the recovery, reflecting
the fact that $s$ is heavily weighted towards the shape at the {\it edge} of
the mass map, rather than at the centre where the stars may dominate the
potential (see \eqnref{eqn:inertia}).

\begin{figure*}
\includegraphics[width=0.33\textwidth]{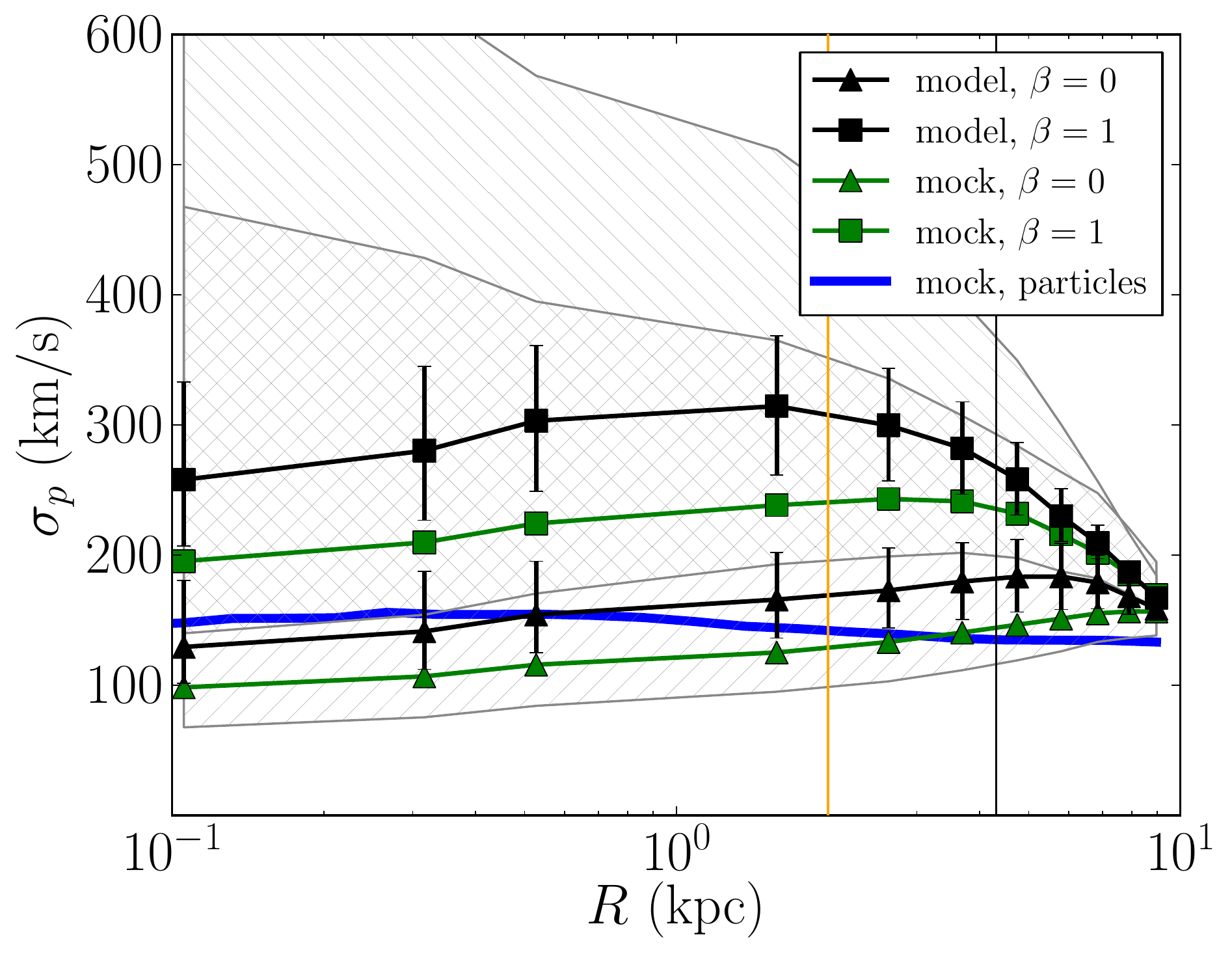}
\includegraphics[width=0.33\textwidth]{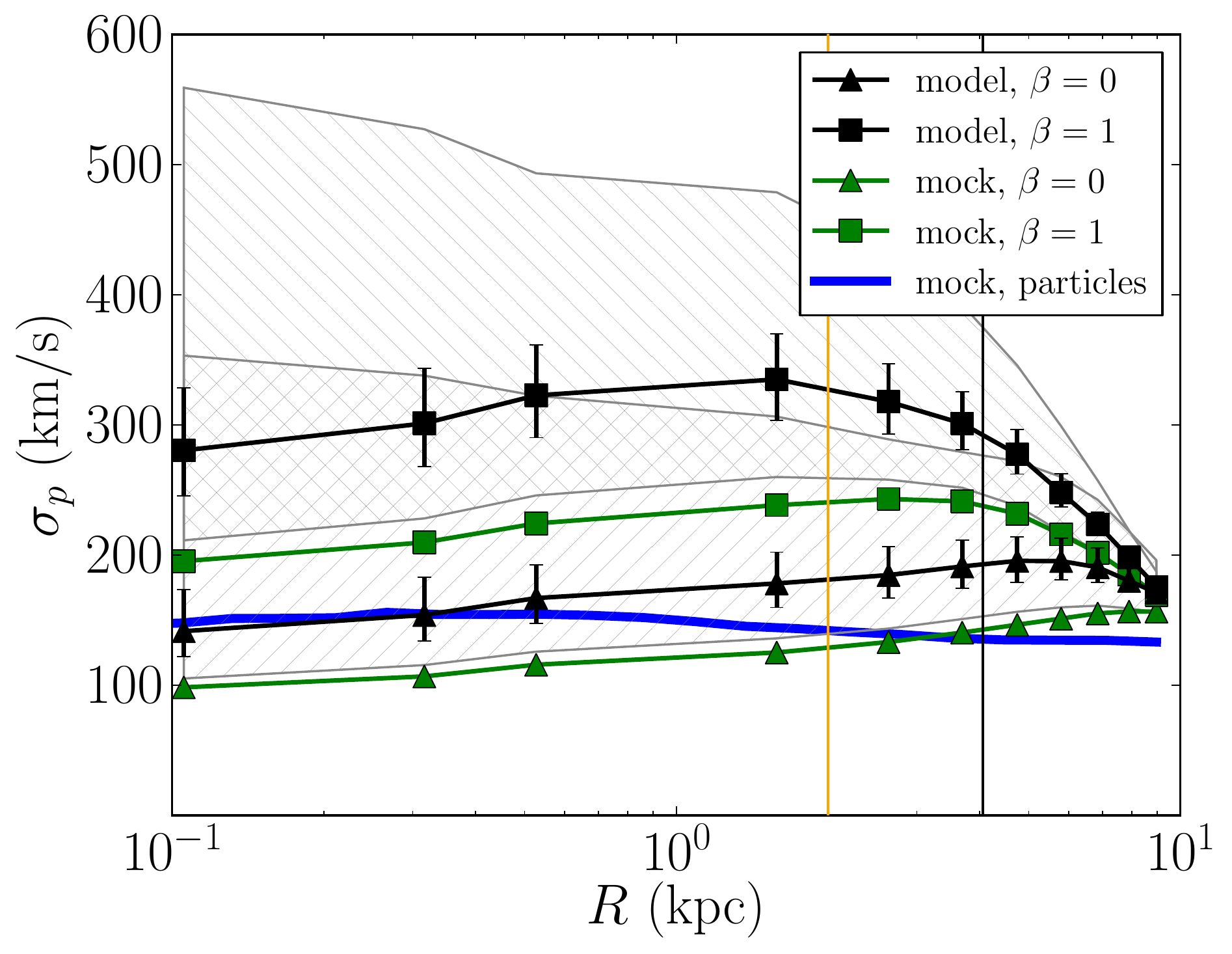}
\includegraphics[width=0.33\textwidth]{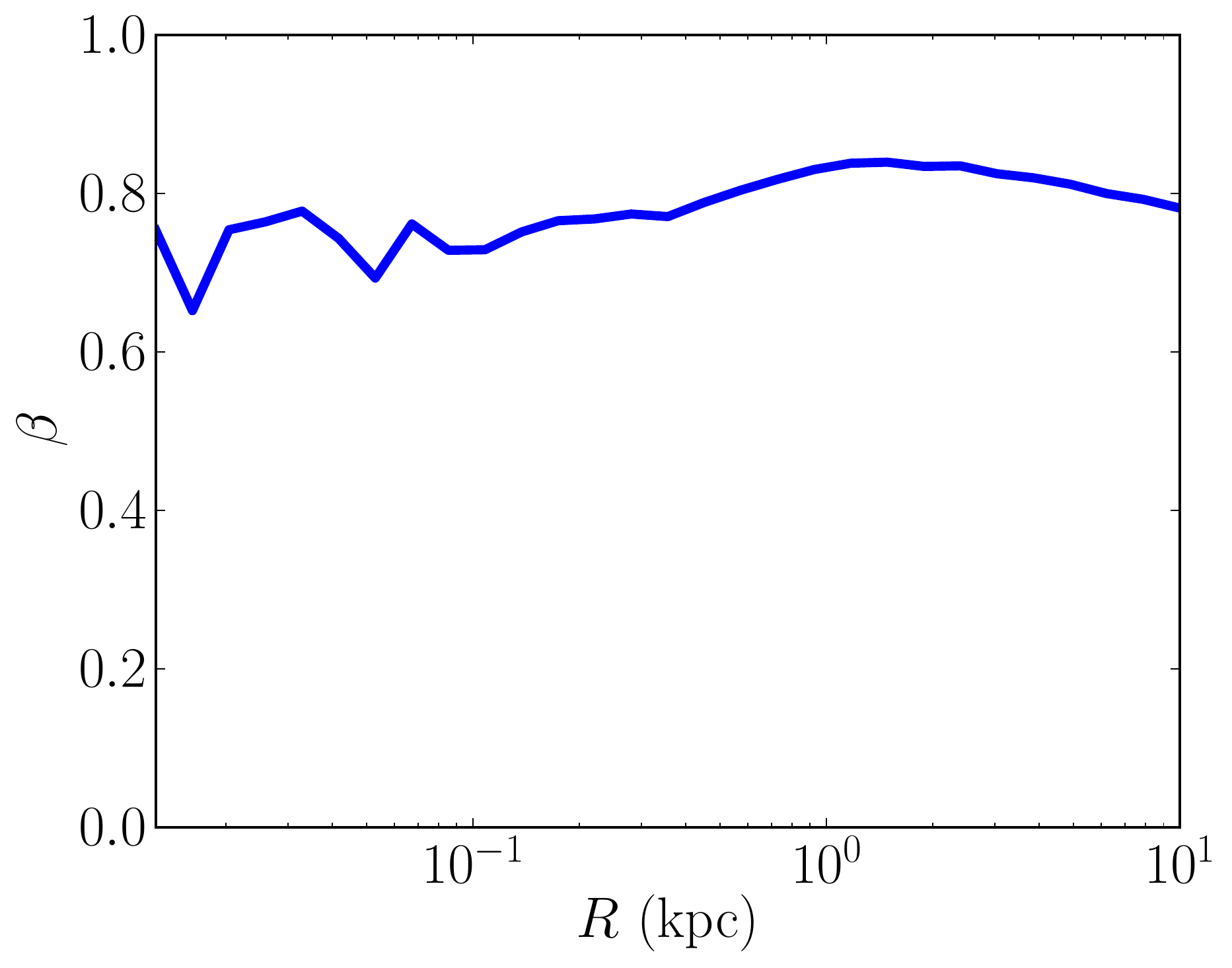}
\caption{ Estimated projected radially averaged velocity dispersion $\sigma_p$
\eqnrefp{eqn:sphericaljeans} for a single quad from the \mockBC{} mock galaxy
without stellar mass (\textbf{left}) and with stellar mass (\textbf{middle})
assuming an anisotropy $\beta=0$ (black triangles) and $\beta=1$ (black squares).
Error bars are $1\sigma$ and two overlapping hatched areas indicate the full range of models.
The equivalent curves are also shown for the projected mock data after using
the same analysis routines (green). The solid blue line is the actual
cylindrically averaged velocity dispersion of the original mock particle data.
The stellar half mass radius (orange) Einstein radius (black) are marked by
vertical lines. For this configuration, these two radii are well-separated. The
actual variation in $\beta(r)$ is also shown (\textbf{right}).}
\label{fig:sigp} \end{figure*}
\subsection{Stellar mass}
\label{stellar mass}

The stellar mass distribution gives a lower bound on the total mass. Where the
stars dominate the central potential, it can provide a powerful constraint
extra to the strong lensing data. We took the stellar mass directly from the
generated galaxies and projected the particles onto the pixels. \Glass{} also
offers an option to interpolate any map of stellar mass (e.g., from an
observation) onto the pixels. The linear constraint is added to \Glass{} by
writing $\kappa_n = \kappa_{dm,n} + \kappa_{s,n}$ as the sum of the dark matter
and stellar mass components in the potential \eqnrefp{discrete potential}.
Since each $\kappa_{s,n}$ is just a constant we do not add new, separate
equations for each pixel. Although we assume a perfect recovery of the stellar
mass with no error on the lower mass bound, it is straightforward to add errors
as the stellar mass constraint remains linear: $\kappa_n = \kappa_{dm,n} +
\epsilon \kappa_{s,n}$, where $\epsilon \sim 1$ is an additional error
parameter. 

With the stellar mass lower bound, there is a significant improvement of the
reconstruction quality shown in \figref{main results} and \figref{main results
pixel-wise} for the doubles in the steepest mock galaxies (\mockAC{} and
\mockBC). This is because these models are dominated by stars in the inner
region. By contrast, the other two galaxies -- where the stars contribute
negligibly to the potential -- are largely unaffected.

\subsection{Stellar kinematics}\label{sec:results_stellar_kinematics}

As outlined in \secref{sec:glass}, \Glass{} can also run post processing
routines on the model ensemble which can be used to apply non-linear
constraints. As an example, we consider here constraints from stellar
kinematics. The models in the \Glass{} ensemble are processed as described in
\S\ref{sec:glasskinematics}.  To illustrate the power of stellar kinematic
constraints, in \figref{fig:sigp}, we plot the projected velocity dispersion
calculated for one model model (extracted from the full ensemble) of the
\mockBC{} Quad with time delays and no stellar mass (left), and the same but
with stellar mass (middle).  In both cases, we calculate curves for two
extrema velocity anisotropies: $\beta=0$ (green) and $\beta=1$ (red).
Over-plotted is the correct answer for the \mockBC{} model (black). The stellar
half mass radius (yellow) Einstein radius (black) are marked by vertical lines.
For this configuration, these two radii are well-separated.

Without even sweeping through the model ensemble and formally
accepting/rejecting models, \figref{fig:sigp} already illustrates what we can
obtain from stellar kinematics. The left plot shows the radially averaged
projected velocity dispersion $\sigma_p(R)$ \eqnrefp{eqn:sphericaljeans} for a
single quad from the \mockBC\ galaxy without the stellar mass constraint. The
blue data points show the 1$\sigma$ distribution from the ensemble assuming
$\beta=0$ (solid) and $\beta = 1$ (dashed); the grey bands show the full
distributions. Also marked are the $\sigma_p(R)$ calculated from the mock data
assuming $\beta=0$ (solid purple) and $\beta =1$ (dashed purple); and the true
$\sigma_p(R)$ measured directly from the stars (black). This latter has a
non-constant $\beta(r)$ (right panel) and differs also from the purple and blue
curves in that these all assume spherical symmetry, whereas the stellar
distribution is really triaxial. Such triaxiality and varying $\beta(r)$
explains why the purple curves do not match the black one. However, they do
largely bracket the correct solution. More interestingly, the curves
approximately cross for $\beta = 0$ at the stellar half light radius (yellow
vertical line). This demonstrates, as has previously been reported in the
literature, that $\sigma_p(R)$ gives a good estimate of the mass enclosed
within $\sim$ the half light radius $M_{1/2}$
\citep[e.g.,][]{2009ApJ...704.1274W,2010MNRAS.406.1220W}. The mass {\it
profile}, however, depends on $\beta$ which is poorly constrained by these
data. If we add stellar mass constraints (middle panel), the situation is
little-improved. The true answer already lay close to the bottom of the
ensemble distribution; it now is forced to lie right at the edge. 

From \figref{fig:sigp}, it is clear that $\sigma_p(R)$ provides two useful
pieces of information. Firstly, it is a powerful probe of $M_{1/2}$. Given a
measurement of $\sigma_p(r_1/2) \sim 150$\,km/s, we could usefully reject many
models in the ensemble as being overly steep in the centre. We would not,
however, obtain a strong constraint on $\beta(r_{1/2})$. We could rule out
$\beta(r_{1/2}) = 1$ (blue dashed line), but since our $\beta=0$ model crosses
the true $\beta \sim 0.5$ line at $r_{1/2}$ it is clear that many $\beta(r)$
profiles will be consistent with the data. On the other hand, if we have a
situation where $r_{1/2} \sim r_E$ (i.e. the vertical yellow and black lines in
\figref{fig:sigp} overlap), then we will obtain tight constraints on $\beta$
since we then have two strong constraints on $M(r_{1/2})$ that become
redundant. This latter situation of redundancy is also exactly what we would
like to constrain cosmological parameters. In this case, we require a third
piece of redundant information -- in this case in the form of strong lensing
time delays. We will discuss such cosmological constraints in a forthcoming
paper. 

The results for stellar kinematics match our expectations from
\secref{sec:kinematics}. Where the lens data already constrain the mass
distribution at $r \sim r_{1/2}$, stellar kinematics provide valuable
information about the velocity anisotropy of the stars, $\beta$ (see
\figref{fig:sigp}). Where the lens data poorly constrain the mass distribution
at $r_{1/2}$, we may `integrate out' the effect of unknown $\beta$ to obtain a
robust measure of $\Mddd(r_{1/2})$ from the stellar kinematics. This latter is
robust to both uncertainties in $\beta(r)$ and to our assumption of spherical
symmetry in the kinematic models \citep{2012ApJ...754L..39A}.

\section{Conclusions}\label{sec:conclusions}

We have introduced a new gravitational lens modelling tool -- Glass{} -- and
used it to test the recovery of the mass profile and shape of mock strong
lensing galaxies. Our key findings are as follows: 

\begin{enumerate}
\item For pure lens data, multiple sources with wide redshift separation give
    the strongest constraints as this breaks the well-known mass-sheet or
    steepness degeneracy;

\item A single quad with time delays also performs well, giving a good recovery
    of both the mass profile and its shape; 

\item Stellar masses -- for lenses where the stars dominate the central
    potential -- can also break the steepness degeneracy, giving a recovery for
    doubles almost as good as having a quad with time delay data, or multiple
    source redshifts; 

\item If the radial density profile is well-recovered, so too is the shape of a lens; 

\item Stellar kinematics provide a robust measure of the mass at the half light
    radius of the stars $M(r_{1/2})$ that can also break the steepness
    degeneracy if $r_{1/2} \neq r_E$ -- the Einstein radius; and

\item If $r_E \sim r_{1/2}$, then stellar kinematic data can be used to probe
    the stellar velocity anisotropy $\beta$ -- an interesting quantity in its
    own right. 

\end{enumerate}

Where information on the mass distribution from lensing and/or other probes
becomes redundant, this opens up the possibility of using strong lensing to
constrain cosmological models. We will study this, and present the first
results from \Glass{} applied to real data, in forthcoming papers.

\section{Acknowledgments}\label{sec:Acknowledgements}

The authors would like to thank Sarah Bryan and Walter Dehnen for creating the
particle distributions for the mock galaxies, and the anonymous referee for
many useful suggestions which has improved the manuscript. JIR would like to
acknowledge support from SNF grant PP00P2\_128540/1.

\bibliographystyle{mn2e}
\bibliography{ms}

\appendix

\section{Implementation Details}

Since we want to model the density distribution with a computer it is
convenient to choose units that make the relevant quantities of order unity.
We therefore measure lengths in light years, time in years, positions in
arcseconds, and choose $c=1$ and $4\pi G = N^2$, where $N^2 \equiv 206,265$
arcsec/rad. The mass unit is then $11.988\ \Msun$. It will also be useful to
define a proxy to the Hubble constant $\zeta \equiv N^2 H_0$. We now express
the equations from \secref{sec:theory} in terms of these new units and
introduce some other useful quantities.

The lens equation in its complete form becomes:
\begin{eqnarray}
N^2ct(\vec\theta) & = & (1+z_L)\frac{D_{L}D_{S}}{D_{LS}}\frac12 |\vec\theta - \vec\beta|^2 \nonumber \\
& & - (1+z_L)\frac{4GD_{L}^2}{c^2}\int \Sigma(\vec\theta') \ln |\vec\theta-\vec\theta'| d^2\vec\theta'
\label{eqn:full_arrival_time}
\end{eqnarray}
where the factor of $D_L^2$ in the second term comes from the fact that
$\Sigma$ has units of \Msun/lyr$^2$. We can clean this up by first writing down
a dimensionless time delay
\begin{equation}
\tau = \left[ (1+z_L) d_L\right]^{-1}\zeta t
\label{tau}
\end{equation}
in terms of our previous definitions and defining $D_L \equiv (c/H_0)d_L$. We
further define a dimensionless density
\begin{equation}
\kappa_\infty = \frac{4\pi G}{c^2}\frac{c}{H_0}d_L\Sigma
              = \frac{d_L}{\zeta}\Sigma
\end{equation}
and a lensing potential
\begin{equation}
\psi(\vec\theta) = \frac1\pi \int \kappa_\infty(\vec\theta') \ln|\vec\theta - \vec\theta'| d^2\vec\theta'\
\label{lensing potential}
\end{equation}
Now we can express \eqnref{eqn:full_arrival_time} very compactly as
\begin{equation}
\tau(\vec\theta) = \frac12 \xi |\vec\theta-\vec\beta|^2 - \psi(\vec\theta)
\label{arrival time}
\end{equation}
where $\xi=d_S / d_{LS}$.  We explicitly write $\kappa_\infty$ to remind
ourselves that there is no source distance factor involved. This is useful when
we consider multiple sources.

\section{Derivation of pixelated density coefficients}
\label{Q derivation}
When the lens plane is pixelized we need a discrete form of the integral
\[\int \kappa(\vec\theta') \ln |\vec\theta-\vec\theta'| d^2\vec\theta' \]
In particular we want
\[\sum_n \kappa_n Q_n(\vec\theta)\]
where $Q_n$ is the logarithm evaluated over the $n$th pixel at position
$\vec\theta_n = (x_n, y_n)$. Let the pixel side length be $a$.  Instead of
working with a position vector $\vec\theta$ we work in Cartesian coordinates
where 
$|\vec\theta| = r = \sqrt{x^2 + y^2}$. The integral now becomes
\[Q_n(x,y) = \frac12 \int_{y_-}^{y_+}\int_{x_-}^{x_+} \ln (x'^2+y'^2) dx' dy'\]
where $x_\pm = x + x_n \pm (a/2)$ and similarly for $y_\pm$.
Using the identity
\[\int \ln(x^2+y^2) dx = x \ln(x^2+y^2) - 2x + 2y\arctan(x/a) \]
we can express $Q_n$ as the sum of four parts
\begin{align*}
Q_n(x,y) = \frac12 [\tilde Q_n(x_+,y_+)
                  + \tilde Q_n(x_-,y_-) & 
\\                - \tilde Q_n(x_-,y_+)
                  - \tilde Q_n(x_+,y_-) ]
\end{align*}
where
\[\tilde Q_n(x,y) = xy(\ln r^2 - 3) + x^2\arctan(y/x) + y^2\arctan(x/y)\]

\section{No radial symmetry prior}\label{no_symm_prior}

\begin{figure*}
\includegraphics[width=0.24\textwidth]{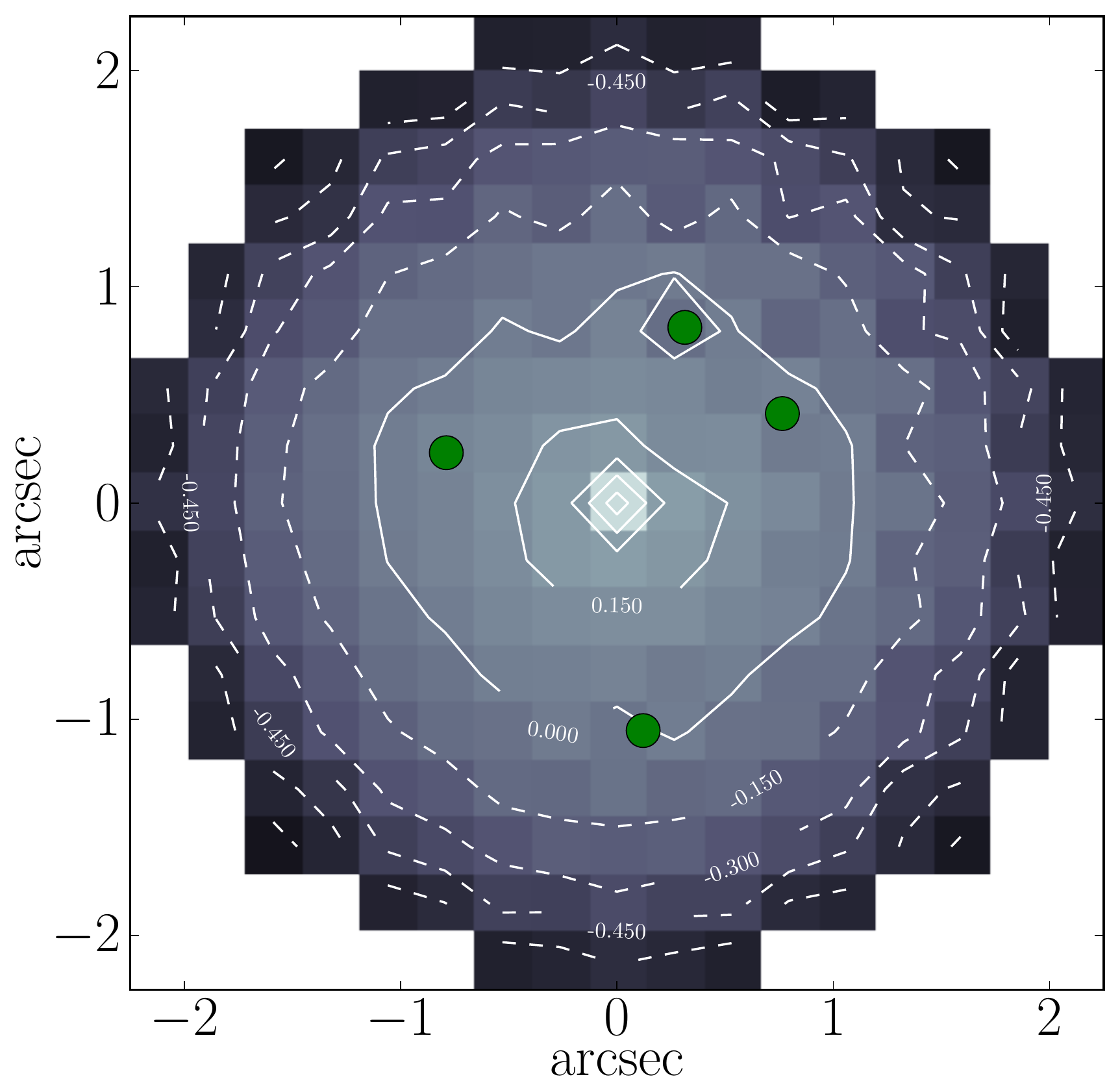}
\includegraphics[width=0.24\textwidth]{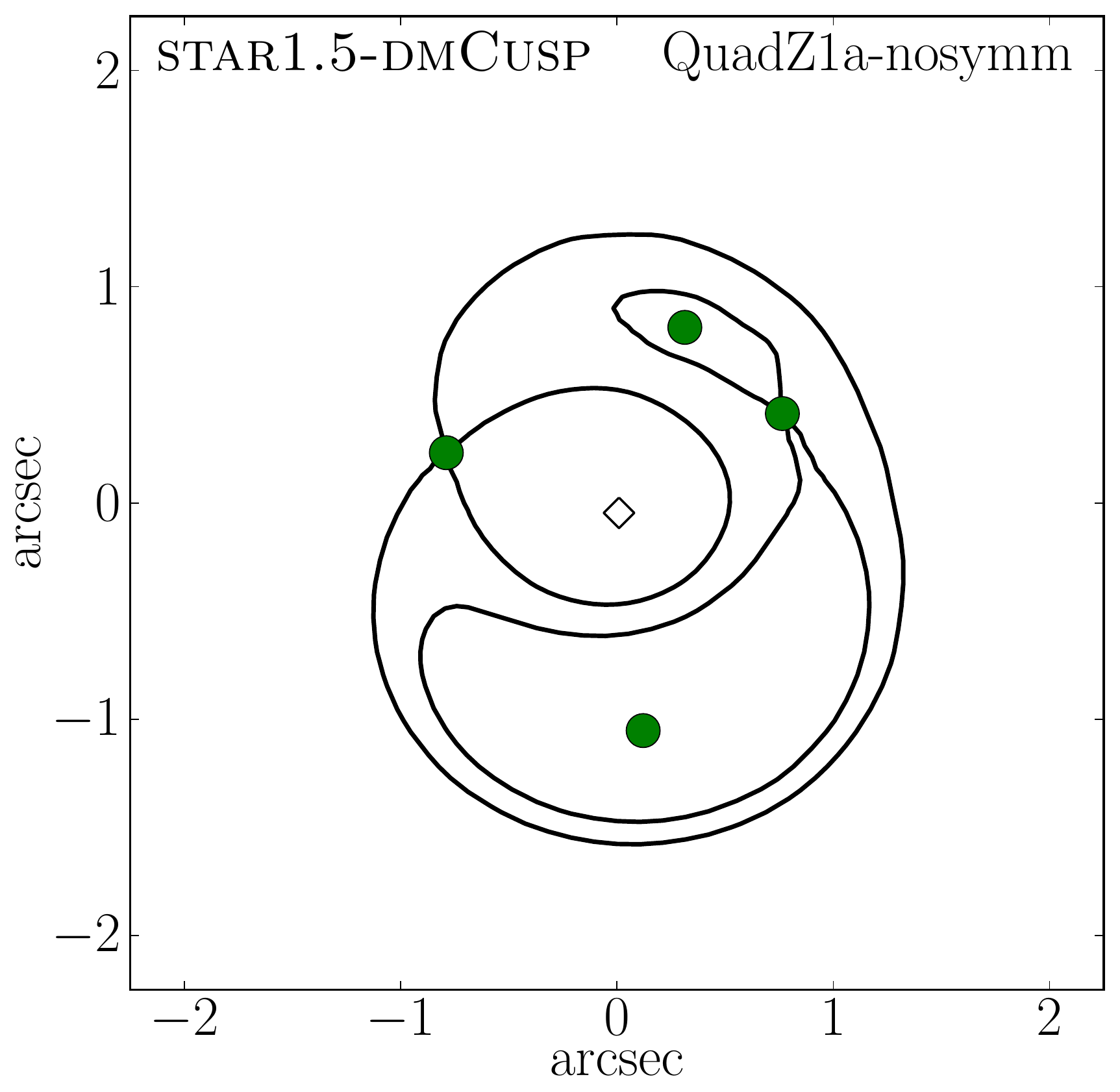}
\includegraphics[width=0.24\textwidth]{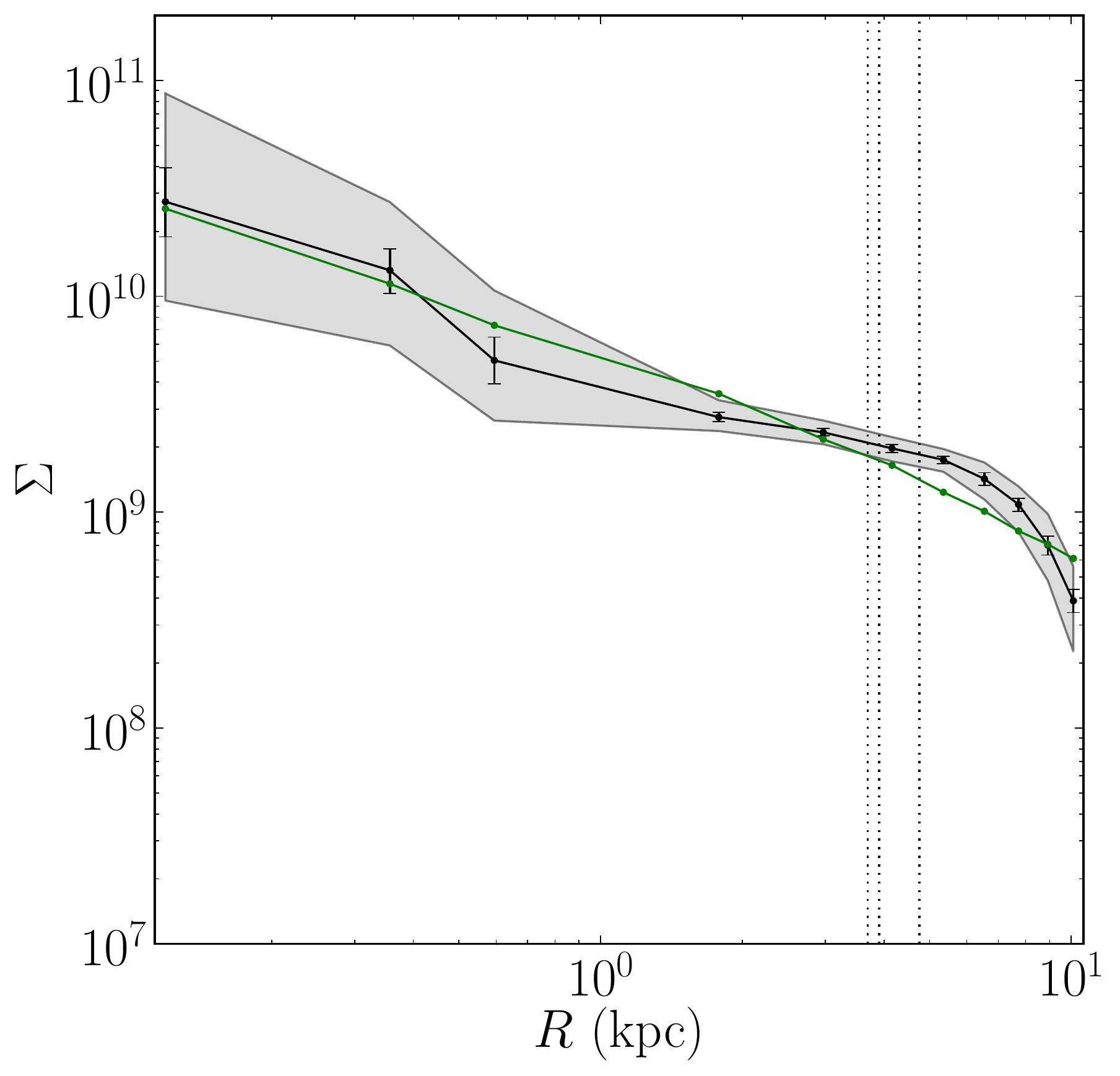}
\includegraphics[width=0.24\textwidth]{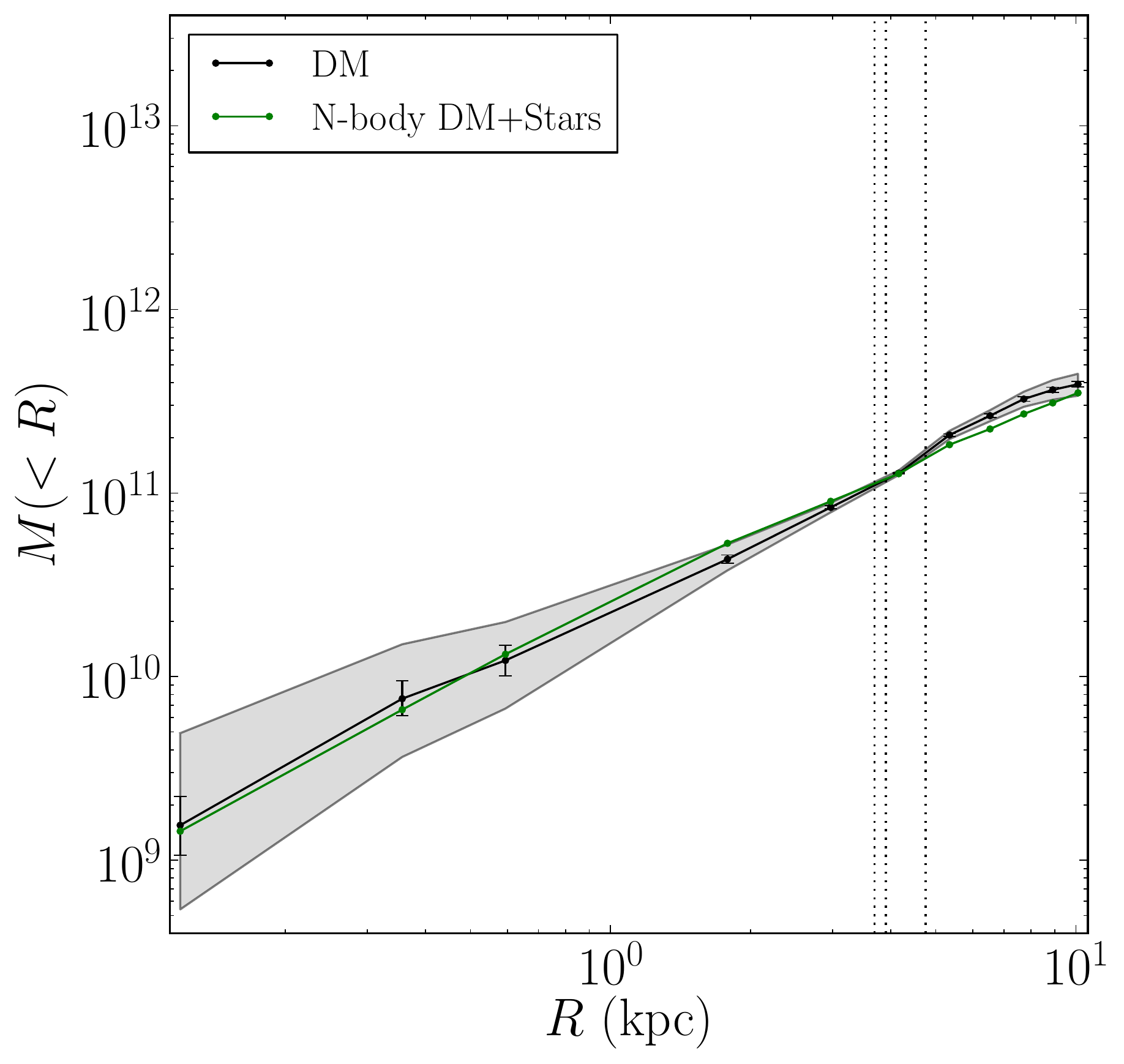}\\
\includegraphics[width=0.24\textwidth]{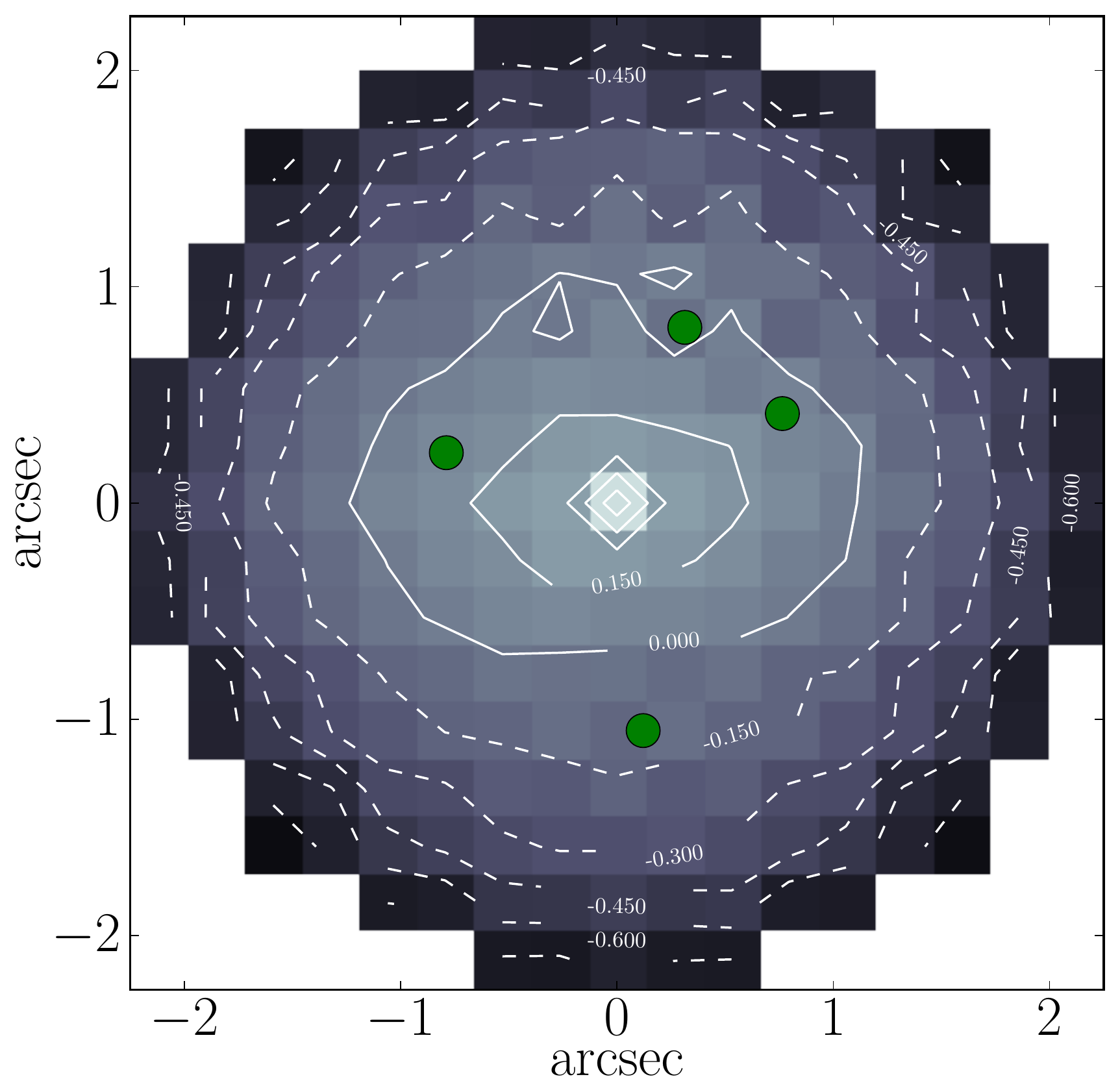}
\includegraphics[width=0.24\textwidth]{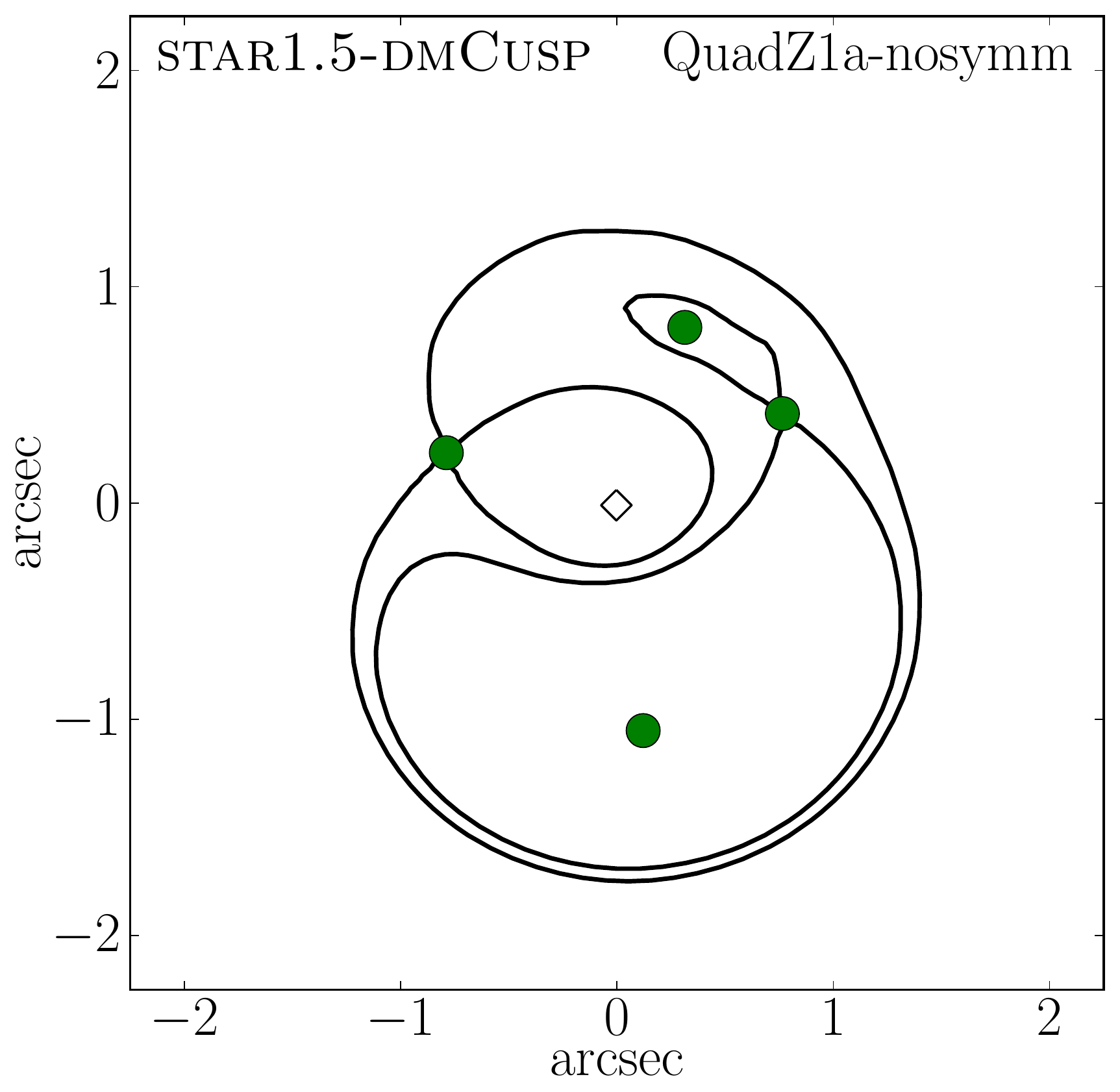}
\includegraphics[width=0.24\textwidth]{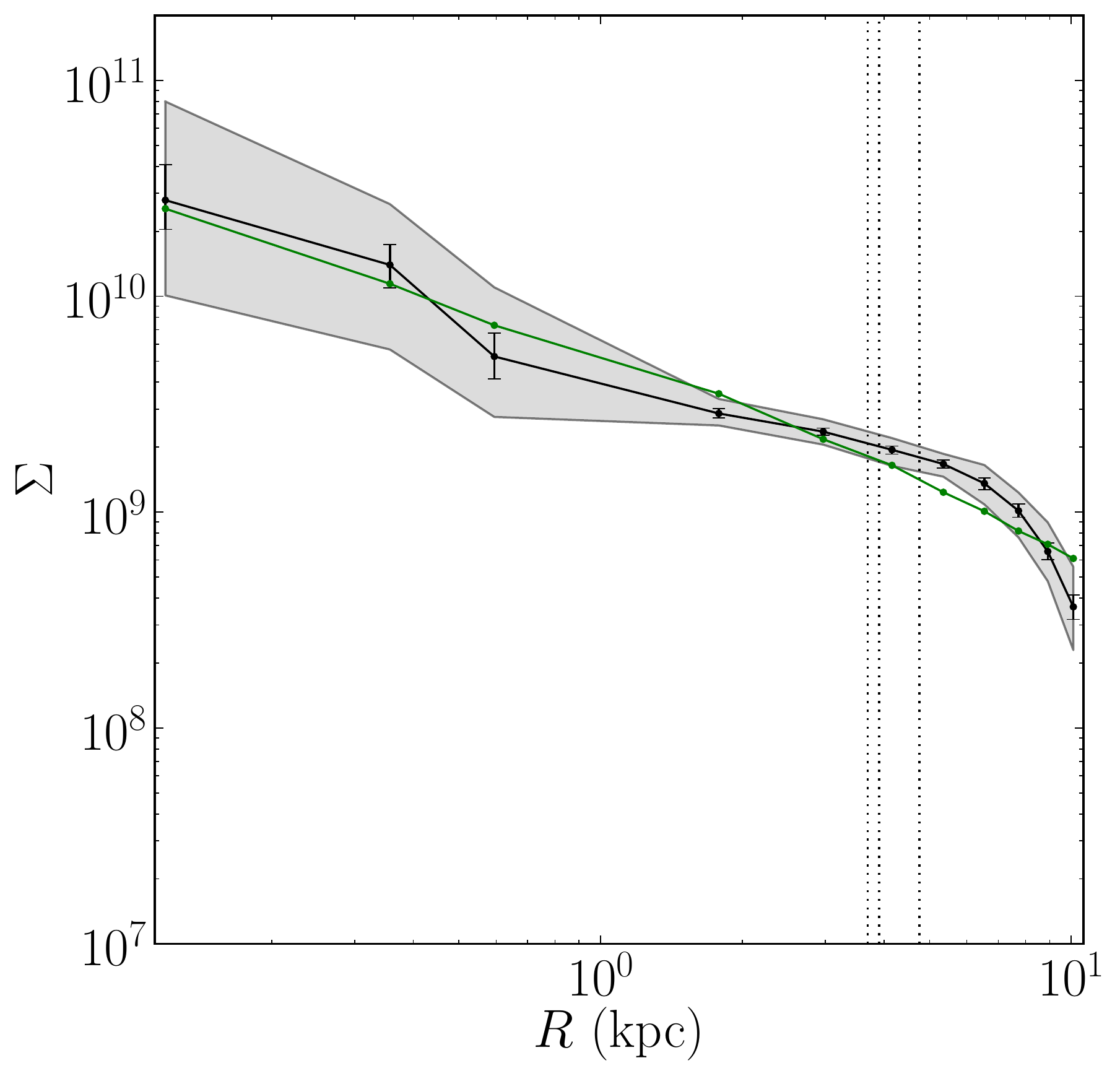}
\includegraphics[width=0.24\textwidth]{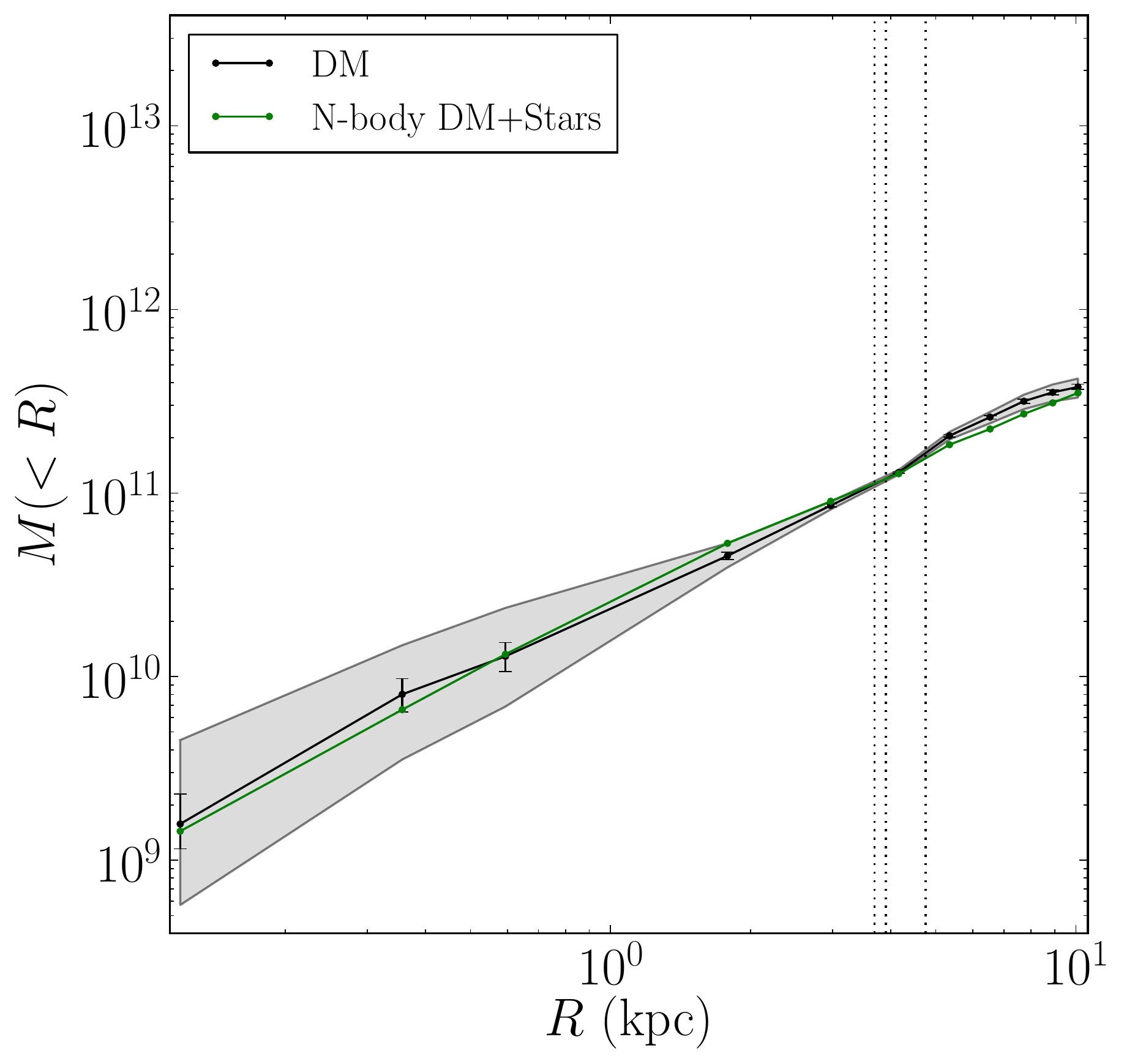}\\
\includegraphics[width=0.24\textwidth]{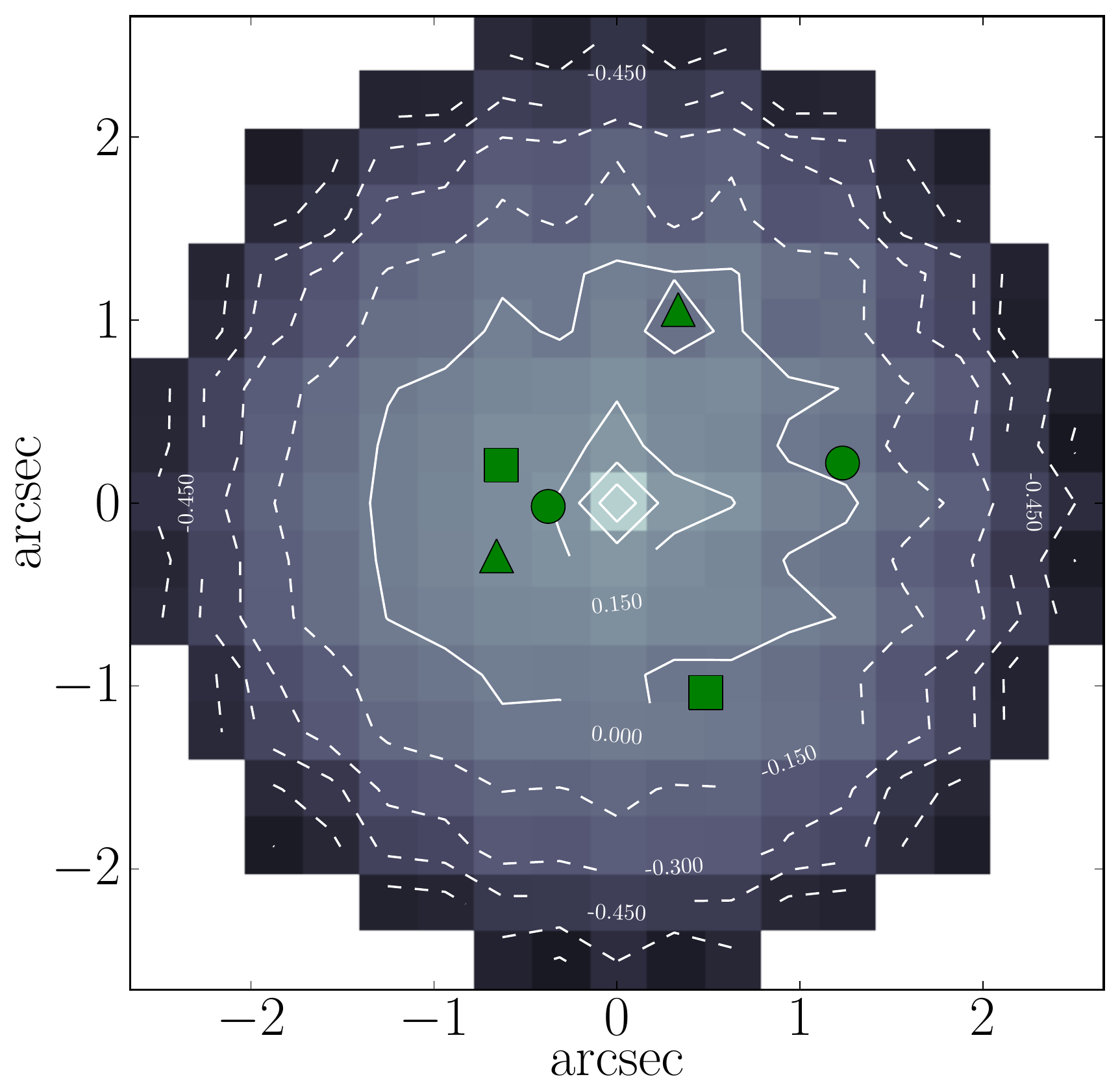}
\includegraphics[width=0.24\textwidth]{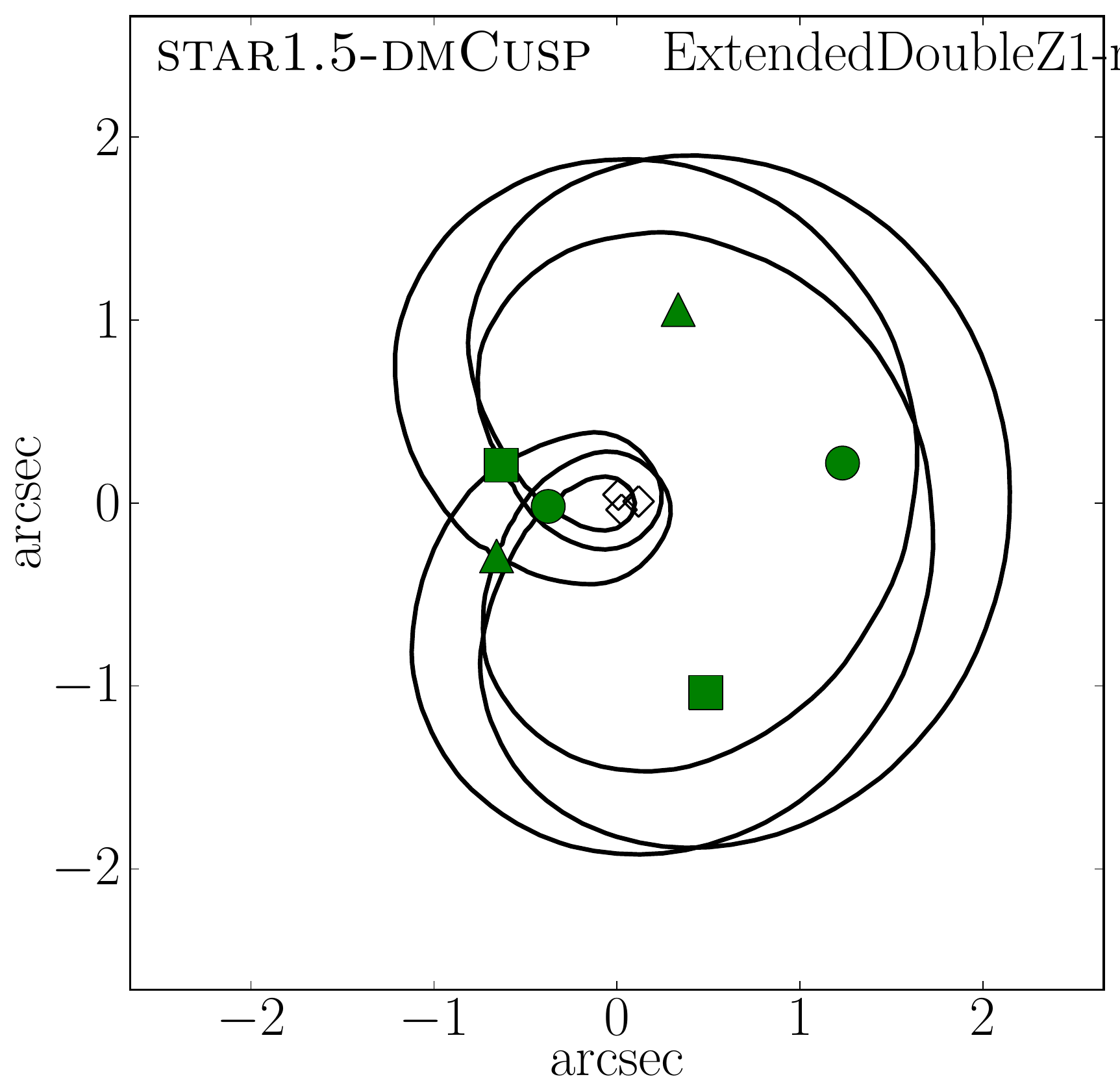}
\includegraphics[width=0.24\textwidth]{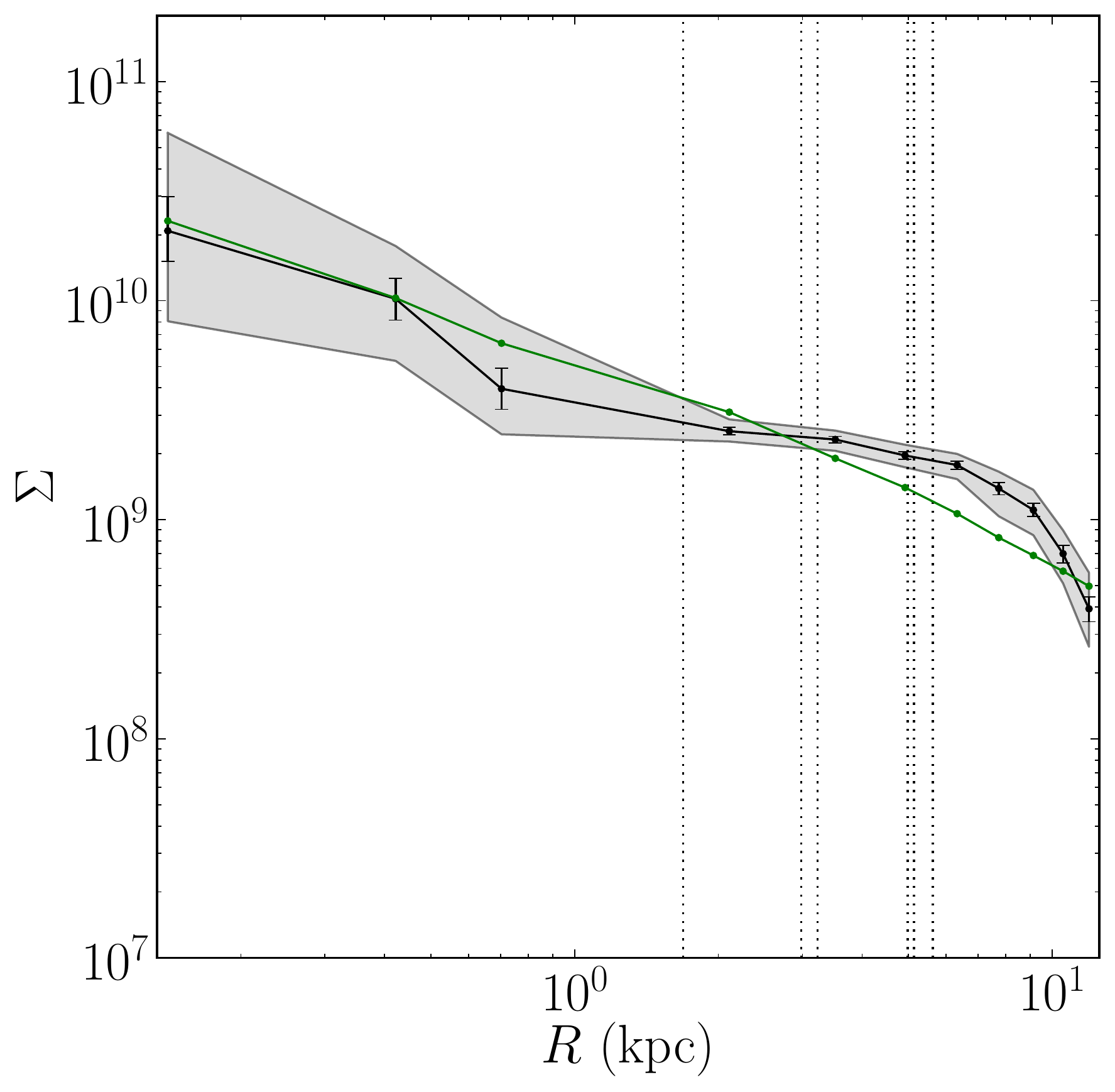}
\includegraphics[width=0.24\textwidth]{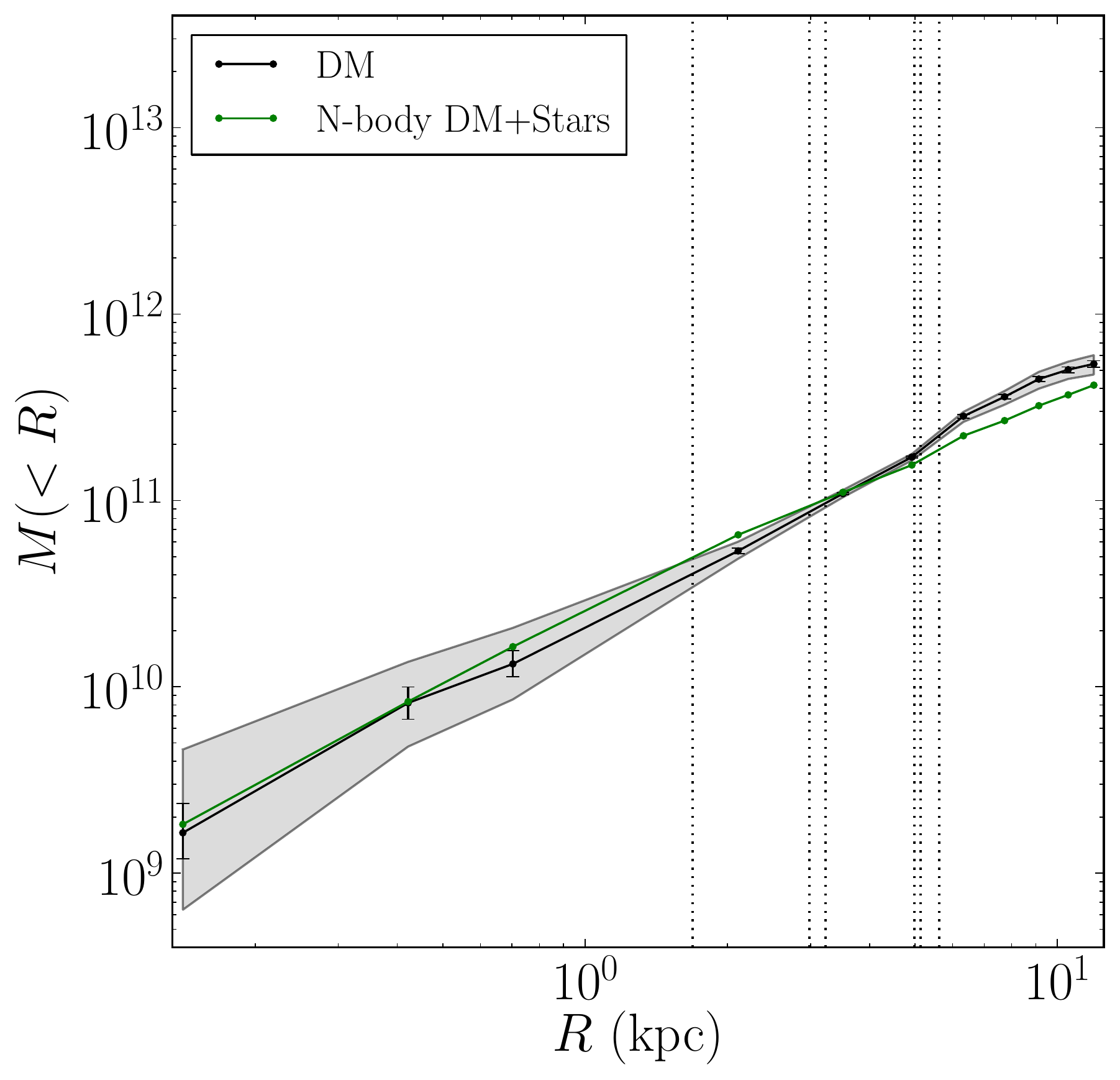}
\includegraphics[width=0.24\textwidth]{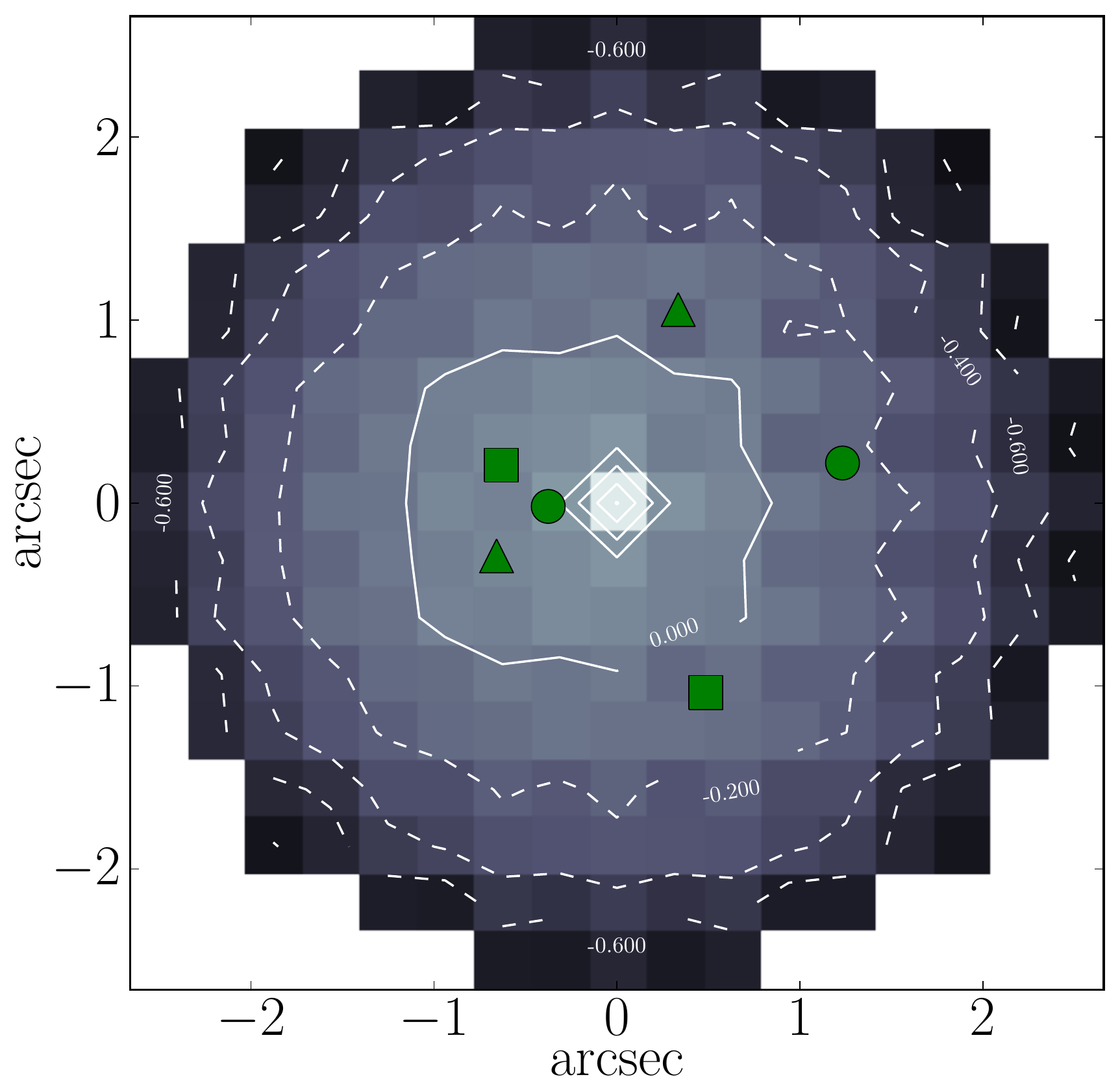}
\includegraphics[width=0.24\textwidth]{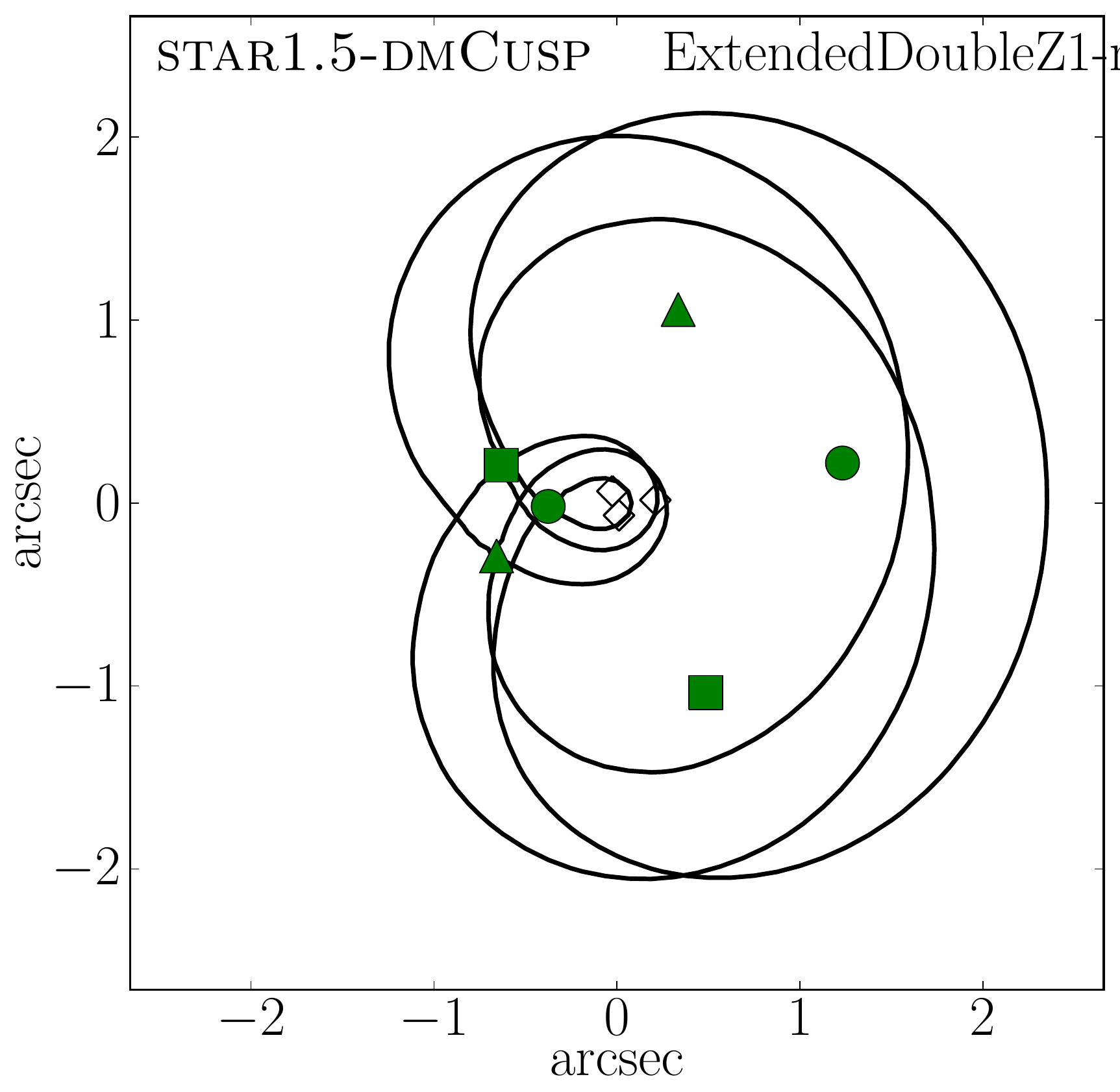}
\includegraphics[width=0.24\textwidth]{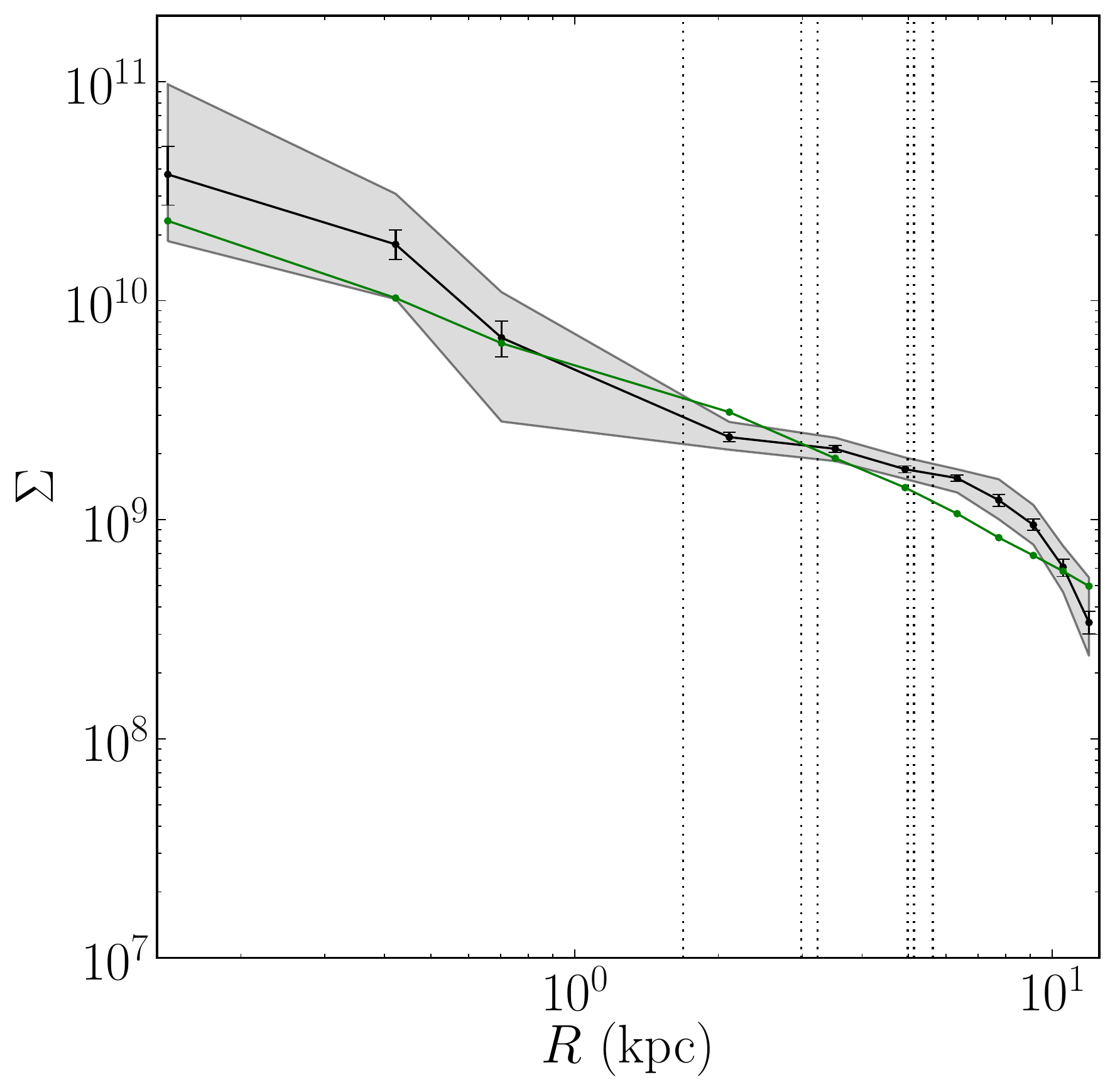}
\includegraphics[width=0.24\textwidth]{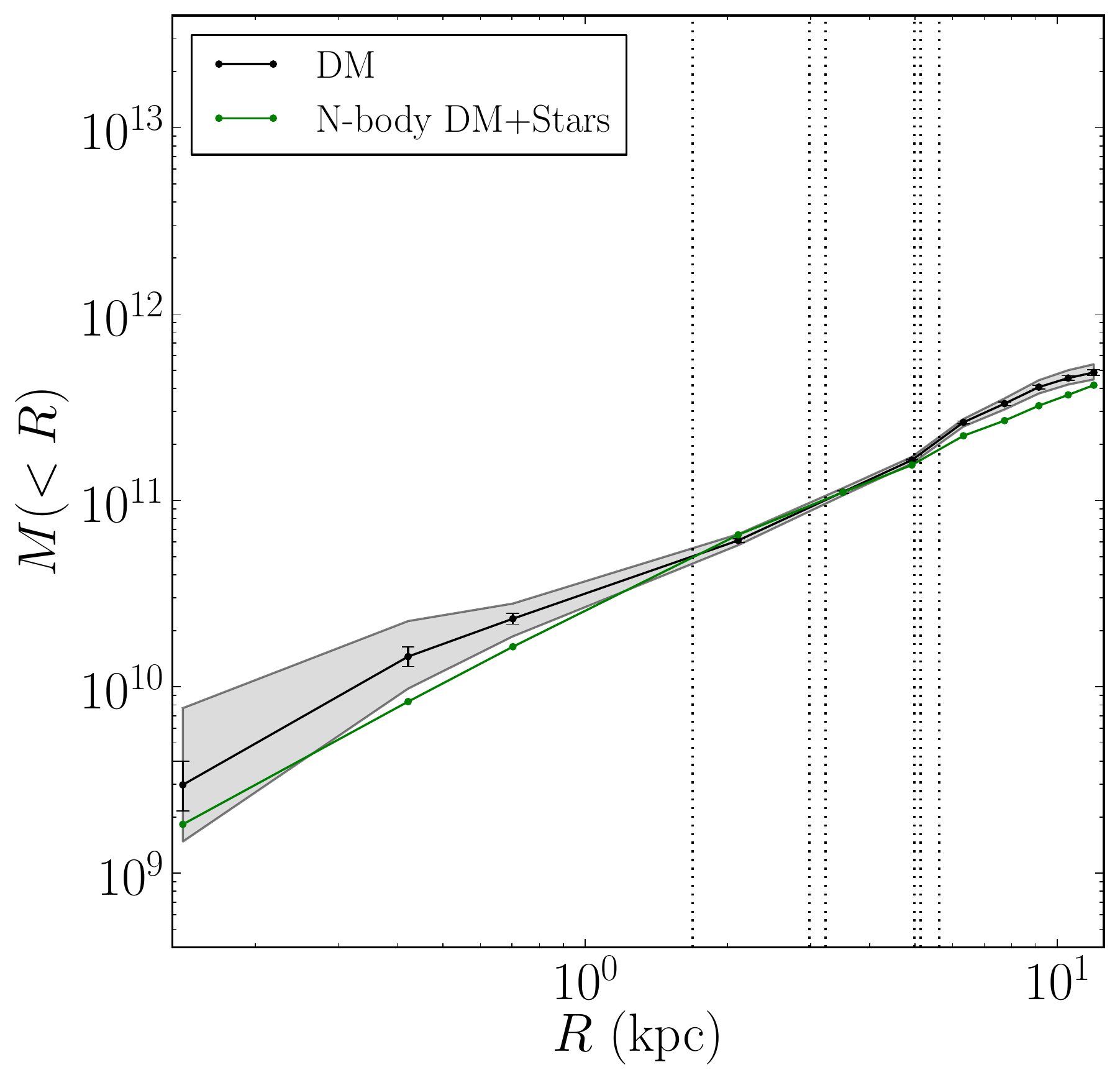}
\caption{Results for a single quad (upper two rows) and an extended double (lower two rows)
    without the radial symmetry prior; the bottom row of each group uses time delay data. The stellar
    mass constraint has not been used in any of these examples. Figures and symbols are as in \figref{reconstruction} and
    \figref{2d mass reconstruction}.}
\label{fig:no_symm_prior}
\end{figure*}

In this appendix, we explore the effect of the radial symmetry prior.
\figref{fig:no_symm_prior} shows results for a single quad (top two rows) and
an extended double (bottom two rows) without the radial prior; in both cases,
we do not use the stellar mass constraints. The bottom row of each group uses
time delay data. Without time delay data or the radial symmetry prior, the
results for the quad are poor -- particularly the shape recovery. Including
time delays, the results are similar to the case with the radial prior
(\figref{reconstruction} and \figref{2d mass reconstruction}). Similarly, for
the extended double the results without time delays are poor. Even with time
delays, the shape is not well recovered without the radial prior, as expected.

\section{Pixel resolution convergence test}\label{pix_convergence_test}

\begin{figure*}
\includegraphics[width=0.24\textwidth]{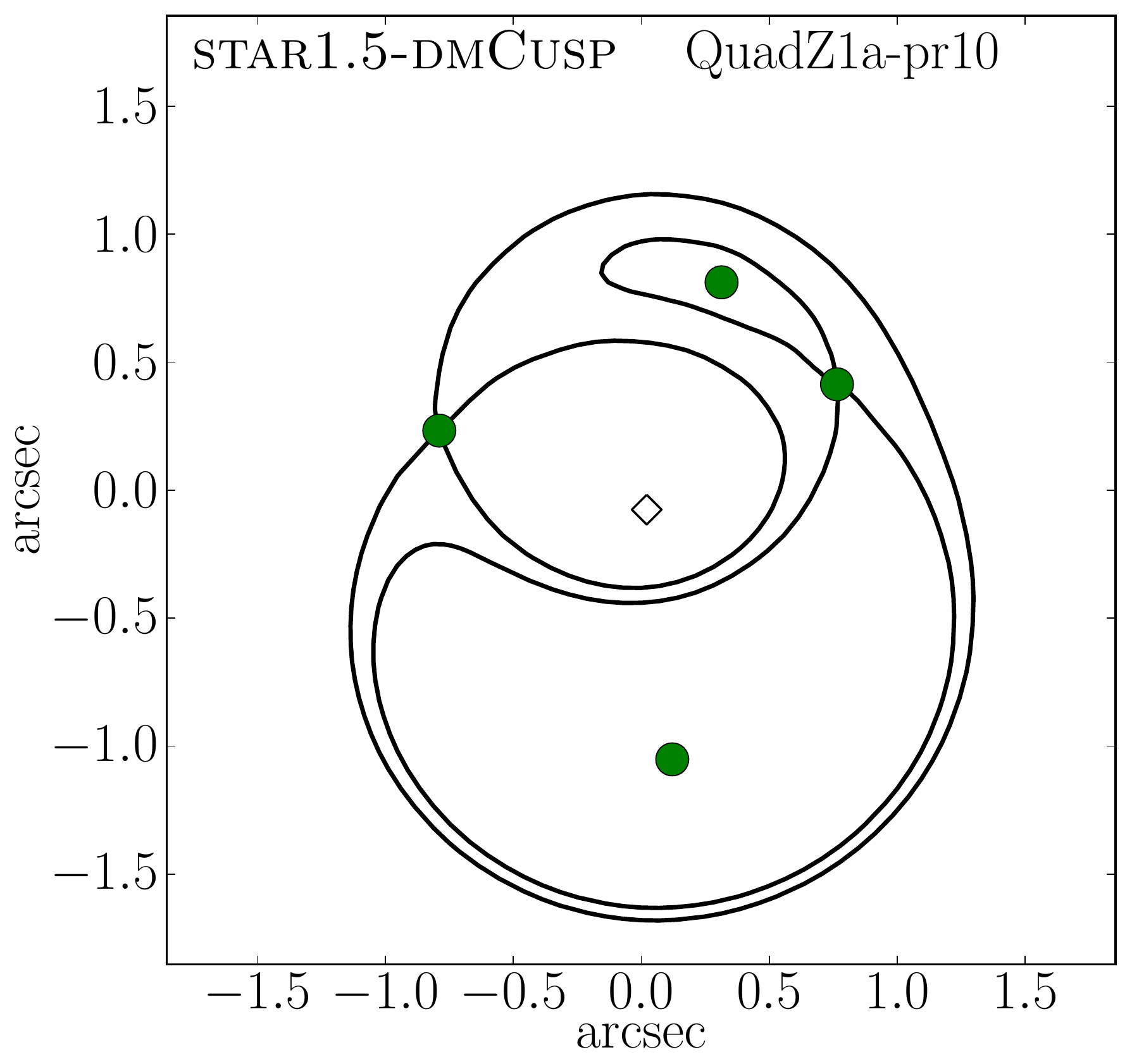}
\includegraphics[width=0.24\textwidth]{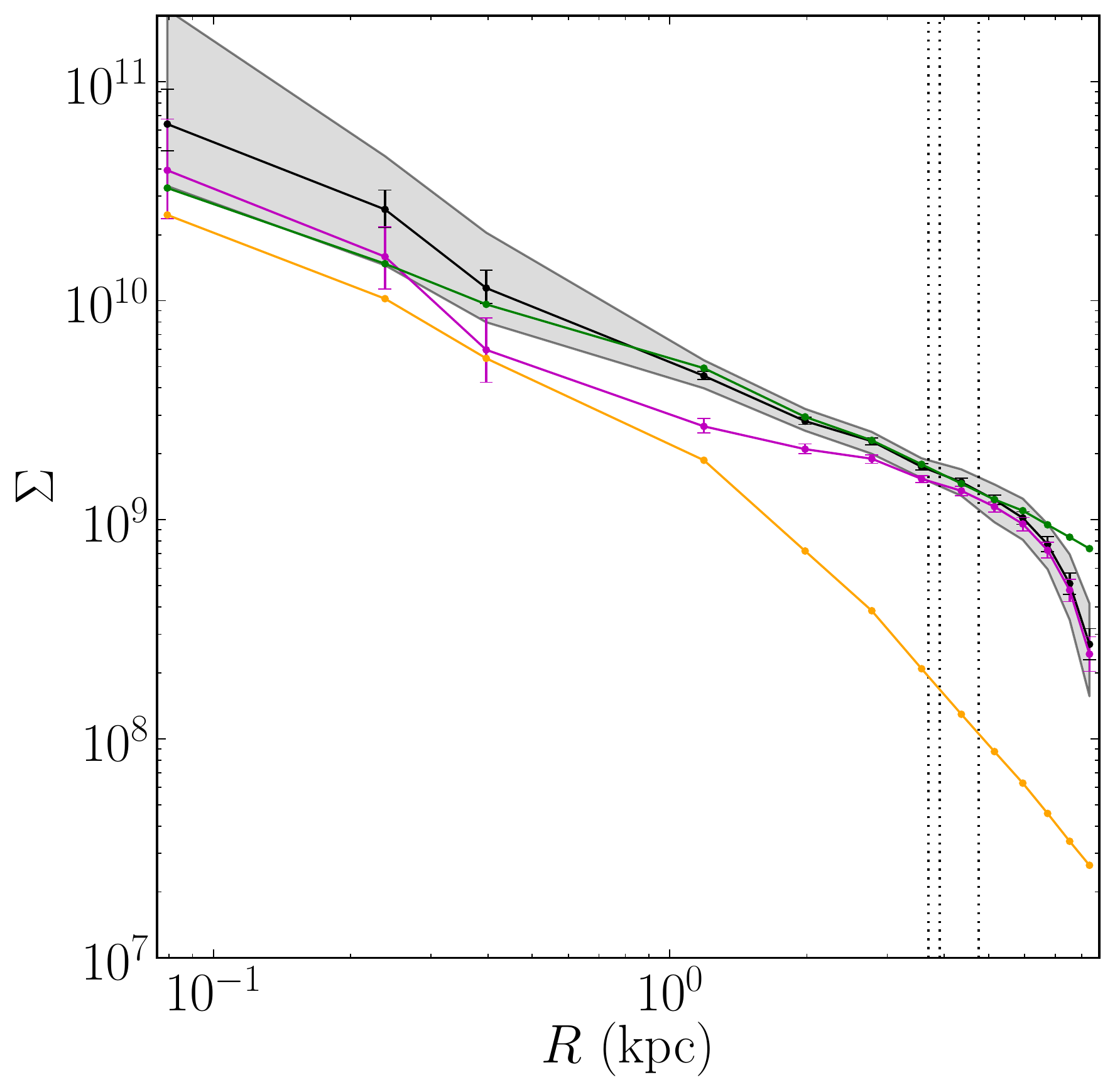}
\includegraphics[width=0.24\textwidth]{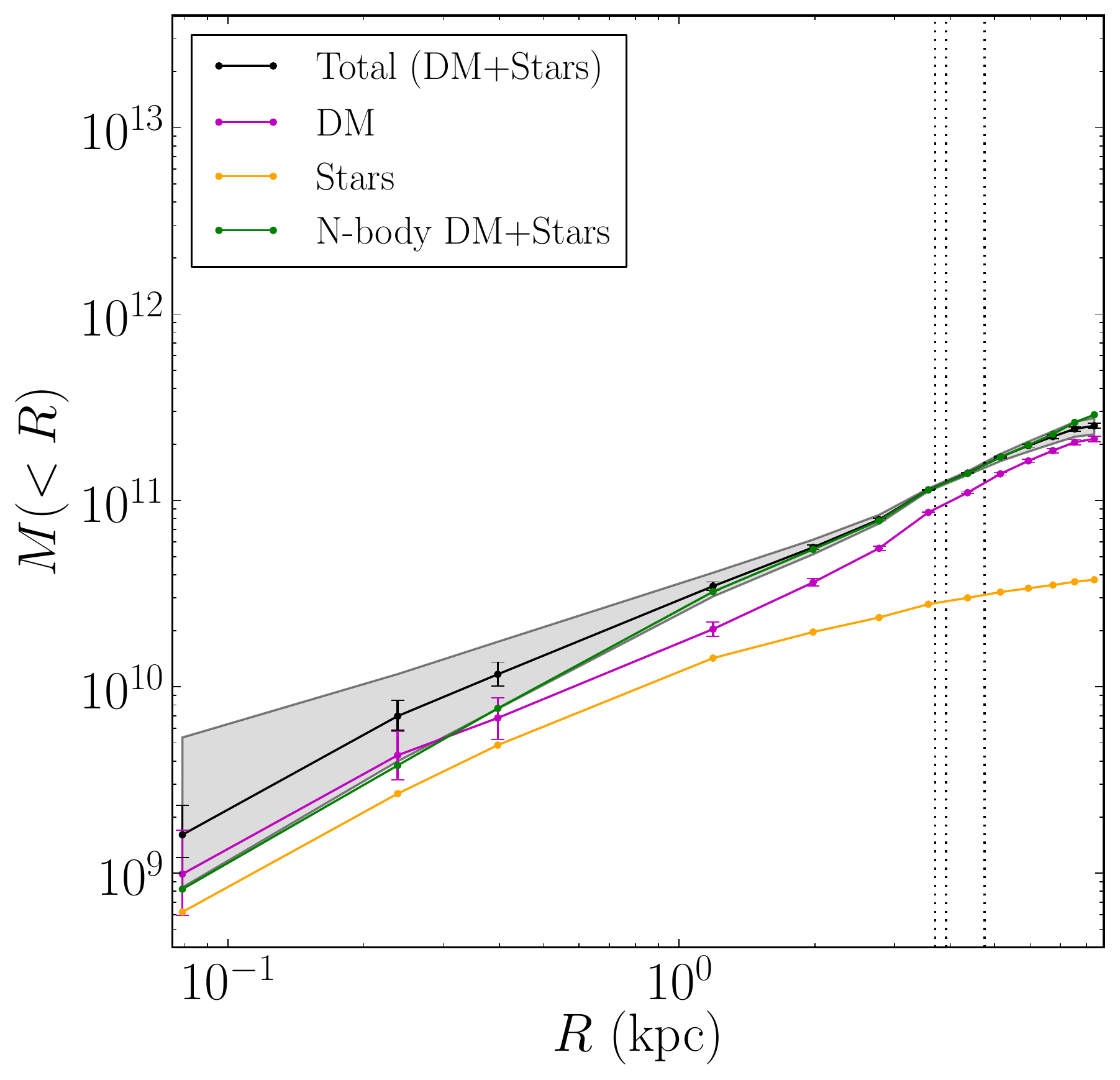}\\
\includegraphics[width=0.24\textwidth]{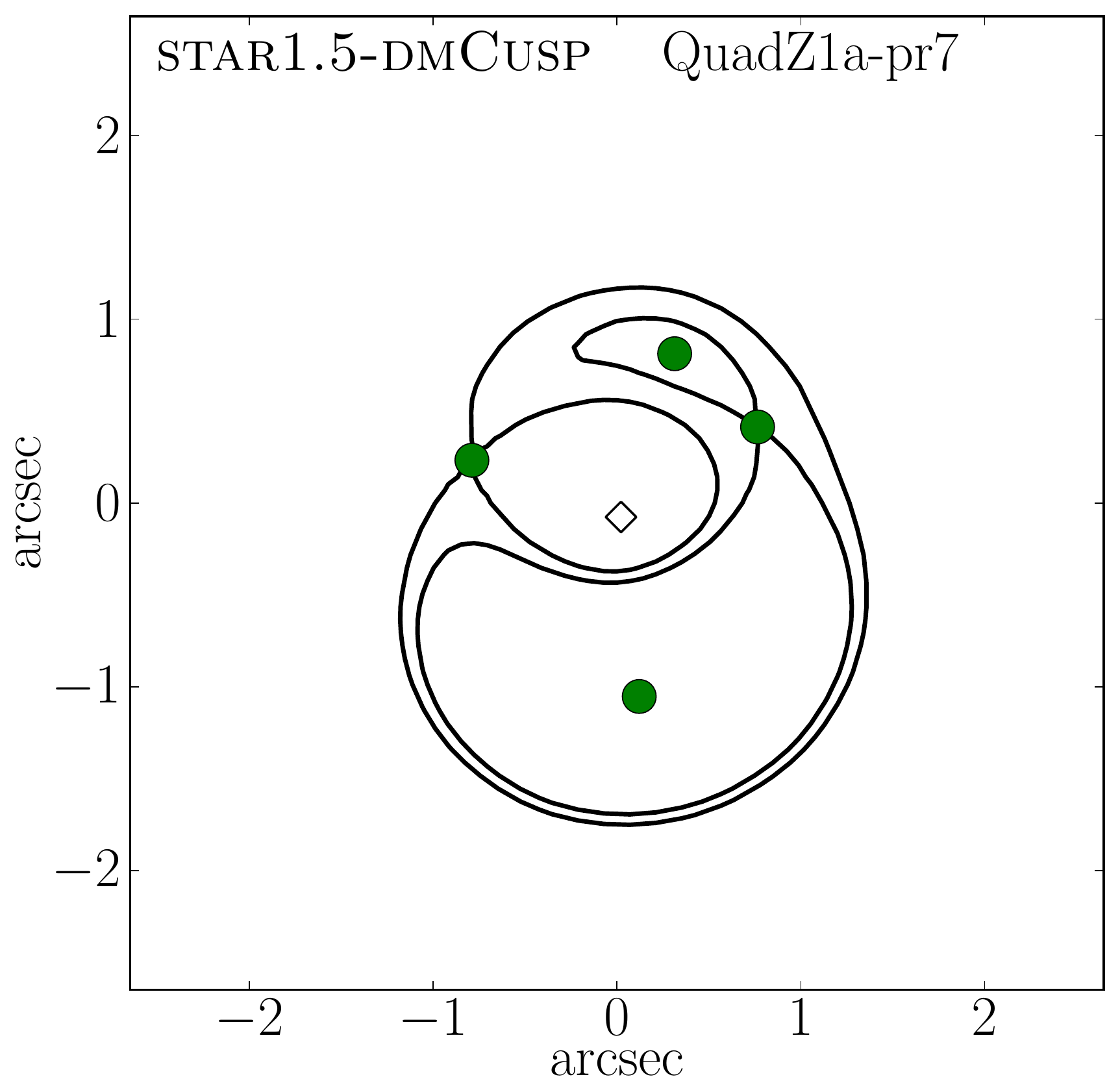}
\includegraphics[width=0.24\textwidth]{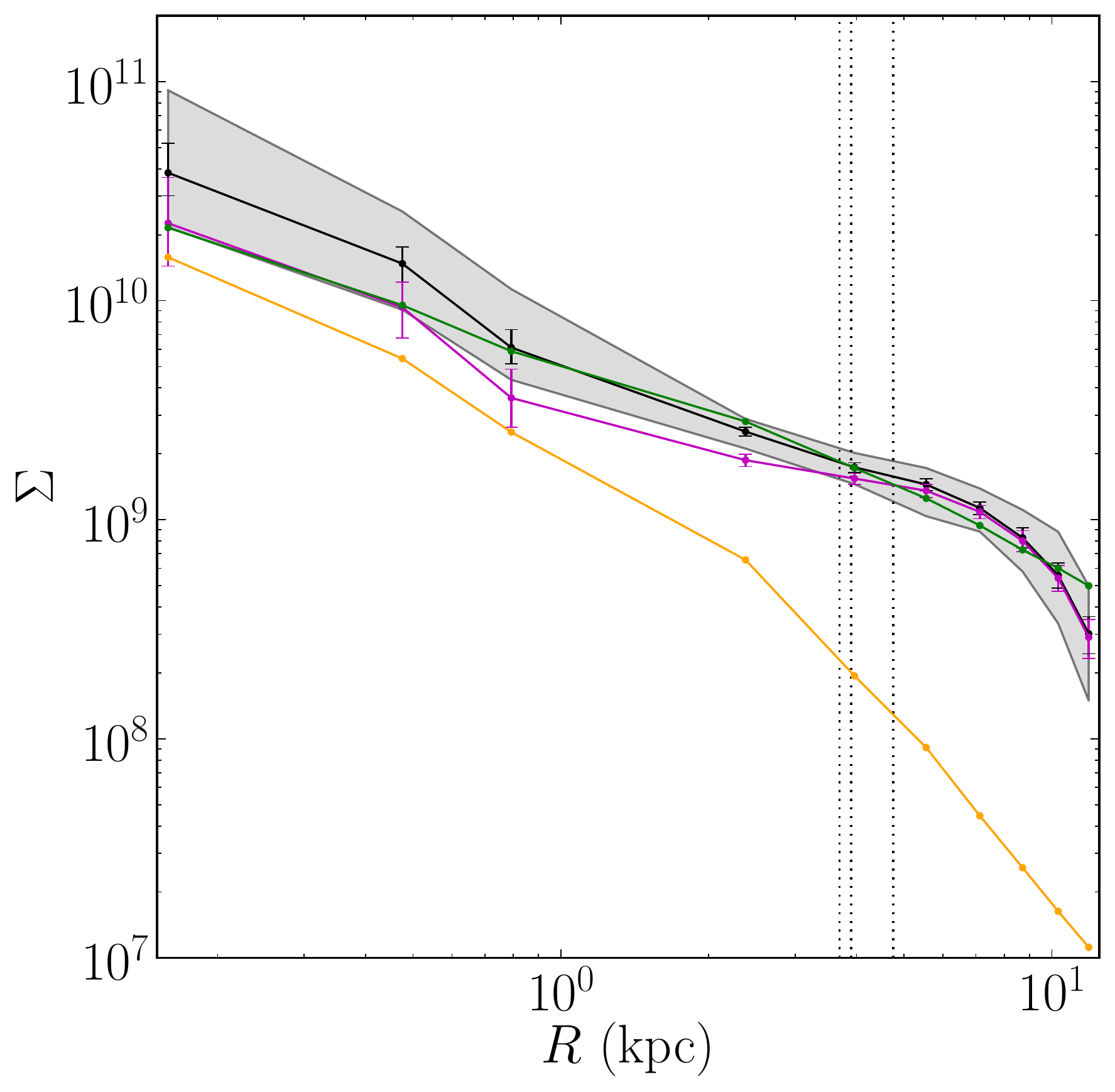}
\includegraphics[width=0.24\textwidth]{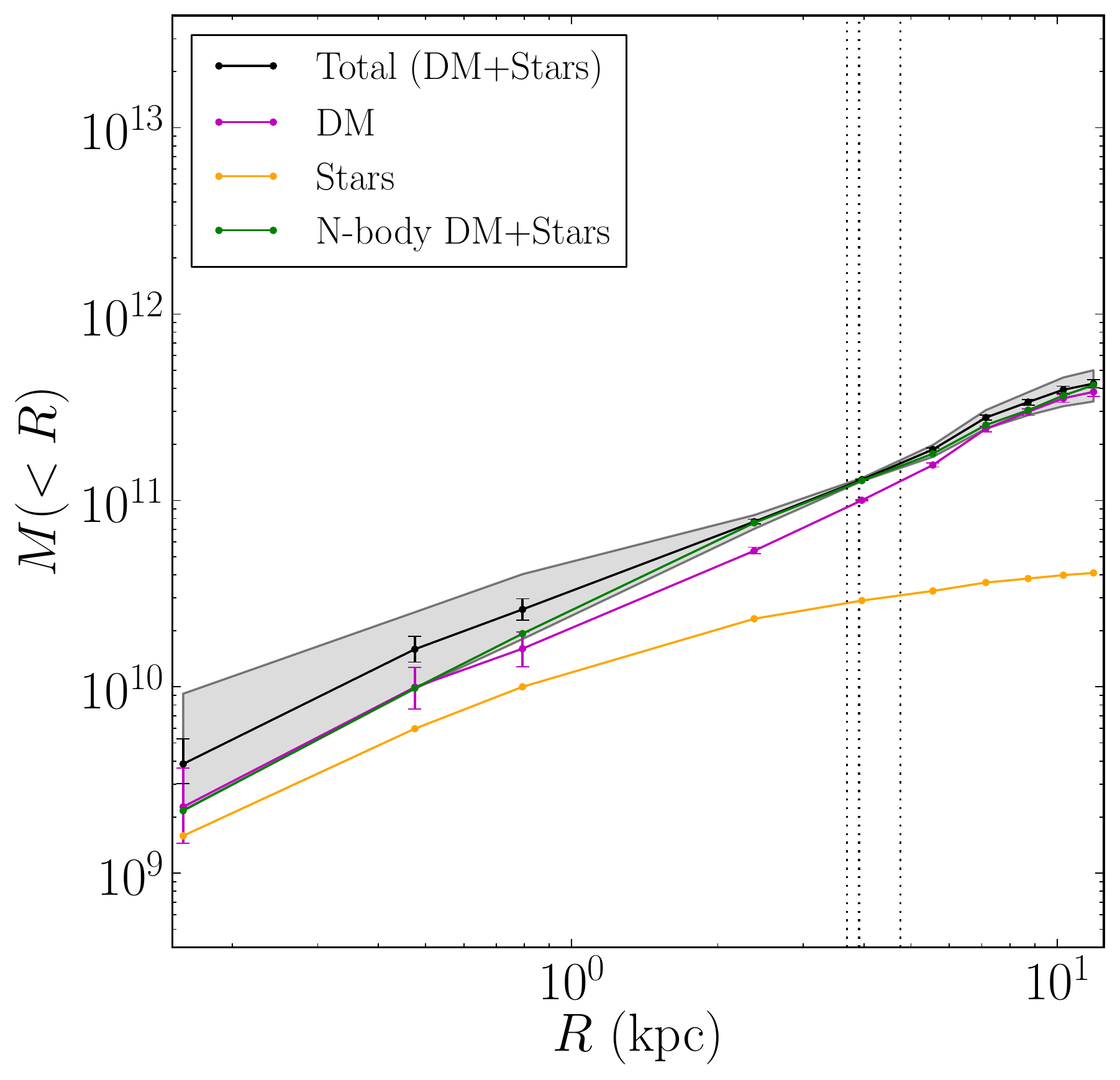}\\
\includegraphics[width=0.24\textwidth]{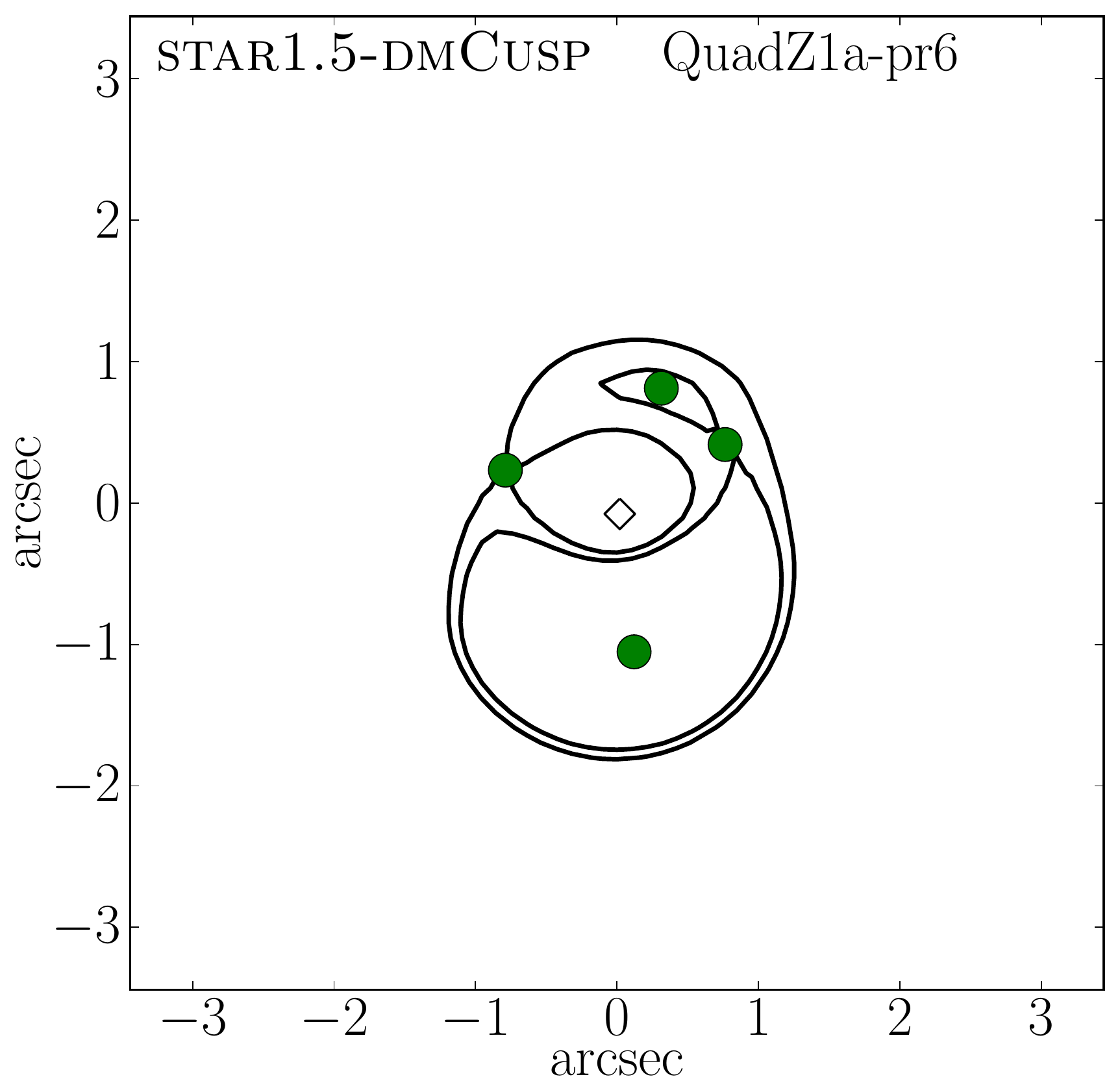}
\includegraphics[width=0.24\textwidth]{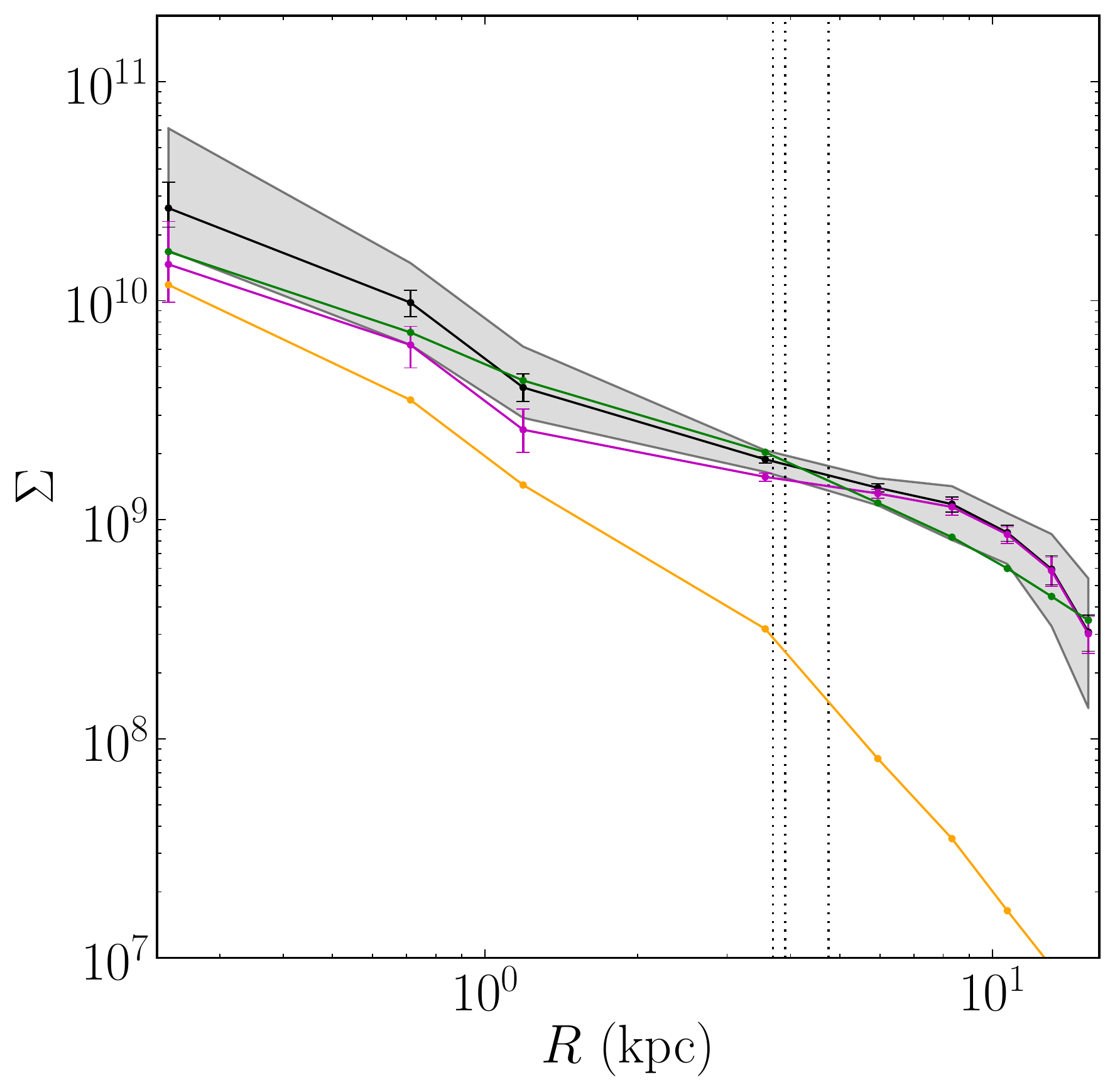}
\includegraphics[width=0.24\textwidth]{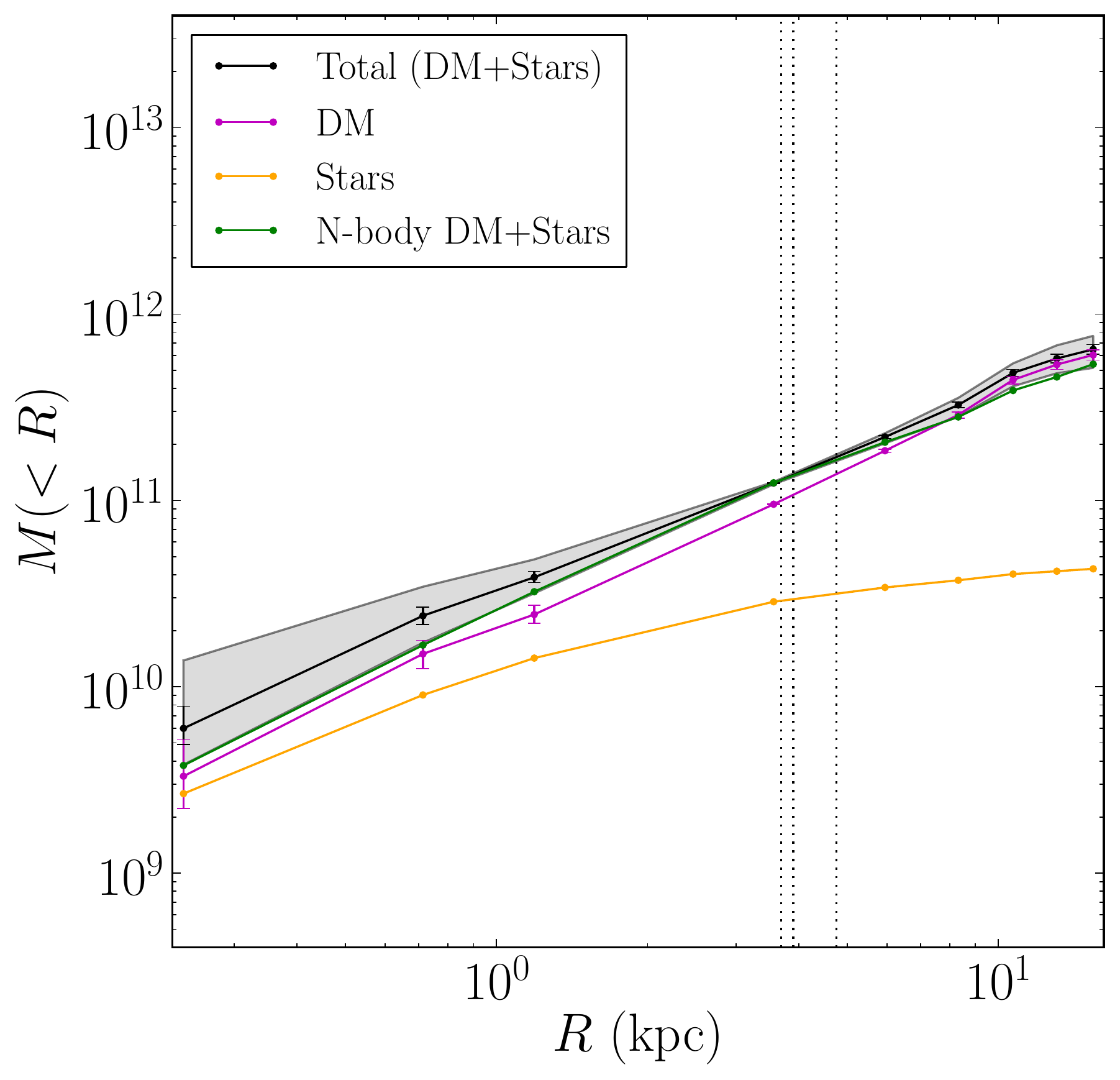}\\
\includegraphics[width=0.24\textwidth]{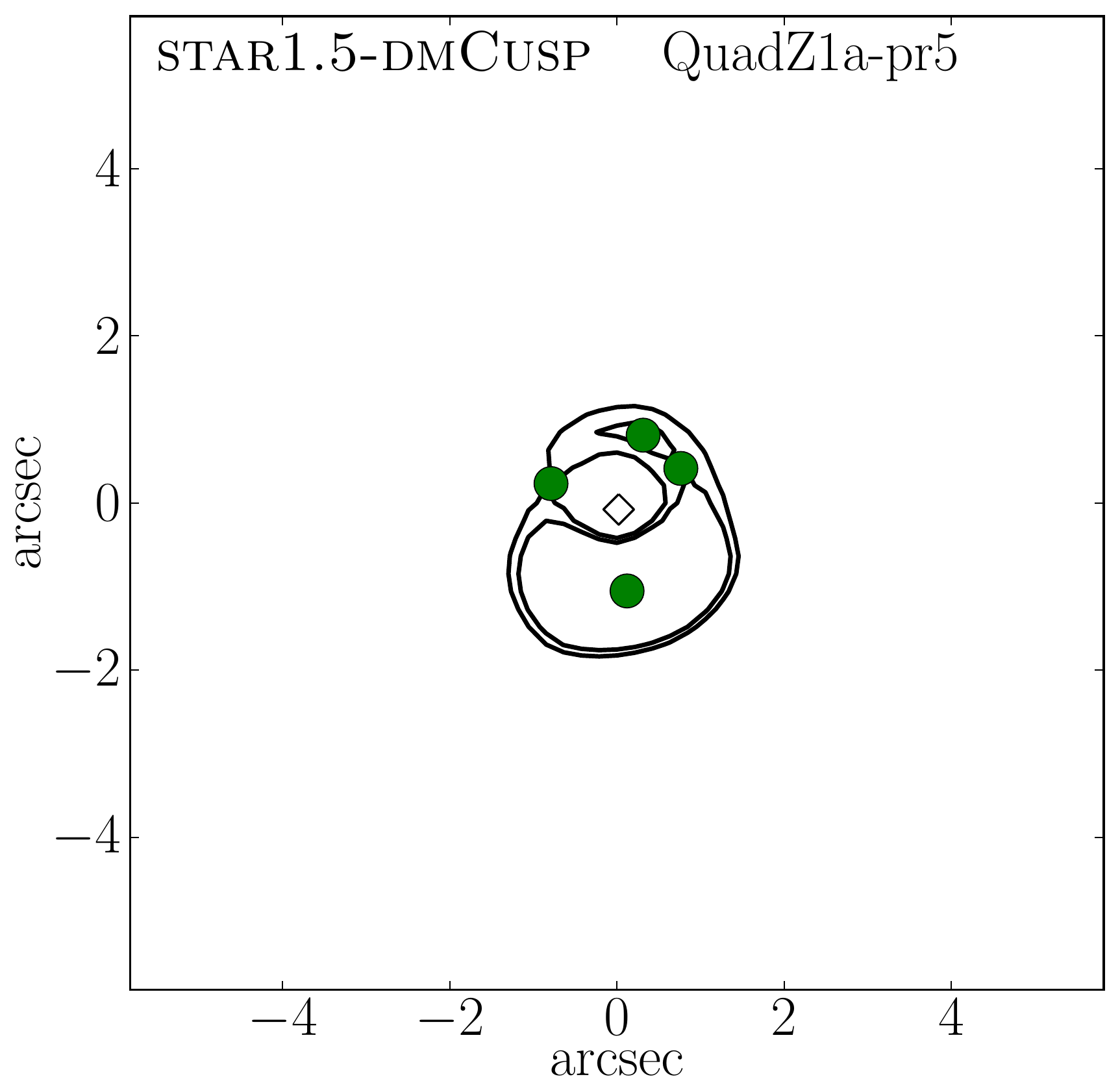}
\includegraphics[width=0.24\textwidth]{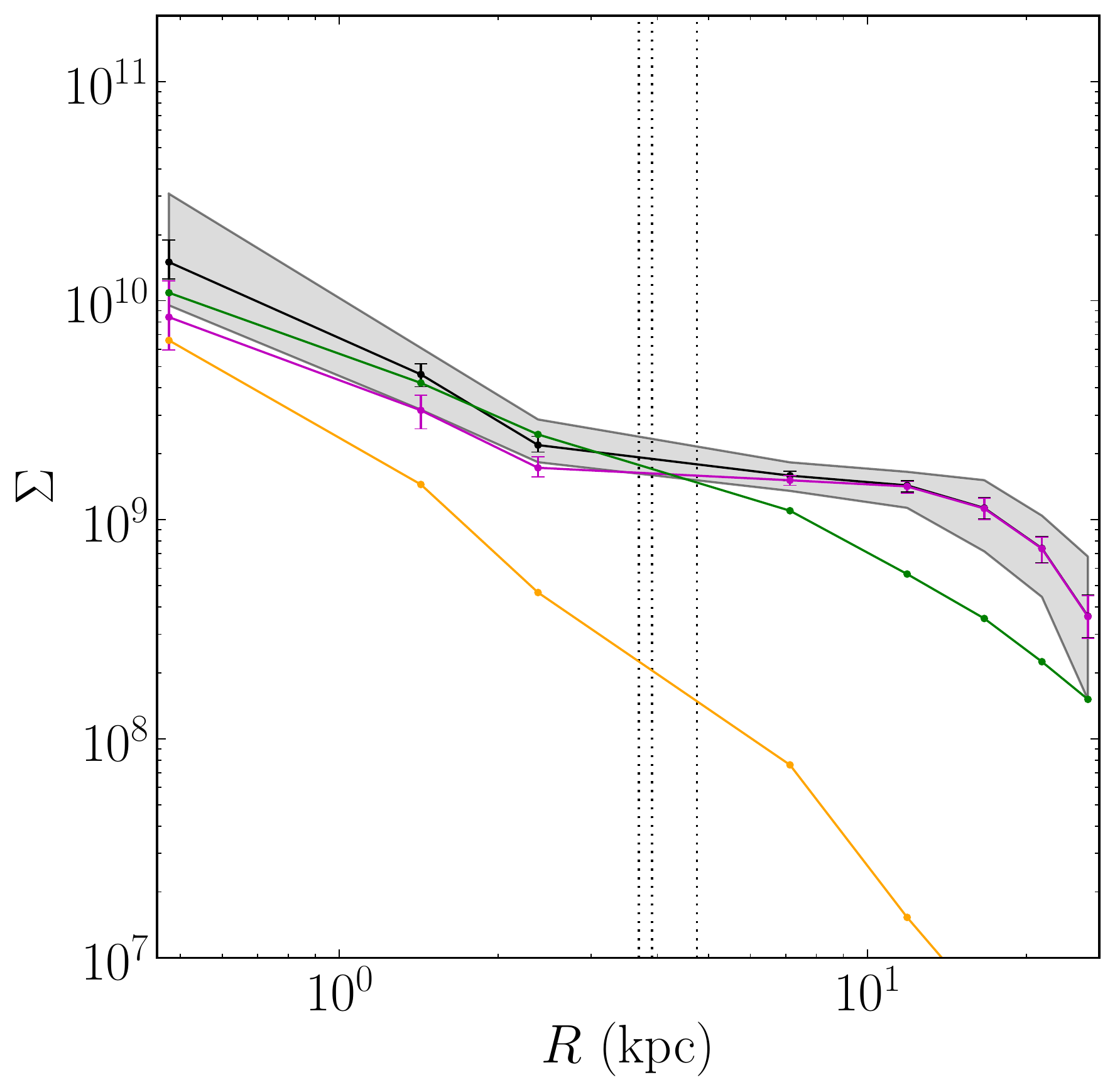}
\includegraphics[width=0.24\textwidth]{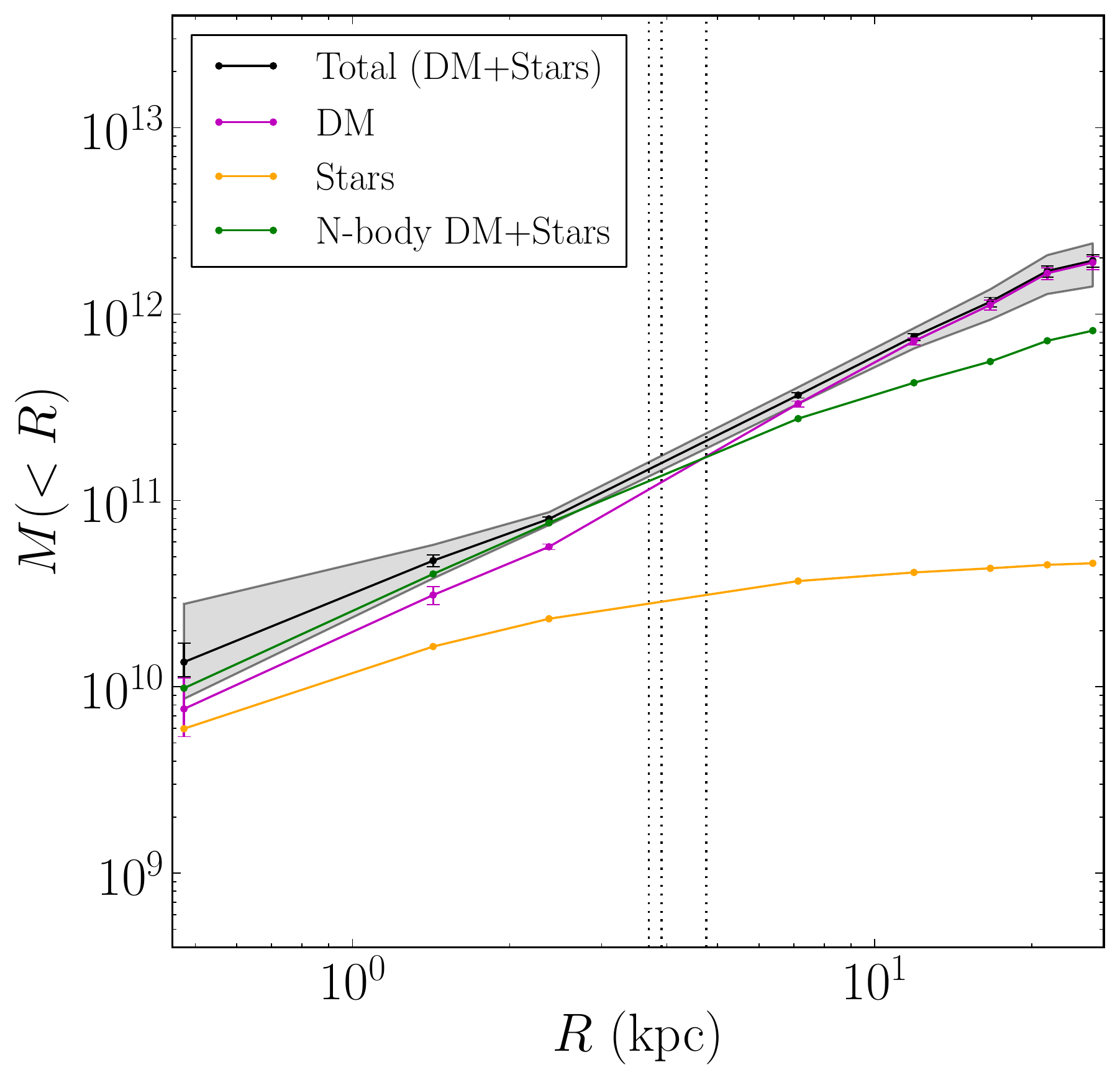}\\
\caption{The effect of changing the grid resolution parameter \pixrad. From top
    to bottom, the panels show results for a single quad with time delays and
    with stellar mass constraints using $\pixrad = 10,7,6,5$,
    respectively. We always demand that there are four radial bins outside the
    outermost image, which causes the total mass to increase with decreasing
    \pixrad, and the plot to shrink with increase pixel size.  In this paper we have used $\pixrad = 8$ in all the
tests.}

\label{fig:pix_convergence_test}
\end{figure*}

In this Appendix, we present a convergence test of our results with the grid
resolution -- \pixrad. By default, this is set to 8 pixels from the centre of
the mass map to the edge. As can be seen from
\figref{fig:pix_convergence_test}, our results are typically well-converged for
$\pixrad > 5$. The results for $\pixrad = 5$ become systematically biased away
from the central regions (where we have the higher resolution adaptive mesh),
because our regularisation prior combined with a low \pixrad{} biases us
towards shallow models. This effect diminishes with increasing resolution and
is already negligible by $\pixrad = 7$. Notice that the mass increases in size
with decreasing resolution. This is because we always demand that there are
four pixels beyond the outermost image. 

\end{document}